\documentclass{jfm}
\usepackage{graphicx}
\usepackage{epstopdf, epsfig}
\usepackage{newtxtext}
\usepackage{newtxmath}

\usepackage{tikz}
\usetikzlibrary{arrows.meta}

\usepackage{hyperref}
\usepackage{wrapfig}
\usepackage{amsfonts}
\usepackage[ruled,vlined]{algorithm2e}
\newsavebox{\tempfig}
\usepackage[utf8]{inputenc}
\usepackage[english]{babel}
\usepackage{amsmath}
\usepackage{amssymb}
\usepackage{amsfonts} %maths
\usepackage{mathrsfs}
\usepackage{mathtools}
\usepackage{physics}
\usepackage{xfrac}
\usepackage[utf8]{inputenc}
\usepackage[english]{babel}
\usepackage[table,usenames,dvipsnames]{xcolor}
 \usepackage{textcomp}
\usepackage[ruled,vlined]{algorithm2e}
 \usepackage[table]{xcolor}
 \usepackage{multirow}
 \usepackage{mathrsfs,amsmath}
\usepackage[square,numbers]{natbib}
\usepackage{arydshln}
\usepackage[makeroom]{cancel}
\usepackage{tikz}
\usetikzlibrary{plotmarks}
\usepackage{etoolbox}
\newcommand{\be}{\begin{equation}}
\newcommand{\ee}{\end{equation}}
\usepackage{graphicx}
\usepackage{etoolbox}
\usepackage{comment}
\DeclareRobustCommand\full  {\tikz[baseline=-0.6ex]\draw[blue,thick] (0,0)--(0.5,0);}
\DeclareRobustCommand\fullgreen{\tikz[baseline=-0.6ex]\draw[green!60!black,thick] (0,0)--(0.5,0);}
\DeclareRobustCommand\fullgreenlight{\tikz[baseline=-0.6ex]\draw[green!100!black,thick] (0,0)--(0.5,0);}
\DeclareRobustCommand\fullblack  {\tikz[baseline=-0.6ex]\draw[black,thick] (0,0)--(0.5,0);}
\DeclareRobustCommand\fullred  {\tikz[baseline=-0.6ex]\draw[red,thick] (0,0)--(0.5,0);}
\DeclareRobustCommand\fullpurple  {\tikz[baseline=-0.6ex]\draw[purple,thick] (0,0)--(0.5,0);}
\DeclareRobustCommand\fullthickred  {\tikz[baseline=-0.6ex]\draw[red,very thick] (0,0)--(0.5,0);}

\DeclareRobustCommand\dashedgreen{\tikz[baseline=-0.6ex]\draw[green!60!black,thick,dashed] (0,0)--(0.54,0);}

\DeclareRobustCommand\dashedblue{\tikz[baseline=-0.6ex]\draw[blue,thick,dashed] (0,0)--(0.54,0);}

\newrobustcmd*{\mycircle}[1]{\tikz{\filldraw[draw=#1,fill=#1] (0,0) circle [radius=0.08cm];}}
\newrobustcmd*{\mytriangle}[1]{\tikz{\filldraw[draw=#1,fill=#1] (0,0)--(0.2cm,0) -- (0.1cm,0.2cm);}}
\newrobustcmd*{\mysquare}[1]{\tikz{\filldraw[draw=#1,fill=#1] (0,0)--(0.15cm,0) -- (0.15cm,0.15cm) -- (0,0.15cm);}}

\definecolor{arrowgreen}{RGB}{0,128,0}

\DeclareRobustCommand{\captionarrow}[2][]{%
  \tikz[baseline=-0.5ex]{%
    \draw[
      draw=#2,
      line width=0.8pt,
      -{Stealth[length=3pt,width=3pt]},
      #1
    ] (0,0) -- (1.8em,0);
  }%
}

\DeclareRobustCommand{\blackarrow}{\captionarrow{black}}
\DeclareRobustCommand{\bluearrow}{\captionarrow[line width=0.5pt]{blue}}
\DeclareRobustCommand{\redarrow}{\captionarrow{red}}
\DeclareRobustCommand{\greenarrow}{%
  \captionarrow[line width=1.2pt,dash pattern=on 3pt off 2pt]{arrowgreen}%
}
\newcommand{\codex}[1]{{\color{red}#1}}
\newcommand{\del}[1]{{}}

\shorttitle{Predictability for surface waves}
\shortauthor{Y. Du et al}

\title{A quantitative measure of predictability for surface gravity waves}

\author{Yifan Du\aff{1}\corresp{\email{yifandu@umich.edu}}, Kevin Maki\aff{1}, Jing Sun\aff{1} 
 \and Yulin Pan \aff{1}}

\affiliation{\aff{1} University of Michigan, Ann Arbor}

\begin{document}

\maketitle

\begin{abstract}
We introduce a Bayesian inference framework to characterize the predictability of a one dimensional ocean surface wave field given surface elevation measurements from a sensor. Unlike the classical theories that only sketch the boundary of a predictable zone based on group velocity, we use the Bayesian posterior uncertainty of surface elevation to quantify the local predictability at every spatiotemporal location. 
% YD might not be very precise for the new formulation
Our approach first constructs a prior distribution to encode existing knowledge of the wave field such as the spectrum and propagation direction. By conditioning on the (possibly noisy) measurement data, we infer the posterior distribution of the wave field, from which a spatiotemporal uncertainty map is extracted. This map agrees well with the group-velocity-based classical predictable zone theory when the frequency spectrum has bounded support. Meanwhile, it also demonstrates predictability even for a spectrum with unbounded frequency support where the classical predictable zone degenerates. 
A refined analysis of posterior uncertainty using observability Gramian matrix reveals a `data-informed' part and an `un-informed' part of the wave field, with the latter being fundamentally un-predictable. 
Furthermore, nonlinear effects can be naturally introduced into the Bayesian framework by incorporating the nonlinear ocean surface wave equations. This is the first time that nonlinearity is rigorously treated in the predictable zone problem. We perform numerical tests using both bounded, piecewise constant spectrum and unbounded JONSWAP spectrum. The numerical results confirm the proposed advantages of our method for linear waves and reveal how wave steepness affects the predictability of nonlinear waves. Nonlinear modal interactions redistribute uncertainty between the inside and the outside of the predictable zone with an amount proportional to the wave steepness. Such scaling is consistent with the triad model interaction due to quadratic nonlinearity. 
\end{abstract}

\vspace*{12pt}

%-----------------------------------------------------------------------
%                  Sec. 1 Introduction
%-----------------------------------------------------------------------

\section{Introduction}
\begin{figure}
    \centering    \includegraphics[width=1.0\textwidth]{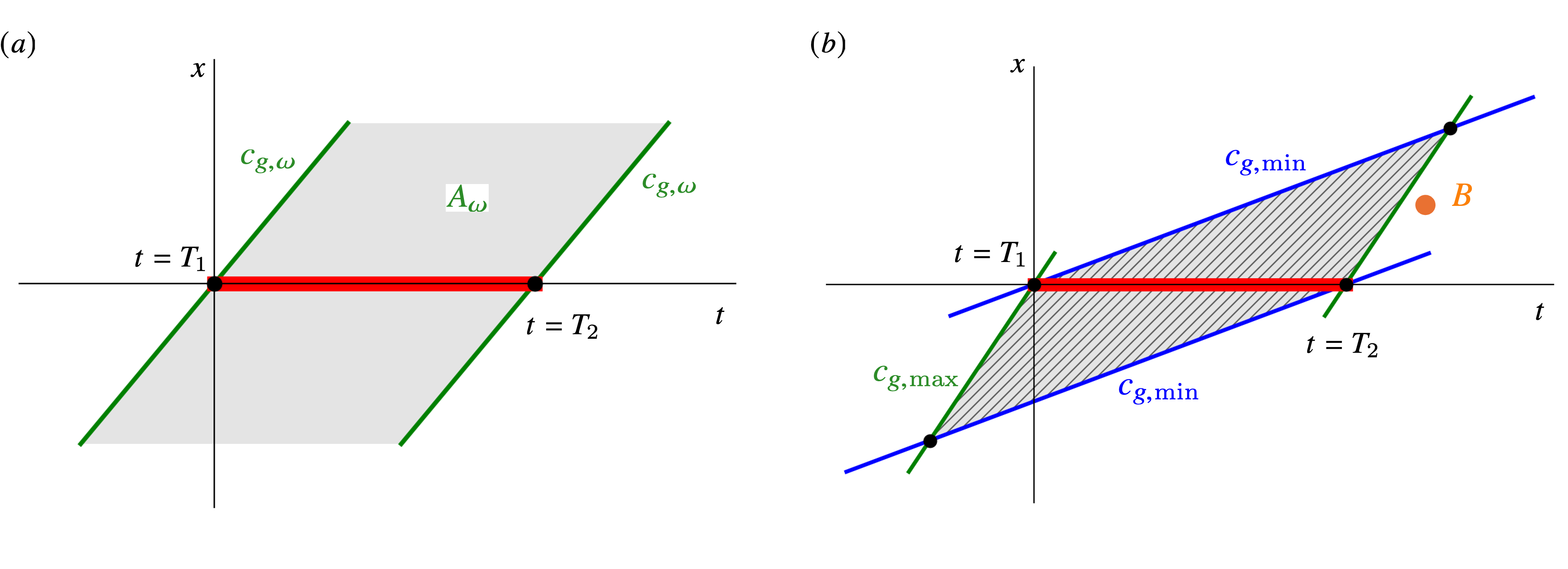}
    \caption{Schematic of the predictable zone (shaded area) in the $x-t$ plane given wave measurements on the red thick line. The waves are assumed to propagate upwards in $x$, with the group speeds $c_{g,\text{min}}$ (\full) and $c_{g,\text{max}}$ (\fullgreen{}) visualized. }
    \label{fig: schm pred zone}
\end{figure}
An important concept named `predictable zone' often appears in the interpretation of ocean surface wave measurements. Consider a one-dimensional ocean surface wave field defined on the $x$-$t$ plane $(x,t)\in\mathbb{R}^2$. Suppose there are some regions where the surface elevation $\eta(x,t)$ is measured as a function of time.
Given the wave measurements, the predictable zone \citep{morris1998short,edgar2000effects} is defined as the region in the spatiotemporal domain where the wave field can be determined using the information carried by the measured data.
%where the available measurements contain sufficient information for the wave field to be determined.
The early dispute of whether the construction of such zones should depend on phase speed \citep{morris1998short,edgar2000effects} or group speed \citep{abusedra2011prediction,naaijen2014limits} is resolved by numerical verifications \citep{blondel2010deterministic,qi2018nonlinear} and analytically constructions \citep{qi2018predictable} that support the latter.
%The early constructions by \citet{morris1998short} and \citet{edgar2000effects} are based on phase speed, while later works by \citet{abusedra2011prediction} and \citet{naaijen2014limits} recognize group speed as the relevant quantity. Although the applicability of stationary phase approximation in \citet{abusedra2011prediction} and \citet{naaijen2014limits} are not well established, the predictable zone based on group speed has been numerically verified by \citet{blondel2010deterministic,qi2018nonlinear} and analytically constructed by \citet{qi2018predictable}.

A schematic of a predictable zone is shown in figure \ref{fig: schm pred zone}, with surface elevation measured at $x=0$ and the one-dimensional waves propagating in the $+x$ direction. The predictable zone  $A_{\omega}$ for a specific frequency $\omega$ is geometrically constructed as the region between two parallel lines determined by the group speed $c_{g,\omega}$.
For a compactly supported wave frequency spectrum in an interval $[\omega_{\text{min}},\omega_{\text{max}}]$, the region where all wave components are predictable is the intersection of $A_{\omega}$ for all $\omega \in[\omega_{\text{min}},\omega_{\text{max}}]$. This intersection produces a region on the spatiotemporal plane bounded by lines representing the maximum and minimum group speed, which is shown in \ref{fig: schm pred zone}$(b)$.

The existing constructions of the predictable zone do not characterize regions where the wave is partially predictable (consider point $B$ in figure \ref{fig: schm pred zone}$(b)$ being apparently predictable for $c_{g,\text{min}}$). In addition, the area of the predictable zone shrinks towards zero if we consider a wave spectrum with unbounded frequency support (where the group velocities $c_{g,\text{min}}\to 0$ and $c_{g,\text{max}}\to \infty$). These limitations motivate a continuous variable describing the zonal predictability rather than the categorical one used in previous works. Such a `continuous predictability variable' should in principle depend on the wave spectrum, since the predictable waves at a certain point is on a subset of the total frequency space, and the percentage of predictable waves in the overall wave spectrum depend on the detailed energy distribution across frequencies. %For an intuitive example, still consider point $B$ in figure \ref{fig: schm pred zone}. If the wave spectrum is mostly concentrated in its high-frequency (low speed) end, then the wave field at point $B$ should be largely predictable. Bayesian
There have been recent effort to define similar generalizations of predictable zones \citep{merigaud2025assessing}. However, \citet{merigaud2025assessing} adopt purely statistical approaches that model the surface elevation fields as spatiotemporal random processes with certain correlations, without taking into account the deterministic wave evolutions governed by fluid dynamical equations.

Here we introduce a Bayesian inversion framework to address the aforementioned issue. We introduce a prior wave distribution prescribed by a wave spectrum. The wave spectrum is not limited in its form or band width. Conditioned on the measurement data (possibly noisy), we formulate the inference of the posterior distribution of the wave solution as a Bayesian inverse problem. Each sample solution from the posterior distribution is governed by deterministic (linear or nonlinear) wave equations. From this posterior distribution, the most likely wave field and the posterior uncertainty can be extracted. This uncertainty is by nature a continuous quantity that can describe `partial predictability', where smaller (larger) uncertainty can be interpreted as the wave being more (less) predictable.

Compared with the classical predictable zone, this posterior uncertainty has the following advantages. For wave spectra with compact supports in the frequency space, the posterior uncertainty not only reproduces a predictable zone the same as the classical approach, but also quantifies the confidence of wave prediction inside and outside of the predictable zone. For unbounded spectra such as the JONSWAP spectrum, the classical approach fails and produces an empty predictable zone, while the posterior inference still provides an uncertainty map on the $x$-$t$ plane that serves as a quantitative measure of predictability. In both cases, measurement noise can lead to the increase of posterior uncertainty, and is naturally incorporated in the Bayesian inversion framework.

%With the Bayesian posterior uncertainty established as a quantitative measure of predictability, we can re-define `predictable solution' more rigorously than before. The works by \citet{morris1998short, abusedra2011prediction, qi2018predictable} decompose the wave solution into predictable and un-predictable parts into separated spatiotemporal regions (see figure \ref{fig: schm pred zone}). 
%Such definition inevitably face difficulty because of the `partially predictable' locations we have discussed. Alternatively, we can consider the predictable and unpredictable solution to co-exist at a certain spatiotemporal locations, and their summation to be the total wave field. 

With the Bayesian posterior uncertainty established as a quantitative measure of predictability, we can re-define the `predictable solution' more rigorously than before. One interpretation of the earlier works of the predictable zones by \citet{qi2018predictable, morris1998short} (and many others) is a split of the full wave field into predictable and un-predictable parts. In their works such split is in terms of zonal locations, where the solutions in one region is predictable while those in other regions are not. Such split inevitably face difficulty because of the `partially predictable' locations we have discussed. Alternatively, we can define the predictable and unpredictable solution to be the parts of solution that have zero and nonzero posterior uncertainty, respectively, where these `parts' are defined in terms of linear superposition rather than zonal locations. These two parts usually co-exist at a spatiotemporal location, and their relative size determines whether the wave field near this is more or less predictable, i.e., a larger portion of predictable part lead to more predictable wave field at that location. Such re-definition is not only mathematically more rigorous, 
but also necessary for analyzing the zero-noise limit of the problem, as we will show.    %As we will show, these two parts of solution usually co-exist at a certain spatiotemporal locations, and their summation is the total wave field. 

In addition to the advantages and capabilities of the Bayesian framework discussed before, the effect of the nonlinearity can also be introduced into the framework by adopting nonlinear wave models. The Bayesian inversion methods have been developed for both linear \citep{bui2013computational} and nonlinear \citep{petra2014computational} partial differential equations, and applied in various physical problems including global seismic inversions and ice-sheet flows. Unlike the linear systems where the posterior uncertainty can be analytically formed \citep{williams2006gaussian}, the posterior covariance for nonlinear systems require approximations such as Markov Chain Monte Carlo \citep{petra2014computational}, variational inference \citep{che2025stable}, and Laplace approximation \citep{kalmikov2014hessian, yoko2024adjoint}. The Laplace approximation is computational appealing and provides accurate estimation of the posterior covariance for weakly nonlinear system, therefore is particularly appropriate for the predictable zone problem.

With this aforementioned advantage, we can investigate the effect of nonlinearity on the predictability of the wave field more rigorously than earlier works. The theoretical constructions \citep{qi2018predictable} does not take into account the effect of nonlinearity. A later study \citep{stuhlmeier2021deterministic} identifies a change of the group-velocity based boundary of the predictable-zone due to the nonlinear correction of the dispersion relation. Such correction typically leads to a slight shift of the predictable zone with the increase of wave steepness. 
However, even if the nonlinear correction of group velocity is included, the construction of predictable zone itself is still based on the independent traveling of different wave components at different speeds, which is a linear assumption. Consequently, this incorporation of nonlinearity by \citet{stuhlmeier2021deterministic} is still incomplete. 
Some more fundamental question to ask are: (1) does the predictability inside the predictable zone change with the level of nonlinearity? (2) does the concept of predictable zone itself become less well-defined as the nonlinearity level increases? These are the questions we rigorously address using the Bayesian inversion framework.

The structure of this work is arranged as follows.
The formal definition of the predictability problem is first introduced in section \S\ref{sec: problem setup}. The Bayesian inversion for linear wave fields is then demonstrated in \S\ref{sec: linear waves} including the theory (\S\ref{subsec: linear wave dynamics}-\ref{subsec: uncertainty decomposition}) and the results (\S\ref{subsec: results linear}). We then present the extension of the Bayesian inversion methods to nonlinear wave dynamics in \S\ref{subsec: nonlinear dynamics}-\ref{subsec: method nonlinear}, the numerical implementations in \S\ref{subsec: numerical methods} and the results in \S \ref{subsec: results nonlinear}.

%------------------------------------------------------------------------------
\section{Problem setup}\label{sec: problem setup}

\begin{figure}
    \centering    \includegraphics[width=1.0\textwidth]{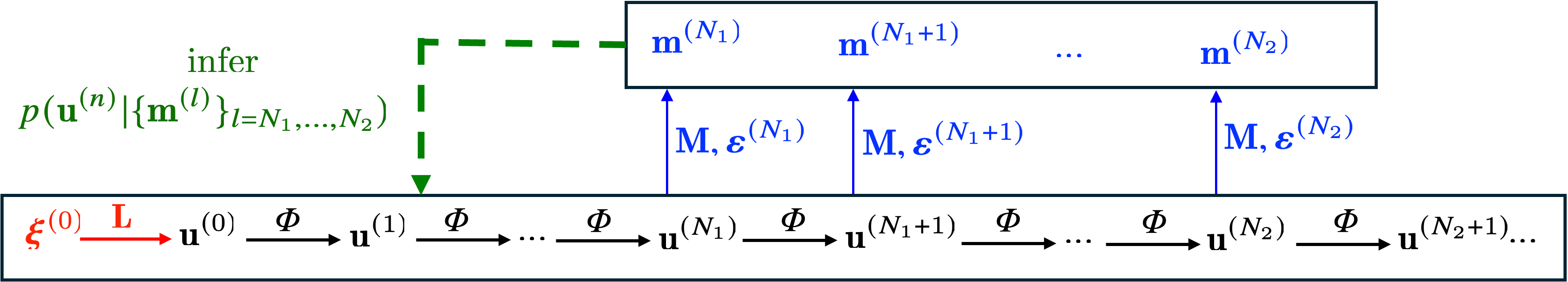}
    \caption{Schematic of the state space model, including the forward mapping of the wave dynamics (\blackarrow),
measurement operator (\bluearrow),
the mapping $\mathbf{L}$ from initial Fourier coefficients $\boldsymbol{\xi}^{(0)}$ to initial wave state $\mathbf{u}^{(0)}$ (\redarrow, see \S \ref{subsec: Bayesian linear}),
and the Bayesian inference problem (\greenarrow).}
    \label{fig: schm state space}
\end{figure}

Consider a one-dimensional wave field with an energy spectrum $S(k)$.
%The surface elevation $\eta(x,t)$ is continuously measured as functions of time in $t\in[T_1,T_2]$ at a set of measurement locations $\mathcal{M}=\left\{x_{m,1},...,x_{m, M }\right\}\subseteq \mathbb{R}$.
The temporal evolution of the wave field and the measurements can be compactly written in the following discretized system,
\begin{eqnarray}\label{eq:nonlinear state space model}
    \label{eq:state}
    \mathbf{u}^{(n)}&=&\Phi(\mathbf{u}^{(n-1)})  \\
	\label{eq:obs}
	\mathbf{m}^{(n)}&=&\mathbf{M}\mathbf{u}^{(n)}+\boldsymbol{\varepsilon}^{(n)},
\end{eqnarray}
where the wave state vector $\mathbf{u}^{(n)} = \left[\eta(x_1,t^{(n)}), ..., \eta(x_{J},t^{(n)}), \psi(x_1,t^{(n)}), ..., \psi(x_{J},t^{(n)})\right]^{\dagger}$ contains the values of surface elevation $\eta$ and surface velocity potential $\psi$ on a set of uniformly distributed grid points $\mathcal{G} = \left\{x_j\in [0,L]|j=1,2,...,J\right\}$ at time $t^{(n)}$ with $n=1,2,...$. The operator $\Phi$ is a mapping that defines the wave dynamics from $t^{(n-1)}$ to $t^{(n)}$. 
The measurement operator $\mathbf{M}$ extracts the clean measurements from the state $\mathbf{u}^{(n)}$, such that $\mathbf{M}\mathbf{u}^{(n)}=\left[\eta(x_{m,1},t^{(n)}),...,\eta(x_{m, M },t^{(n)})\right]^{\dagger}$ contains the surface elevations at the sensor locations $\mathcal{M}=\left\{x_{m,1},...,x_{m, M }\right\}\subseteq [0,L]$. The measurements $\mathbf{m}^{(n)}$ are available in a time window $N_{1}\leq n\leq N_{2}$ and are contaminated by independent, identically distributed Gaussian noise $\boldsymbol{\varepsilon}^{(n)}\sim N(\mathbf{0},\mathbf{R})$. A schematic demonstration of the system (\ref{eq:state}-\ref{eq:obs}) is shown in figure \ref{fig: schm state space}.
Given this setup, our goal is to infer the probability distribution $p(\mathbf{u}^{(n)}|\{\mathbf{m}^{(l)}\}_{l=N_{1},...,N_{2}})$ conditioned on the measurement data.

\section{Predictability for linear surface wave fields}\label{sec: linear waves}
\subsection{Linear wave dynamics}\label{subsec: linear wave dynamics}
We start from a simplified setting of linear surface waves. In this case, the wave evolution equation (\ref{eq:state}) is replaced by the linear dynamics $\mathbf{u}^{(n)}=\mathbf{A}\mathbf{u}^{(n-1)}$, where $\mathbf{A}\in\mathbb{R}^{2J}$ is a matrix. We obtain $\mathbf{A}$ by first deriving the one-time-step propagator of the Fourier coefficients $\hat{\eta}_k^{(n)}$ and $\hat{\psi}_k^{(n)}$ with wave number $k$, and then transform it back to the physical space.
The continuous time evolution of a Fourier coefficient pair $\hat{\eta}_k(t), \hat{\psi}_k(t)$ is
\begin{equation}\label{eq:linear wave dynamics}
\frac{\mathrm{d}}{\mathrm{d}t}
\begin{pmatrix}
\hat{\eta}_k \\
\hat{\psi}_k
\end{pmatrix}
=
\begin{pmatrix}
0 & |k| \\
-g & 0
\end{pmatrix}
\begin{pmatrix}
\hat{\eta}_k \\
\hat{\psi}_k
\end{pmatrix},  
\end{equation}
where the gravitational constant is set to $g=1$. The one-step propagator during a time step $\Delta t$ for the Fourier coefficients $\hat{\eta}_k^{(n)}$ and $\hat{\psi}_k^{(n)}$ is,
\begin{equation}\label{eq: propagator single k}
\begin{pmatrix}
\hat{\eta}_k^{(n+1)} \\
\hat{\psi}_k^{(n+1)}
\end{pmatrix}
=
\exp\left[\begin{pmatrix}
0 & |k| \\
-g & 0
\end{pmatrix}\Delta t\right]\begin{pmatrix}
\hat{\eta}_k^{(n)} \\
\hat{\psi}_k^{(n)}
\end{pmatrix}
=
\begin{pmatrix}
c_k & b_k \\
d_k & c_k
\end{pmatrix}
\begin{pmatrix}
\hat{\eta}_k^{(n)} \\
\hat{\psi}_k^{(n)}
\end{pmatrix},
\end{equation}
where $g$ is the gravitational constant and the coefficients $c_k$, $b_k$ and $d_k$ are computed by evaluating the matrix exponential (see appendix \ref{app: matrix exponential}), 
\begin{equation}\label{eq: linear evolution coeffs}
c_k = \cos(\omega_k\Delta t),
\qquad
b_k =
\frac{|k|}{\omega_k}
\sin(\omega_k\Delta t),
\qquad
d_k =
-\frac{g}{\omega_k}
\sin(\omega_k\Delta t).
\end{equation}
The frequency $\omega_k=\sqrt{g|k|}$ satisfies the deep-water dispersion relation. The evolutions of all Fourier modes are decoupled and described by (\ref{eq: propagator single k})-(\ref{eq: linear evolution coeffs}). These formulae of all individual wavenumber $k$ implicitly define an operator $\hat{\mathbf{A}}$ that evolves all Fourier modes together for one time step. 
\begin{equation}\label{eq: linear propagator Fourier physical}
    \hat{\mathbf{A}}
=
\begin{pmatrix}
\mathbf{C} & \mathbf{B} \\
\mathbf{D} & \mathbf{C}
\end{pmatrix}, 
\end{equation}
where the matrices $\mathbf{B}$, $\mathbf{C}$ and $\mathbf{D}$ are diagonal matrices containing the coefficients $b_k$,  $c_k$ and $d_k$ on their diagonals, respectively. Subsequently, an evolution operator $\mathbf{A}$ in the physical space can be defined by transforming $\hat{\mathbf{A}}$ into physical space using the Fourier transform matrix $\mathbf{F}$,
\begin{equation}
    \mathbf{A}
=
\begin{pmatrix}
\mathbf{F}^{-1} & 0 \\
0 & \mathbf{F}^{-1}
\end{pmatrix}
\hat{\mathbf{A}}
\begin{pmatrix}
\mathbf{F} & 0 \\
0 & \mathbf{F}
\end{pmatrix}. 
\end{equation}
%\codex{[Check: Specify the Fourier-series/DFT convention for $\mathbf{F}$: $k_m=m\Delta k$, $\Delta k=2\pi/L$, its normalization, and the retained index range relative to $J$. These choices determine the units of $\hat{\eta}_{k_m}$ and the spectral factors in $\mathbf{L}$.]}

\subsection{Bayesian framework for linear wave dynamics}\label{subsec: Bayesian linear}
%We describe the basic physical setup and the predictability problem in \S \ref{subsec: problem setup}.
To solve the predictability problem defined in \S \ref{sec: problem setup} for linear waves, a Bayesian inversion method is first introduced with its main ingredients.
%We then present an additional component of the framework in \S\ref{subsec: Treatment of singular prior} to incorporate singular priors that often appear in applications.
A refined analysis of the structure of the posterior uncertainty and its dependence on noise level is then demonstrated in \S \ref{subsec: uncertainty decomposition}.
%\subsubsection{Bayesian posterior inference for linear wave dynamics}\label{subsec: Gramian and Bayesian posterior}
%A Bayesian inversion framework is adopted to estimate the posterior distribution of $\mathbf{u}^{(0)}$ conditioned on the measurement data, from which the \textit{maximum a posterior} (MAP) initial condition and the posterior covariance can be extracted and propagated to the whole time window using the wave dynamics.

We introduce a Bayesian inversion framework to estimate the posterior distribution $p(\mathbf{u}^{(n)}|\{\mathbf{m}^{(l)}\}_{l=N_{1},...,N_{2}})$ for all $n>0$. Since the wave field evolves in time according to the dynamics (\ref{eq:state}), any solution trajectory $\{\mathbf{u}^{(n)}\}_{n> 0}$ can be determined by the initial condition $\mathbf{u}^{(0)}$, and therefore determined by the initial Fourier coefficients of surface elevation $\hat{\eta}_k^{(0)}$. For a wave spectrum $S(k)$ which is nonzero on an interval $0\leq k_{\mathrm{min}}\leq k\leq k_{\mathrm{max}}$, we can define a control vector $\boldsymbol{\xi}^{(0)}$ in the following form, 
\begin{equation}\label{eq: control vector}
    \boldsymbol{\xi}^{(0)}=[a_{k_{\mathrm{min}}}, a_{k_{\mathrm{min}}+1}, ..., a_{k_{\mathrm{max}}}, a_{k_{\mathrm{min}}}^{*}, a_{k_{\mathrm{min}}+1}^{*}, ..., a_{k_{\mathrm{max}}}^{*}], 
\end{equation}
where $a_k=\hat{\eta}_k^{(0)}/\sqrt{S(k)/2}$ are the normalized Fourier coefficients and $a_k^*$ are their complex conjugates.
%contains all of the `normalized' Fourier coefficients $a_k=\hat{\eta}_k/\sqrt{S(k)}$ and their complex conjugate $a_k^{*}$ for $k\in\mathcal{K}$ at initial time $n=0<N_1$, where $\mathcal{K}$ contains all wavenumbers such that $S(k)\neq 0$. 
We use $r_0=2(k_{\mathrm{max}}-k_{\mathrm{min}}+1)$ to denote the number of dimensions of this control vector $\boldsymbol{\xi}^{(0)}$, and each active wave mode corresponds to two dimensions. The rest of the Fourier coefficients $\hat{\eta}_k^{(0)}$ where $S(k)=0$ is not considered in the control vector $\boldsymbol{\xi}^{(0)}$, since these modes do not exist in the wave system and have no effect on the uncertainty of the wave field. 
In general, $S(k)$ is not limited to such a `compactly supported' form. For a general $S(k)$, $\boldsymbol{\xi}^{(0)}$ could still be constructed to contain the $a_k$ and $a_k^{*}$ for all wavenumbers $k$ with $S(k)\neq 0$. 
Under this setting, the problem of inferring the distribution of the whole trajectory $p(\mathbf{u}^{(n)}|\{\mathbf{m}^{(l)}\}_{l=N_{1},...,N_{2}})$ (for all $n$) is reduced to inferring the distribution of the control vector $p(\boldsymbol{\xi}^{(0)}|\{\mathbf{m}^{(l)}\}_{l=N_{1},...,N_{2}})$. We can first estimate $p(\boldsymbol{\xi}^{(0)}|\{\mathbf{m}^{(l)}\}_{l=N_{1},...,N_{2}})$, and then propagate it along the wave dynamics to obtain $p(\mathbf{u}^{(n)}|\{\mathbf{m}^{(l)}\}_{l=N_{1},...,N_{2}})$
%\codex{[Check: Once physical wavenumbers are written as $k_m=m\Delta k$, express the control dimension using the integer indices, $r_0=2(m_{\max}-m_{\min}+1)$. The displayed expression in $k_{\min}$ and $k_{\max}$ is valid only when $\Delta k=1$.]}

Bayes' formula, $\text{posterior} = \text{prior} \times \text{likelihood}/\text{normalization}$, estimates the posterior distribution of $\boldsymbol{\xi}^{(0)}$ (and $\mathbf{u}^{(n)}$) through two steps. First, a prior distribution is constructed that encodes existing knowledge of the wave field, here specifically the energy spectrum and the wave traveling direction. Second, a likelihood function updates the prior knowledge into the posterior distribution using the measurement data.
We present these key components in this section, with the full derivation collected in appendix \ref{app:posterior_gaussian}.

We first construct a prior Gaussian distribution for the control vector $\boldsymbol{\xi}^{(0)}$. Since the coefficients $a_k$ in $\boldsymbol{\xi}^{(0)}$ are the normalized Fourier coefficients with $O(1)$ size, we can specify a standard (complex) Gaussian distribution as the prior distribution of $\boldsymbol{\xi}^{(0)}$,
\begin{equation}\label{eq: prior}
    p(\boldsymbol{\xi}^{(0)})=\frac{1}{C_0}\mathrm{e}^{-\mathcal{J}_0(\boldsymbol{\xi}^{(0)})}, \qquad \mathcal{J}_0(\boldsymbol{\xi}^{(0)})=\frac{1}{2}\left\Vert\boldsymbol{\xi}^{(0)}\right\Vert^2_{\left(\mathbf{\Sigma}^{(0)}_{\xi}\right)^{-1}}
\end{equation}
where $\left\Vert\mathbf{a}\right\Vert_{\mathbf{B}}^2=\mathbf{a}^{\dagger}\mathbf{B}\mathbf{a}$ represents the squared weighted norm of a complex vector $\mathbf{a}$ by a Hermitian matrix $\mathbf{B}$, and we choose the prior covariance $\mathbf{\Sigma}^{(0)}_{\xi}=\mathbf{I}_{r_0\times r_0}$ to be an identity matrix. The constant $C_0$ guarantees unit mass of the distribution $p(\boldsymbol{\xi}^{(0)})$. Although the prior distribution $p(\boldsymbol{\xi}^{(0)})$ in (\ref{eq: prior}) does not explicitly contain $S(k)$ and wave traveling direction, such information is indeed incorporated in the prior distribution $p(\mathbf{u}^{(n)})$ of the wave state $\mathbf{u}^{(n)}$. To derive  $p(\mathbf{u}^{(n)})$, we first construct the mapping from control vector $\boldsymbol{\xi}^{(0)}$ to $\mathbf{u}^{(n)}$,
%\codex{[Check: The main text calls $\boldsymbol{\xi}^{(0)}=[a_k,a_k^*]$ a complex Gaussian vector, whereas appendix \ref{app: nonlinear posterior derivation} evaluates derivatives in $r_0$ independent real coordinates. State this representation change explicitly and use its covariance and inner product consistently.]}
\begin{equation}\label{eq: xi to u}
    \mathbf{u}^{(n)} = \mathbf{A}^n\mathbf{u}^{(0)}=\mathbf{A}^n\mathbf{L}\boldsymbol{\xi}^{(0)}.   
\end{equation}
Here the superscript $n$ denotes a matrix power, whereas the parenthesized superscript $(n)$ denotes a time-step index. This expression (\ref{eq: xi to u}) first constructs the full initial wave state $\mathbf{u}^{(0)}=\mathbf{L}\boldsymbol{\xi}^{(0)}$, then marches it $n$ times forward to a later state $\mathbf{u}^{(n)}$. The linear operator $\mathbf{L}$ from $\boldsymbol{\xi}^{(0)}$ to $\mathbf{u}^{(0)}$ is defined implicitly by, 
\begin{equation}\label{eq: L definition}
\begin{aligned}
    \eta(x_j,0)
    &=
    \sum_{k=k_{\mathrm{min}}}^{k_{\mathrm{max}}}
    \sqrt{\frac{S(k)}{2}}\,
    \left\{a_k\mathrm{e}^{\mathrm{i}kx_j}+a_k^{*}\mathrm{e}^{-\mathrm{i}kx_j}\right\},\\
    \psi(x_j,0)
    &=
    \sum_{k=k_{\mathrm{min}}}^{k_{\mathrm{max}}}
    -\frac{i\omega_k}{|k|}\sqrt{\frac{S(k)}{2}}\,
    \left\{ a_k\mathrm{e}^{\mathrm{i}kx_j}+a_k^*\mathrm{e}^{-\mathrm{i}kx_j}\right\},   
\end{aligned}
\end{equation}
which encodes the energy spectrum $S(k)$  and $+x$ propagation direction. 
%\codex{[Check: Specify whether $S(k)$ is a one-sided discrete modal variance or a continuous one-sided wavenumber density. Under the density convention, write $k_m=m\Delta k$, use $\sqrt{S(k_m)\Delta k/2}$ in both lines of (\ref{eq: L definition}), and define $a_{k_m}=\hat{\eta}_{k_m}^{(0)}/\sqrt{S(k_m)\Delta k/2}$ in (\ref{eq: control vector}); then $\mathbb{E}[\eta^2]=\sum_{m>0}S(k_m)\Delta k$. If $S$ is already modal variance, retain the present factors and state that convention.]}
Using the linear mapping (\ref{eq: xi to u}), we transform the Gaussian prior distribution $p(\boldsymbol{\xi}^{(0)})=N(0,\mathbf{\Sigma}^{(0)}_{\boldsymbol{\xi}})$ to a Gaussian prior distribution $p(\mathbf{u}^{(n)})=N(0,\mathbf{\Sigma}^{(n)})$ with a covariance $\mathbf{\Sigma}^{(n)} = \mathbf{A}^n\mathbf{L} \mathbf{\Sigma}^{(0)}_{\xi}\mathbf{L}^{\dagger}\mathbf{A}^{n,\dagger}=\mathbf{A}^n\mathbf{L}\mathbf{L}^{\dagger}\mathbf{A}^{n,\dagger}=\mathbf{L}\mathbf{L}^{\dagger}$. The last equality naturally arises due to the invariance, or stationarity of the prior covariance along the linear wave evolution. 
%YD Might need to show the invariance in appendix. 
By this construction, the information of energy spectrum $S(k)$  and propagation direction is also encoded in the prior covariance $\mathbf{\Sigma}^{(n)}$ of the wave state $\mathbf{u}^{(n)}$. 
Since the wave state $\mathbf{u}^{(n)}$ has two parts representing the surface elevation field $\eta$ and velocity potential $\psi$ respectively, its covariance $\mathbf{\Sigma}^{(n)}$ has a $2\times2$ structure that contains $\eta\eta$, $\eta\psi$ and $\psi\psi$ blocks.

By the Wiener-Khinchin theorem \citep{brockwell2002introduction}, the constructions (\ref{eq: prior}-\ref{eq: L definition}) implies that the fields $\eta(x,t)$ and $\psi(x,t)$ are both modeled as stationary Gaussian processes in $x$ direction at any time. We therefore use a prior standard deviation $\varrho=\sqrt{\sum_{k>0}S(k)}=\sqrt{\mathbf{\Sigma}^{(n)}_{jj}}$ (for $1\leq j\leq J$, $n\ge0$) which is constant in both space and time, to quantify the uncertainty of the surface elevation field $\eta(x,t)$ at any $(x,t)$ before taking any measurements.

We move forward to the next step of the Bayesian framework, which is the construction of the likelihood function for $\boldsymbol{\xi}^{(0)}$. For a given initial vector of Fourier coefficients $\boldsymbol{\xi}^{(0)}$, the wave system predicts the noise-free measurement $\mathbf{M}\mathbf{A}^{n}\mathbf{L}\boldsymbol{\xi}^{(0)}$ (which we subsequently call `model prediction') at the $n^{\text{th}}$ time step. 
%Here, $\mathbf{L}$ maps $\boldsymbol{\xi}^{(0)}$ to the initial wave state $\mathbf{u}^{(0)}$, $\mathbf{A}^{n}$ advances $\mathbf{u}^{(0)}$ to $\mathbf{u}^{(n)}$, and $\mathbf{M}$ extracts the (noise-free) measurement from $\mathbf{u}^{(n)}$. The mapping $\mathbf{L}$ is defined implicitly by
%\begin{equation}\label{eq: L definition}
%    \eta(x_j,0) = \sum_k\sqrt{S(k)}a_k\mathrm{e}^{ikx_j},\qquad \psi(x_j,0)=\sum_k\frac{1}{|k|}\sqrt{S(k)}a_k\mathrm{e}^{ikx_j}.
%\end{equation}
Measurement noise causes measurements $\mathbf{m}^{(n)}$ to fluctuate around the model predictions $\mathbf{M}\mathbf{A}^{n}\mathbf{L}\boldsymbol{\xi}^{(0)}$ (see equation (\ref{eq:obs})). Thus, even for a fixed $\boldsymbol{\xi}^{(0)}$, the measurements are random, and their probability density is determined by the noise distribution as, 
\begin{equation}\label{eq: linear likelihood}
    p(\{\mathbf{m}^{(l)}\}_{l=N_{1},...,N_{2}}|\boldsymbol{\xi}^{(0)})\propto\mathrm{e}^{-\mathcal{J}_{\text{m}}(\boldsymbol{\xi}^{(0)})}, \quad \mathcal{J}_\text{m}(\boldsymbol{\xi}^{(0)}) = \sum_{n=N_{1}}^{N_{2}}\frac{1}{2}\left\Vert\mathbf{M}\mathbf{A}^n\mathbf{L}\boldsymbol{\xi}^{(0)}-\mathbf{m}^{(n)}\right\Vert^2_{\mathbf{R}^{-1}}.
\end{equation}
The expression $p(\{\mathbf{m}^{(l)}\}_{l=N_{1},...,N_{2}}|\boldsymbol{\xi}^{(0)})$ 
% is viewed as a function of $\boldsymbol{\xi}^{(0)}$ and 
is called the likelihood of $\{\mathbf{m}^{(l)}\}_{l=N_{1},...,N_{2}}$ conditioned on $\boldsymbol{\xi}^{(0)}$. The likelihood quantifies how well the model predictions generated from a certain Fourier coefficient vector $\boldsymbol{\xi}^{(0)}$ agree with the observed measurements $\{\mathbf{m}^{(l)}\}_{l=N_1,...,N_2}$: a smaller misfit $\mathcal{J}_{\mathrm{m}}$ corresponds to a higher likelihood. 
%Since the predicted measurements depend linearly on $\boldsymbol{\xi}^{(0)}$, $\mathcal{J}_{\mathrm{m}}$ is quadratic in $\boldsymbol{\xi}^{(0)}$.

By Bayes formula, the posterior distribution of $\boldsymbol{\xi}^{(0)}$ given measurements $\{\mathbf{m}^{(l)}\}_{l=N_{1},...,N_{2}}$ is constructed by combining the prior (\ref{eq: prior}) and the likelihood (\ref{eq: linear likelihood}) using $\text{posterior} = \text{prior} \times \text{likelihood}/ \text{normalization}$,
\begin{equation}\label{eq:linear posterior}
    p(\boldsymbol{\xi}^{(0)}|\{\mathbf{m}^{(l)}\}_{l=N_{1},...,N_{2}}) = \frac{1}{C}\mathrm{e}^{-\mathcal{J}_{\mathrm{m}}(\boldsymbol{\xi}^{(0)})-\mathcal{J}_0(\boldsymbol{\xi}^{(0)})}. 
\end{equation}
%Since the prior and the noise satisfy Gaussian distributions and the wave dynamics are linear,
Since $\mathcal{J}_0$ and $\mathcal{J}_\text{m}$ are both quadratic with respect to $\boldsymbol{\xi}^{(0)}$, the exponential part $-\mathcal{J}_{\mathrm{m}}(\boldsymbol{\xi}^{(0)})-\mathcal{J}_0(\boldsymbol{\xi}^{(0)})$ in (\ref{eq:linear posterior}) remains quadratic. Consequently, we can obtain the following explicit expressions of the posterior Gaussian distribution by a `completion of square' derivation (see appendix \ref{app:posterior_gaussian}),
\begin{align}\label{eq: linear posterior centered}
p(\boldsymbol{\xi}^{(0)}|\{\mathbf{m}^{(l)}\}_{l=N_{1},...,N_{2}})
=
\frac{1}{\tilde C}
\exp\left[
-\frac12
(\boldsymbol{\xi}^{(0)}-\tilde{\boldsymbol{\xi}}^{(0)})^{\dagger}
(\tilde{\mathbf{\Sigma}}^{(0)}_{\boldsymbol{\xi}})^{-1}
(\boldsymbol{\xi}^{(0)}-\tilde{\boldsymbol{\xi}}^{(0)})
\right].
\end{align}
The normalization constant $\tilde{C}$ guarantees unit mass of the posterior distribution.
This distribution (\ref{eq: linear posterior centered}) is centered around the maximum \textit{a posteriori} (MAP) solution $\tilde{\boldsymbol{\xi}}^{(0)}$ and has posterior covariance $\tilde{\mathbf{\Sigma}}^{(0)}_{\boldsymbol{\xi}}$ (see full derivation in appendix \ref{app:posterior_gaussian}),
\begin{equation}
\label{eq: posterior mean cov}
\begin{aligned}
\tilde{\mathbf{\Sigma}}^{(0)}_{\boldsymbol{\xi}}
&=
\left(\mathbf{I}_{r_0\times r_0}+\mathbf{G}\right)^{-1},
\\
\tilde{\boldsymbol{\xi}}^{(0)}
&=
\tilde{\mathbf{\Sigma}}^{(0)}_{\boldsymbol{\xi}}
\sum_{n=N_{1}}^{N_{2}}
\mathbf{L}^{\dagger}(\mathbf{A}^n)^{\dagger}\mathbf{M}^{\dagger}
\mathbf{R}^{-1}\mathbf{m}^{(n)},
\end{aligned}
\end{equation}
where we use the tilde symbol $\tilde{(\cdot)}$ to denote the parameters (mean and covariances) of the posterior distribution. 
The matrix $\mathbf{G}$ has the following expression
%, and is the $2^{nd}$ order derivative of $\mathcal{J}_m$ with respect to $\mathbf{u}^{(0)}$,
\begin{equation}\label{eq: linear observability Gramian}
    \mathbf{G}
    =
    \sum_{n=N_{1}}^{N_{2}}
    \mathbf{L}^{\dagger}(\mathbf{A}^n)^{\dagger}\mathbf{M}^{\dagger}\mathbf{R}^{-1}\mathbf{M}\mathbf{A}^n\mathbf{L}.
\end{equation}
%In control theory, this matrix $\mathbf{G}$ is originally defined as the observability Gramian that describes whether a dynamical system is observable given a sensor layout \citep{chen1984linear}. 
In control theory, this matrix $\mathbf{G}$ is originally defined as the observability Gramian that characterizes how accurately each element in the state vector in a dynamical system can be determined given the constraint of measurement data \citep{chen1984linear}. When a certain part of state vector can be determined with high accuracy, their uncertainty after taking the measurement must have been reduced significantly compared to that before taking the measurement.  
Consequently, we use $\mathbf{G}$ to characterize how the measurements contribute to the reduction of uncertainty in the wave field. Since $\mathbf{G}$ is positive semidefinite, the expressions in (\ref{eq: posterior mean cov}) show that the posterior covariance $\tilde{\mathbf{\Sigma}}^{(0)}_{\boldsymbol{\xi}}$ is always smaller or equal to the prior covariance $\mathbf{I}_{r_0\times r_0}$ \footnote{This `smaller' is in the sense of semidefinite order, i.e., $\mathbf{A}\preceq\mathbf{B}$ is defined by $\mathbf{B}-\mathbf{A}$ being positive semi-definite.}. 
%-----------------Explanation of the matrix comparison-----------------------------
%To understand what this `smaller' means, we can consider any unit vector $\mathbf{e}\in\mathbb{R}^{2J}$ and the projection $\mathbf{u}^{(0)}\cdot\mathbf{e}$ (a scalar quantity). The `smaller' is in the sense that the posterior variance of this projection $\mathbb{E}[(\mathbf{u}^{(0)}\cdot\mathbf{e})^2|\{\mathbf{m}^{(l)}\}_{l=N_{1},...,N_{2}}]= \mathbf{e}^{\dagger}\tilde{\mathbf{\Sigma}}^{(0)} \mathbf{e}$ is always reduced or equal to the prior variance $\mathbb{E}[(\mathbf{u}^{(0)}\cdot\mathbf{e})^2]=\mathbf{e}^{\dagger}\mathbf{\Sigma}^{(0)} \mathbf{e}$ . One can consider, for example, $\mathbf{e}$ to be a unit vector with the $j$-$th$ ($1\le j \le J$) element being equal to $1$ and other elements being zero. In such a case, the inequality becomes $\mathbb{E}[(\eta(x_j,0))^2|\{\mathbf{m}^{(l)}\}_{l=N_{1},...,N_{2}}] \le \mathbb{E}[(\eta(x_j,0))^2]$, i.e., the posterior variance of initial surface elevation $\eta(x_j,0)$ at $x_j$ is smaller than its prior variance. The arbitrariness of $\mathbf{e}$ means that this inequality holds for the surface elevation and velocity potential at all grid points as well as all of their possible linear combination. This is the meaning of the posterior covariance $\tilde{\mathbf{\Sigma}}^{(0)}$ being smaller or equal to the prior covariance $\mathbf{\Sigma}^{(0)}$, and this reduction is caused by the observability Gramian $\mathbf{G}$ \citep{tarantola2005inverse}. 
%-----------------------------------------------------------------------------------
If we define the prior and posterior `precision' to be $\mathbf{I}_{r_0\times r_0}^{-1}$ and $\left(\tilde{\mathbf{\Sigma}}^{(0)}_{\boldsymbol{\xi}}\right)^{-1}$ describing the `inverse' of uncertainty, then the matrix $\mathbf{G}$ is simply the precision gain due to the information provided by the data. This precision gain has the quadratic form $\mathbf{e}^{\dagger}\mathbf{G}\mathbf{e}$ for a direction $\mathbf{e}$ (in the space of initial Fourier coefficients) and leads to reduction of posterior variance in the same direction \citep{tarantola2005inverse,bui2013computational}. Consequently, directions with large values of $\mathbf{e}^{\dagger}\mathbf{G}\mathbf{e}$ tend to have a larger precision gain and uncertainty reduction. In contrast, directions in which $\mathbf{e}^{\dagger}\mathbf{G}\mathbf{e}$ vanishes do not have precision gain. The nonzero and zero eigenspaces of $\mathbf{G}$ therefore characterize the data-informed and data-uninformed components in the space of initial Fourier coefficients \citep{wang2022observable}.
%Specifically, the precision gain in a certain direction $\Delta\mathbf{u}_0$ to the initial condition changes the misfit by an amount proportional to $\Delta\mathbf{u}_0^{\dagger}\mathbf{G}\Delta\mathbf{u}_0$. Therefore, directions with large values of this quadratic form are strongly constrained by the measurement data and tend to have a larger precision gain and uncertainty reduction. In contrast, directions in which $\mathbf{G}$ vanishes are not sensitive to the measurement misfit and do not have precision gain \citep{chen1984linear,bui2013computational}. The nonzero and zero eigenspaces of $\mathbf{G}$ therefore characterize the data-informed and data-uninformed components of the initial condition space \citep{wang2022observable}.

We now turn back to the Bayesian framework, where the ultimate goal is to obtain $p(\mathbf{u}^{(n)}|\{\mathbf{m}^{(l)}\}_{l=N_{1},...,N_{2}})$ for all $n$.
We already have the posterior distribution for the initial condition $p(\boldsymbol{\xi}^{(0)}|\{\mathbf{m}^{(l)}\}_{l=N_{1},...,N_{2}})$ which is parameterized by the posterior mean $\tilde{\boldsymbol{\xi}}^{(0)}$ and posterior covariance $\tilde{\mathbf{\Sigma}}^{(0)}_{\boldsymbol{\xi}}$ (see (\ref{eq: linear posterior centered})-(\ref{eq: posterior mean cov})). 
To obtain the posterior distribution of the wave state $\mathbf{u}^{(n)}$, the posterior mean $\tilde{\boldsymbol{\xi}}^{(0)}$ and covariance $\tilde{\mathbf{\Sigma}}_{\boldsymbol{\xi}}^{(0)}$ of the Fourier coefficients can be propagated to the posterior mean $\tilde{\mathbf{u}}^{(n)}$ and covariance $\tilde{\mathbf{\Sigma}}^{(n)}$ of the wave states by,
%\codex{[Check: Standardize $\tilde{\boldsymbol{\xi}}^{(0)}$ and the covariance subscript $\boldsymbol{\xi}$; this sentence instead uses $\tilde{\boldsymbol{\xi}}^0$ and $\tilde{\mathbf{\Sigma}}_{\boldsymbol{\xi}}^{(0)}$.]}
\begin{equation}\label{eq: mean cov propagate linear}
    \tilde{\mathbf{u}}^{(n)} = \mathbf{A}^n\mathbf{L}\tilde{\boldsymbol{\xi}}^{(0)}, \qquad\tilde{\mathbf{\Sigma}}^{(n)} = \mathbf{A}^n\mathbf{L}\tilde{\mathbf{\Sigma}}^{(0)}_{\boldsymbol{\xi}} \mathbf{L}^{\dagger}\left(\mathbf{A}^{\dagger}\right)^{n}.
\end{equation}
The posterior distribution $p(\mathbf{u}^{(n)}|\{\mathbf{m}^{(l)}\}_{l=N_{1},...,N_{2}})$ is Gaussian and centered at $\tilde{\mathbf{u}}^{(n)}$ with covariance $\tilde{\mathbf{\Sigma}}^{(n)}$.
The posterior covariance $\tilde{\mathbf{\Sigma}}^{(n)}$ has a $2\times2$ structure similar to the prior covariance $\mathbf{\Sigma}^{(n)}$. 
The posterior standard deviation $\rho_j^{(n)}=\sqrt{(\tilde{\mathbf{\Sigma}}^{(n)})_{jj}}$ (for $1<j<J$) describes the uncertainty level of the surface elevation $\eta(x_j,t^{(n)})$. We can define the following normalized uncertainty to quantify the reduction of posterior standard deviation $\rho_j^{(n)}$ compared with the prior standard deviation $\varrho$,
\begin{equation}\label{eq: normalized uncertainty}
    \vartheta^{(n)}_j = \frac{\rho_j^{(n)}}{\varrho},
\end{equation}
where $\mathbf{\Sigma}^{(0)}$ is the prior covariance of $\mathbf{u}^{(0)}$ obtained using the linear transformation $\mathbf{u}^{(0)}=\mathbf{L}\boldsymbol{\xi}^{(0)}$.  The ratio $\vartheta^{(n)}_j$ is bounded between $0$ and $1$ because posterior uncertainty cannot exceed prior uncertainty. At spatiotempotal locations where $\vartheta^{(n)}_j\approx 0$, the posterior uncertainty is reduced significantly compared to the prior uncertainty by the information from the measurement data, therefore the wave field at these locations are considered more predictable. In comparison, $\vartheta^{(n)}_j\approx 1$ means that the information provided by the data is less useful to reduced the uncertainty at these locations, and the waves there is considered less predictable.
%\codex{[Check: The definition above restricts $j$ to $1<j<J$, but this paragraph calls $\vartheta_j^{(n)}$ arbitrary in $j$. State the endpoint convention consistently.]}
Intermediate values of $\vartheta^{(n)}_j$ quantifies the reduction of $\rho_j^{(n)}$ relative to $\varrho$ due to the precision gain introduced by the observability Gramian $\mathbf{G}$, 
and demonstrates the level of local predictability ranging between $0$ and $1$.
In summary, $\vartheta^{(n)}_j$ is a continuous indicator of how predictable the wave field at different spatiotemporal locations is.

%--------------Discussion of why posterior covariance does not depend on base flow---------------------------------
%From the expressions (\ref{eq: posterior mean cov}) and (\ref{eq: mean cov propagate linear}), the posterior mean $\tilde{\mathbf{u}}^{(n)}$ depends on the measurement data, but the covariance $\tilde{\mathbf{\Sigma}}^{(n)}$ does not.We can think of an uncertain solution being the summation of a true solution that reproduces the true measurements and a random perturbation part representing the uncertainty. Due to the principle of superposition for linear wave dynamics, the random perturbative part evolves autonomously and does not depend on the true solution. Therefore the posterior covariance, as the second order moment of the random perturbative part, does not depend on the measurements $\{\mathbf{m}^{(l)}\}_{l=N_{1},...,N_{2}}$ and the underlying true solution that generates them.
%------------------------------------------------------------------------------------------------------------------

%We now introduce a refined characterization of the posterior covariance by splitting it into two parts according to whether the wave state is sensitive to the data (\S \ref{subsec: uncertainty decomposition}). Such characterization would be especially useful when we consider the zero-noise limit (clean measurements). 

%YD: start a new version v0p9, reframe the decomposition

%\subsection{Data-informed and un-informed parts of $\tilde{\mathbf{\Sigma}}^{(n)}$}\label{subsec: uncertainty decomposition}

\subsection{Split the wave fields into data-informed and uninformed parts}\label{subsec: uncertainty decomposition}
%We have define a new variable $\vartheta$ to quantify the local predictability of the wave field at each spatiotemporal location, which is supposed to be more precise than the earlier predictable zone works by \citet{qi2018predictable, morris1998short}, among others. 
One interpretation of the earlier works of the predictable zones by \citet{qi2018predictable, morris1998short} (and many others) is a split of the full wave field into predictable and un-predictable parts. In their works such split is in terms of zonal locations, where the solution in one region is predictable while the other is not. Such split would inevitably face difficulty since the wave field at many locations could be partially predictable and does not belong to either side. 

We now propose another type of split of the wave field in terms of linear superposition rather than zonal locations. Specifically, we propose a split of the wave state $\mathbf{u}^{(n)} = \mathbf{u}^{(n)}_{I} + \mathbf{u}^{(n)}_{u}$, where $\mathbf{u}^{(n)}_{I}$ is the data-informed part, and $\mathbf{u}^{(n)}_{u}$ is the data-uninformed part of the solution. These two solution co-exist at each spatiotempotal location. We can guarantee that such split is exact, in the sense that at the limit of zero measurement noise, $\mathbf{u}^{(n)}_{I}$ can be exactly determined, and $\mathbf{u}^{(n)}_{u}$ is entirely undetermined. 
We now quantitatively define this split using the observability Gramian $\mathbf{G}$ (\ref{eq: linear observability Gramian}), and discuss the property of this split using the language of Bayesian inference. Both tools have already been introduce in \S \ref{subsec: Bayesian linear}. 
%Equivalently by the language of Bayesian inference in \S \ref{subsec: Bayesian linear}, the posterior uncertainty of $\mathbf{u}^{(n)}_{u}$ is not reduced compared to its prior, while the uncertainty of $\mathbf{u}^{(n)}_{I}$ is reduced to zero (at the limit of zero measurement noise) after the measurement. 

%As seen from the previous section, the measurement data improves the precision, or reduces the uncertainty of the initial Fourier coefficients $\boldsymbol{\xi}^{(0)}$ (and the subsequent wave states). Intuitively, one might expect certain parts of the wave field to be sensitive to the measurement, while the other parts are not. We can obtain a more refine characterization of the predictability of the wave field by considering the posterior uncertainty of the respective `sensitive' and `insensitive' parts. We describe this split quantitatively as follows. 

The observability Gramian $\mathbf{G}$ (see \ref{eq: linear observability Gramian}) describes the precision gain of the Fourier coefficients $\boldsymbol{\xi}^{(0)}$ from measurements. The structure of this matrix $\mathbf{G}$ with rank $r>0$ is characterized by its eigenvalue decomposition (or equivalently, its singular value decomposition, since $\mathbf{G}$ is positive semidefinite), 
\begin{equation}\label{eq: prior preconditioned Gramian eigendecomposition}
    \mathbf{G}
    =
    \begin{bmatrix}
        \mathbf{V}_{I} & \mathbf{V}_{u}
    \end{bmatrix}
    \begin{bmatrix}
        \boldsymbol{\Lambda}_{I} & \mathbf{0}\\
        \mathbf{0} & \boldsymbol{\Lambda}_{u}
    \end{bmatrix}
    \begin{bmatrix}
        \mathbf{V}_{I} & \mathbf{V}_{u}
    \end{bmatrix}^{\dagger},
\end{equation}
where $\mathbf{V}_{I}$ and $\mathbf{V}_{u}$ contain the eigenvectors that together span the space of all possible realizations of initial Fourier coefficients $\boldsymbol{\xi}^{(0)}$, and $\boldsymbol{\Lambda}_{I}$ and $\boldsymbol{\Lambda}_{u}$ are diagonal matrices containing the corresponding eigenvalues in descending orders. 
The matrix $\boldsymbol{\Lambda}_{I}=\operatorname{diag}(\lambda_1,\ldots,\lambda_r)$ contains the positive eigenvalues, and the columns of $\mathbf{V}_{I}$ span the `data-informed' (with the subscript `I') subspace, i.e. the precision gain $\boldsymbol{\xi}^{(0)\dagger}\mathbf{G}\boldsymbol{\xi}^{(0)}$ is positive for any $\boldsymbol{\xi}^{(0)}\in\mathrm{span}(\mathbf{V}_{I})$. The remaining tail $\boldsymbol{\Lambda}_{u}=\operatorname{diag}(\lambda_{r+1},\ldots,\lambda_{r_0})$ contains zero eigenvalues and corresponds to the `uninformed' (with the subscript `u') subspace spanned by the columns of $\mathbf{V}_{u}$, and there is no precision gain in this subspace. This split of the solution space originated in control theory, where the corresponding subspaces are known as observable and unobservable subspaces \citep{wang2022observable,chen1984linear}. More recently, this concept has been adopted in Bayesian uncertainty quantification with the subspaces usually termed data-informed and uninformed \citep{bui2013computational,kalmikov2014hessian}. We adopt the latter terminology because our framework is more closely aligned with Bayesian inference.

Using $\mathbf{V}_{I}$ and $\mathbf{V}_{u}$ described above, we can readily define an orthogonal decomposition of $\boldsymbol{\xi}^{(0)}$ 
\begin{equation}\label{eq: xi decomp}
    \boldsymbol{\xi}^{(0)} = \boldsymbol{\xi}_{I}+\boldsymbol{\xi}_{u},\qquad \boldsymbol{\xi}_{I}=\mathbf{V}_{I}\mathbf{V}_{I}^{\dagger}\boldsymbol{\xi}^{(0)}, \qquad \boldsymbol{\xi}_{u}=\mathbf{V}_{u}\mathbf{V}_{u}^{\dagger}\boldsymbol{\xi}^{(0)}
\end{equation}
where $\boldsymbol{\xi}_{I}$ and $\boldsymbol{\xi}_{u}$ represent the data-informed and uninformed parts of $\boldsymbol{\xi}^{(0)}$. 
This split of $\boldsymbol{\xi}^{(0)}$ leads to the split of the wave state variable $\mathbf{u}^{(n)}$ at any time step into data-informed part $\mathbf{u}^{(n)}_{I}$ and un-informed part $\mathbf{u}^{(n)}_{u}$,   
\begin{equation} \label{eq: u decomp}
    \mathbf{u}^{(n)} = \mathbf{u}^{(n)}_{I} + \mathbf{u}^{(n)}_{u}, \qquad \mathbf{u}^{(n)}_{I} = \mathbf{A}^{n}\mathbf{L}\boldsymbol{\xi}_{I}, \qquad \mathbf{u}^{(n)}_{u} = \mathbf{A}^{n}\mathbf{L}\boldsymbol{\xi}_{u}.  
\end{equation}
As can be seen from (\ref{eq: u decomp}), the two parts $\mathbf{u}^{(n)}_{I}$ and $\mathbf{u}^{(n)}_{u}$ both satisfy the linear wave dynamical equation, and their superposition forms the full wave state $\mathbf{u}^{(n)}$.  Since both $\mathbf{u}^{(n)}_{I}$ and $\mathbf{u}^{(n)}_{u}$ can be seen as linear functions of $\boldsymbol{\xi}^{(0)}$, their respective prior and posterior distributions can be derived straightforwardly from those of $\boldsymbol{\xi}^{(0)}$ using linear transformation rule of Gaussian distributions (as shown in appendix \ref{app: uncertainty decomposition}). We therefore denote $\mathbf{\Sigma}_{I}^{(n)}$, $\mathbf{\Sigma}_{u}^{(n)}$ as the respective prior covariance, and $\tilde{\mathbf{\Sigma}}_{I}^{(n)}$, $\tilde{\mathbf{\Sigma}}_{u}^{(n)}$ as the respective posterior covariance.
Two important feature of the split $\mathbf{u}^{(n)} = \mathbf{u}^{(n)}_{I} + \mathbf{u}^{(n)}_{u}$ are often discussed in this context (derived quantitatively in appendix \ref{app: uncertainty decomposition}). First, 
the posterior covariance $\tilde{\mathbf{\Sigma}}_{I}^{(n)}$ of the informed part $\mathbf{u}_{I}^{(n)}$ is reduced compared to its prior covariance, while posterior covariance $\tilde{\mathbf{\Sigma}}_{u}^{(n)}$ of $\mathbf{u}_{u}^{(n)}$ remains at the prior level. In other words, measurement data helps to reduce uncertainty for $\mathbf{u}_{I}^{(n)}$, but not for $\mathbf{u}_{u}^{(n)}$. Second, as the measurement noise level decrease, $\tilde{\mathbf{\Sigma}}_{I}^{(n)}$ diminishes, while the $\tilde{\mathbf{\Sigma}}_{u}^{(n)}$ remain unchanged. At the limit of clean measurements, $\mathbf{u}_{I}^{(n)}$ becomes exactly predictable (with zero uncertainty), while $\mathbf{u}_{u}^{(n)}$ still contains nonzero uncertainty. In summary, (\ref{eq: u decomp}) defines a novel way of splitting the wave fields into predictable parts and un-predictable parts in the sense of linear superposition instead of spatiotemporal zonal regions. These `predictable' and `unpredictable' concepts in our context are rigorously based on uncertainty characterization. 

We now discuss what aspects of the physical configuration can affect this split (\ref{eq: u decomp}). The relative size of the two spaces $\mathbf{V}_{I}$ and $\mathbf{V}_{u}$ determines the how many dimensions of the total wave state vector goes into either spaces, and is ultimately determined by $\mathbf{G}$. As can be seen from the expression (\ref{eq: linear observability Gramian}), $\mathbf{G}$ depend on the measurement configuration (specified by $\mathbf{M}$, $N_1$, $N_2$) and the number of active wave modes (specified by the matrix $\mathbf{L}$). 
On one hand, more measurement data provides more information to infer the wave system, therefore often lead to a smaller uninformed space. 
On the other hand, with a fixed amount of measurement data (which provides the same amount of information), more active modes to be inferred in the system could lead to a larger uninformed space. These effects are quantitatively examined by numerically computing the decomposition (\ref{eq: prior preconditioned Gramian eigendecomposition}) in the result section.

\subsection{Numerical results}\label{subsec: results linear}
We present the setup of the numerical tests. We consider a spatiotemporal domain specified by $0<t<40$ and $0<x<2\pi$, discretized with $J=256$ grid points and $N_T=2000$ time steps. 
A sensor located at $x_m=\pi$ records the surface elevation $\eta(x_m,t)$ as a function of time in the interval $16<t<24$. For all of the numerical experiments (unless otherwise specified), the measurement noise satisfies a zero-mean Gaussian distribution with covariance matrix $\mathbf{R}=\gamma^2\mathbf{I}$, where $\gamma =0.01\varrho=0.0025H_s$ is the noise standard deviation.
%\codex{[Check: The equality $\beta=0.0025H_s$ assumes $\varrho^2=H_s^2/16$. State whether this is imposed after the finite-$\Delta k$ discretization of both spectra, including truncated JONSWAP, or whether $\gamma=0.01\varrho$ is computed directly.]}
%\codex{[Check: State $J$, $\Delta x=L/J$, $\Delta k=2\pi/L$, the retained Fourier-index range and the Nyquist/dealiasing rule, together with $\Delta t$, $N_T$, $N_1$ and $N_2$. These parameters determine the discrete spectral quadrature and connect the physical windows to the sums and numerical algorithm.]}

Two model spectra are adopted for the prior covariance in the numerical tests. The first one has constant values in a wavenumber interval,
\begin{equation}
S_c(k)=
\begin{cases}
\dfrac{H_s^2}{16\left(k_{\max}-k_{\min}+1\right)},
& k_{\min}\le k \le k_{\max},\\
0, & \text{otherwise}.
\end{cases}
\end{equation}
%\codex{[Check: The displayed normalization treats $S_c(k)$ as a discrete modal variance, since its sum over the 28 retained modes is $H_s^2/16$. If it is instead the one-sided wavenumber density used in a $\Delta k$-weighted $\mathbf{L}$, write $S_c=H_s^2/(16N_m\Delta k)$ with $N_m=m_{\max}-m_{\min}+1$, so that $\sum_m S_c(k_m)\Delta k=H_s^2/16$. The two forms happen to agree here because $L=2\pi$ gives $\Delta k=1$.]}
where $k_{\text{min}}=2$ and $k_{\text{max}}=29$ are the two ends of the interval. This spectrum $S_c(k)$ has compact support, therefore allows direct comparison between the posterior uncertainty and the classical predictable zone \citep{qi2018predictable}. The other one is the JONSWAP spectrum,
%\begin{equation}
%S_j(\omega)
%=
%\alpha\,\frac{g^2}{\omega^5}
%\exp\!\left[-\beta\left(\frac{\omega_p}{\omega}\right)^4\right]\,
%\Gamma^{\,\exp\!\left[-\frac{(\omega-\omega_p)^2}{2\sigma^2\omega_p^2}\right]},
%\label{eq:jonswap}
%\end{equation}
\begin{equation}
S_j(k)
=
\alpha\,\frac{g^2}{\omega(k)^5}
\exp\!\left[-\beta\left(\frac{\omega_p}{\omega(k)}\right)^4\right]\,
\Gamma^{\,\exp\!\left[-\frac{(\omega(k)-\omega_p)^2}{2\sigma^2\omega_p^2}\right]}\frac{g}{2\sqrt{gk}},
\label{eq:jonswap}
\end{equation}
with
\begin{equation}
\beta=\frac{5}{4},\qquad
\sigma=
\begin{cases}
0.07, & \omega \le \omega_p,\\
0.09, & \omega > \omega_p~, 
\end{cases}
\label{eq:jonswap_params}
\end{equation}
The enhancement factor takes $\Gamma=3.0$, and the last factor $\mathrm{d}\omega/\mathrm{d}k=g/(2\sqrt{gk})$ in (\ref{eq:jonswap}) is introduced to transform the JONSWAP spectrum from frequency domain to wavenumber domain. The peak frequency $\omega_p=\sqrt{gk_p}$ is set up by the peak wavenumber $k_p$ case by case in the results.
%\codex{[Check: Equation (\ref{eq:jonswap}) is a one-sided frequency density. Before using it in (\ref{eq: L definition}), define $S_k(k)=S_j(\sqrt{gk})\,g/(2\sqrt{gk})$ for $k>0$ and use the modal variance $S_k(k_m)\Delta k$. State the retained positive-mode range and Nyquist compatibility behind $r_0=256$, and choose $\alpha$ so that $\sum_m S_k(k_m)\Delta k=H_s^2/16$, or report the retained truncated variance. Distinguish this finite numerical spectrum from analytical JONSWAP.]}

\subsubsection{Posterior uncertainties for $S_c$ and $S_j$ }

\begin{figure}
    \centering
    \includegraphics[width=0.8\textwidth]{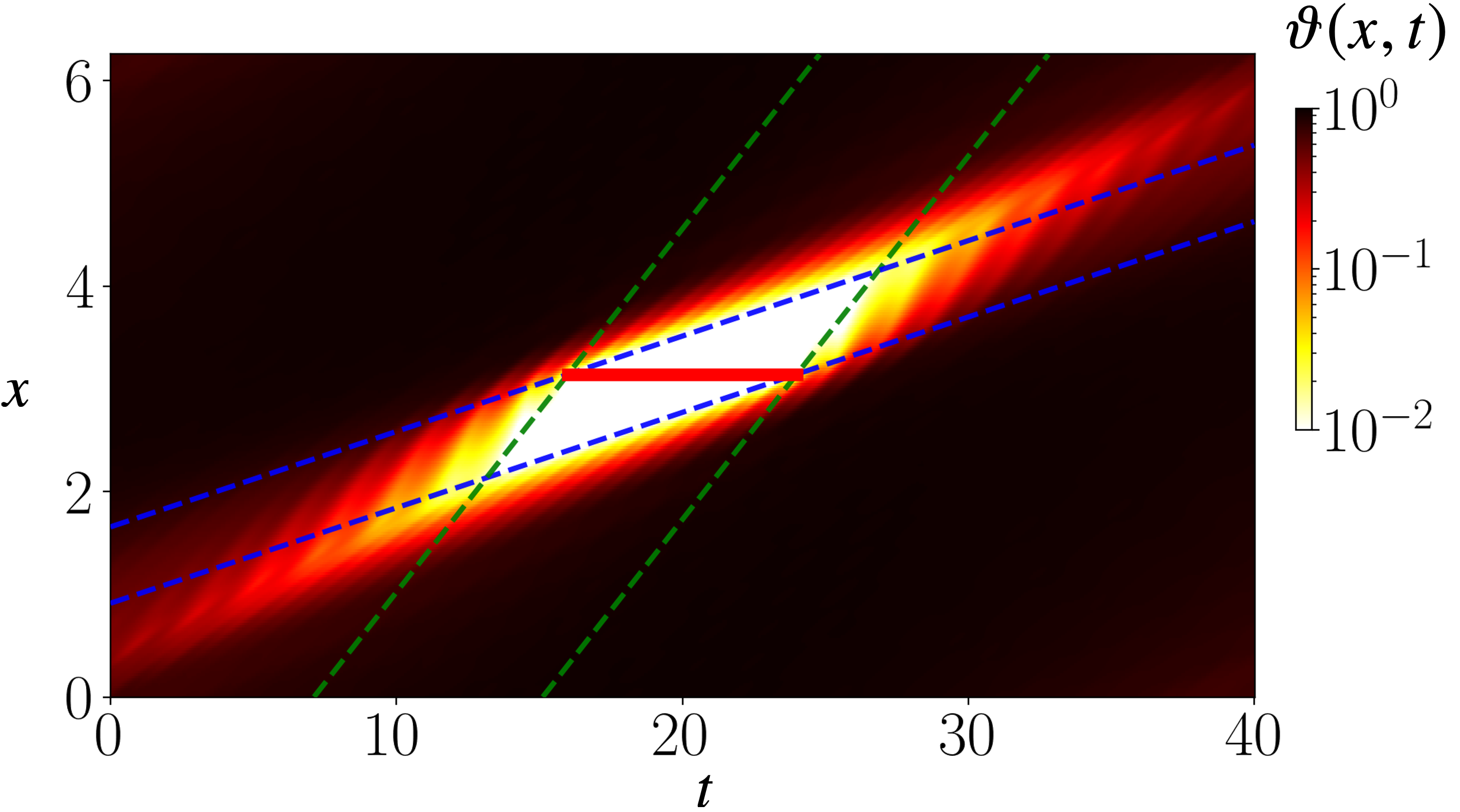}
    \caption{The $x$-$t$ map of $\vartheta$ with energy spectrum $S_c$, visualized together with measurement set (\fullthickred{}), $c_{g,\text{min}}$ (\dashedblue{}) and $c_{g,\text{max}}$ (\dashedgreen{})}
    \label{fig: linear rho S_c x t}
\end{figure}

\begin{figure}
    \centering
    \includegraphics[width=1.0\textwidth]{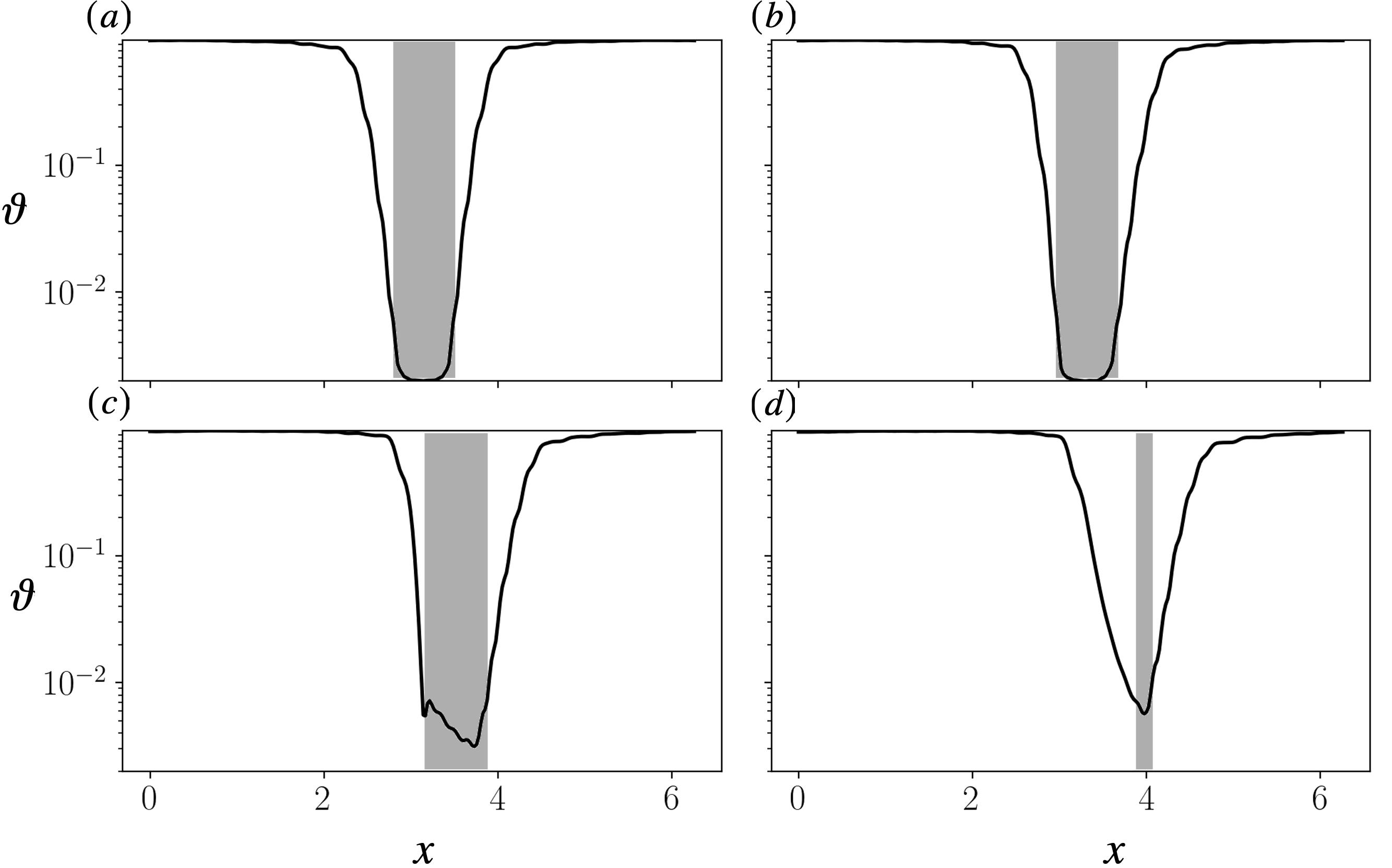}
    \caption{The spatial profile of $\vartheta$ with energy spectrum $S_c$ at $t = \{20, 22,24,26\}$ for $(a$-$d)$. The theoretical predictable zone is denoted by the grey area.  }
    \label{fig: linear rho S_c x 1d}
\end{figure}

We first report the posterior uncertainty for the linear wave model with the piecewise constant spectrum $S_c(k)$. The spatiotemporal iso-contour of the normalized posterior standard deviation $\vartheta(x_j,t^{(n)})=\vartheta_j^{(n)}$ is shown in figure \ref{fig: linear rho S_c x t} and compared with the theoretically constructed predictable zone based on $c_{g,\text{min}}$ and $c_{g,\text{max}}$. The low uncertainty region ($\vartheta \lessapprox 10^{-2}$) agrees well with the predictable zone reproduced by the group velocity argument. 
A more detailed visualization of the spatial profile of $\vartheta$ is shown at $t=\{20,22,24,26\}$ in figure \ref{fig: linear rho S_c x 1d}. Outside of the predictable zone, the normalized uncertainty $\vartheta$ grows gradually with the distance away from the predictable zone. 
The theoretical work by \citet{qi2018nonlinear} is unable to characterize these `partially predictable' locations. In comparison, the normalized uncertainty $\vartheta$ in our work precisely quantifies how predictable the wave field is at these partially predictable locations. 
%\yd{The wave field at these locations outside of the predictable zone is `partially' predictable, yet their predictability can still be quantified by $\vartheta$}
%The theoretically constructed predictable zone no longer characterizes the predictability of the wave field at these locations. 
At the far locations, the $\vartheta$ approaches $1$, which is expected since no information of the wave field can reach here from the sensor, therefore the waves are completely unpredictable.
In summary, our posterior uncertainty $\vartheta$ not only produces a low-uncertainty zone that agrees with earlier theoretical works, but also quantifies the partial predictability outside the zone. 
%Intermediate values of $\vartheta$ correspond to the locations where the wave field is partially predictable, and $\vartheta$ quantitatively characterizes how predictable the wave is.

\begin{figure}
    \centering
    \includegraphics[width=1.0\textwidth]{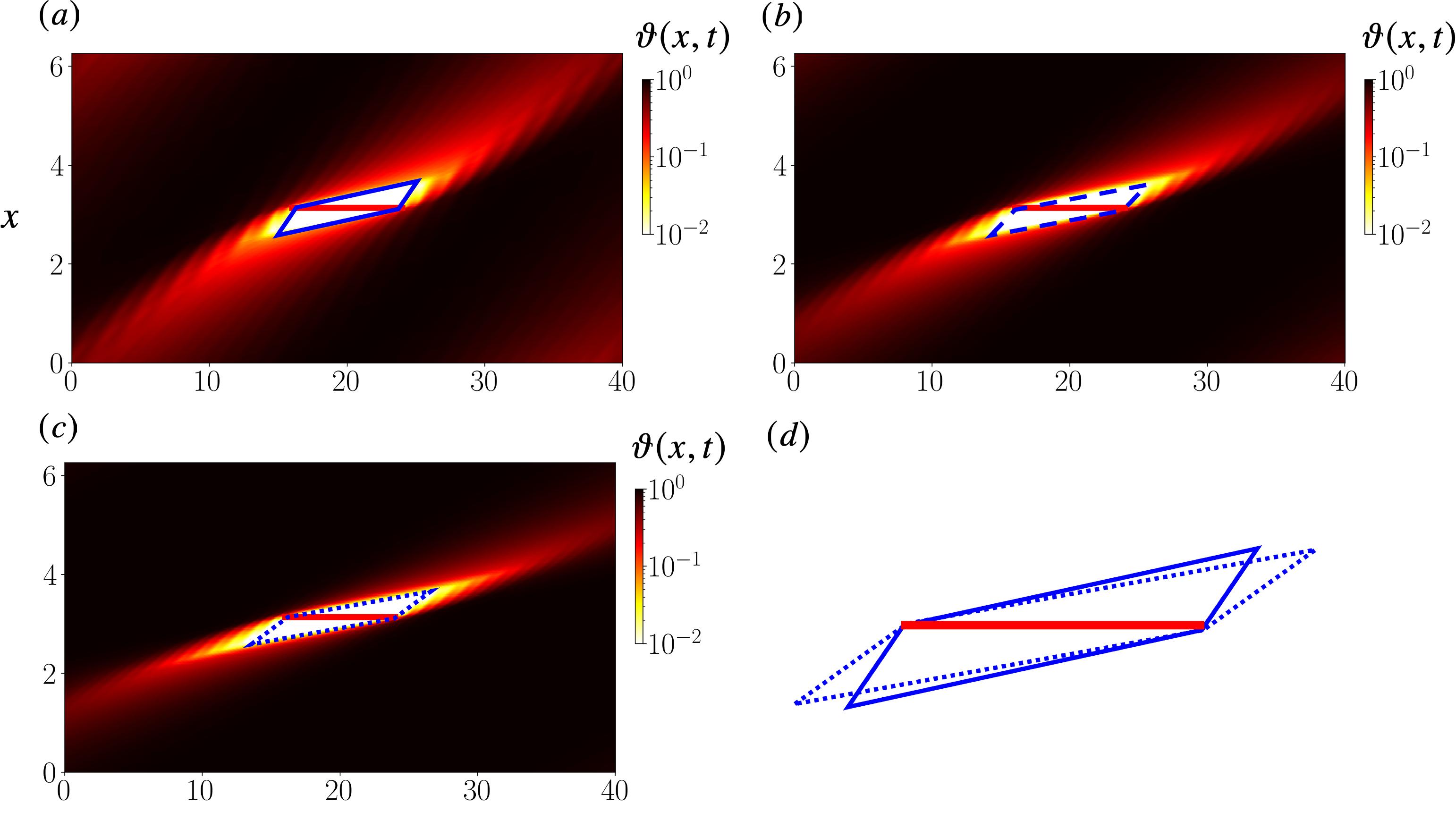}
    \caption{The $x$-$t$ map of $\vartheta$ with energy spectrum $S_j$ and $k_p=\{5,10,20\}$ (panels $(a$-$c)$),  visualized together with measurements (\fullthickred{}). The low uncertainty zone ($\vartheta\leq 0.02$) are marked by the blue line contours. Panel $(d)$ overlaps the low uncertainty zone from $(a)$ and $(c)$. }
    \label{fig: linear rho jonswap x t}
\end{figure}

\begin{figure}
    \centering
    \includegraphics[width=1.0\textwidth]{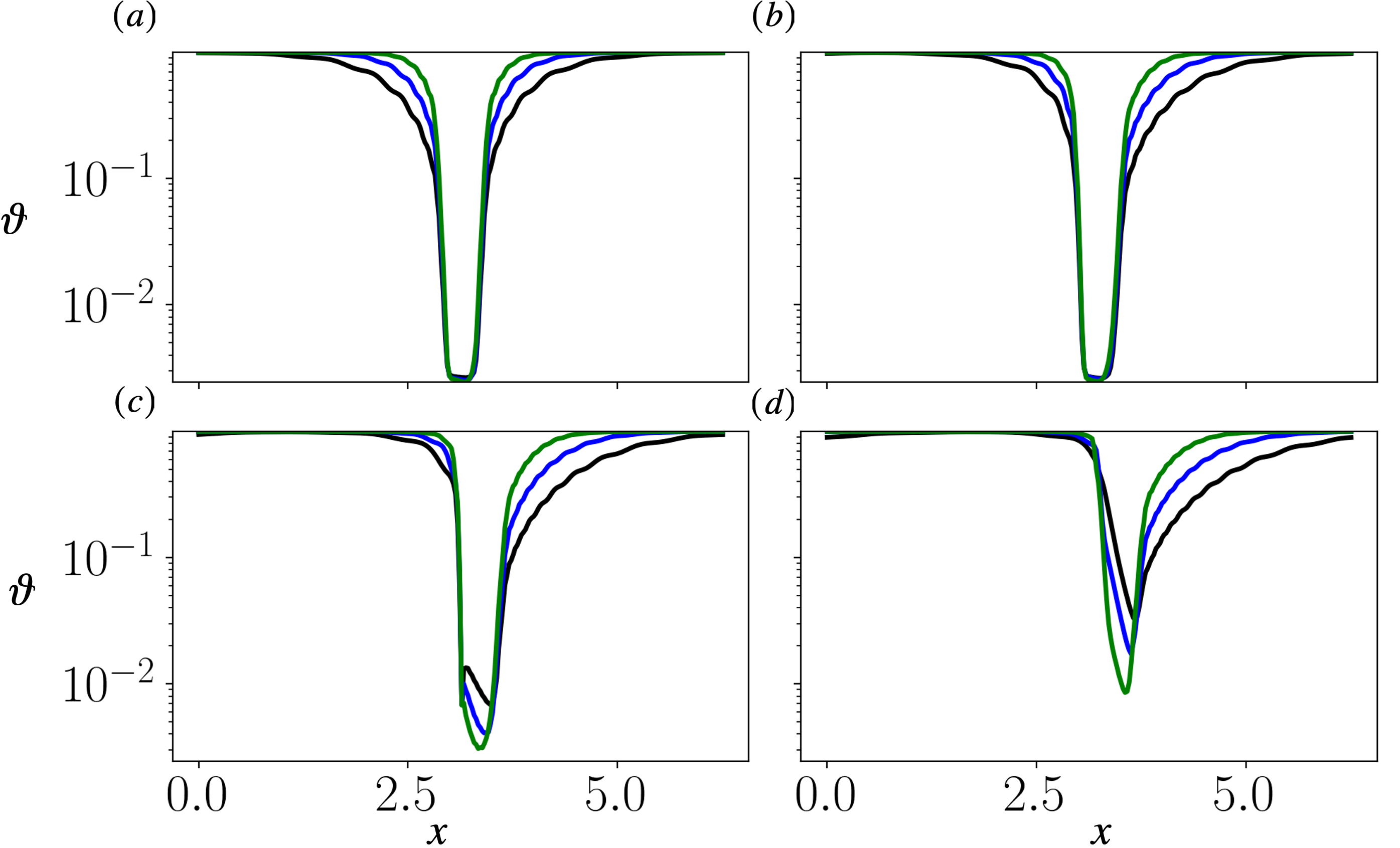}
    \caption{The spatial profile of $\vartheta$ with JONSWAP spectrum $S_j$ and $k_p$ = $5$ (\fullblack{}), $10$ (\full) and $20$ (\fullgreen{}) at $t = \{20, 22,24,26\}$}
    \label{fig: linear rho jonswap x 1d}
\end{figure}

Next we present the results for the JONSWAP spectrum $S_j$ with unbounded frequency support. The spatiotemporal iso-contours of $\vartheta$ are shown in figure \ref{fig: linear rho jonswap x t} for $k_p=\{5,10,20\}$, and their spatial profiles are shown in figure \ref{fig: linear rho jonswap x 1d} at selected time instants $t=\{20,22,24,26\}$. A low-uncertainty region appear near the measurement set, with its width and shape changing with $k_p$. 
Since $k_p$ is related to the peak group velocity by $c_g=\frac{1}{2}\sqrt{g/k_p}$, the increase of $k_p$ corresponds to a shift of the wave towards smaller group velocity. Such shift cause the measurement information to propagate with smaller speeds across the domain. 
%and therefore smaller slopes of the boundaries of the low-uncertainty zone. 
This trend is confirmed in figure \ref{fig: linear rho jonswap x t}, where the low-uncertainty zone extends in time and shrinks in space with the increase of $k_p$. 
%These changes can be more clearly revealed in the spatial profiles of $\vartheta$ at different time, shown in figure \ref{fig: linear rho jonswap x 1d}.
Since JONSWAP spectrum has unbounded frequency support, the classical construction by \citet{qi2018predictable} does not apply, and an analytical predictable zone cannot be marked. Although the cutoff strategies such as \citet{kim2008technical, desmars2020experimental} and \citet{stuhlmeier2021deterministic} can be useful to define a predictable zone in such cases, the slopes of the predictable zone boundaries are usually sensitive to the specific cutoff threshold. In comparison, our approach quantitatively describe the uncertainty point-wise on the spatiotemporal plane, and the results does not depend on any tunable, un-physical parameters,

\subsubsection{The decomposition $\mathbf{u}^{(n)}=\mathbf{u}_{I}^{(n)}+\mathbf{u}_{u}^{(u)}$ and their uncertainty}\label{subsec: result uncertainty decomposition}

\begin{figure}
    \centering
    \includegraphics[width=1.0\textwidth]{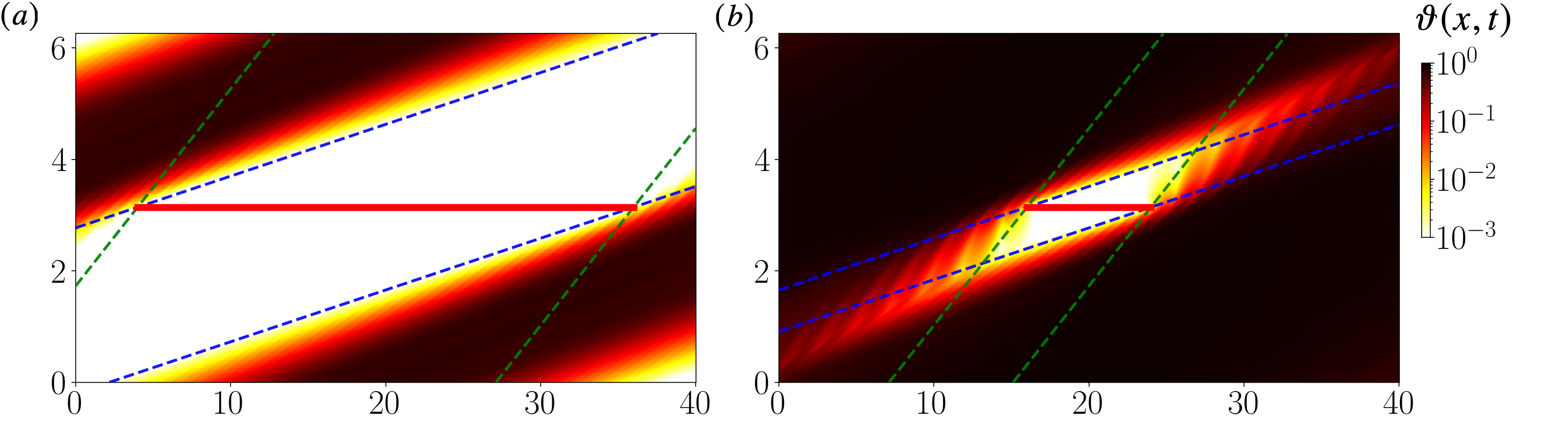}
    \caption{Measurement configurations used for the uncertainty decomposition: the long observation window $4<t<36$ $(a)$ and the short observation window $16<t<24$ $(b)$.}
    \label{fig: short long configuration}
\end{figure}

\begin{figure}
    \centering
    \includegraphics[width=1.0\textwidth]{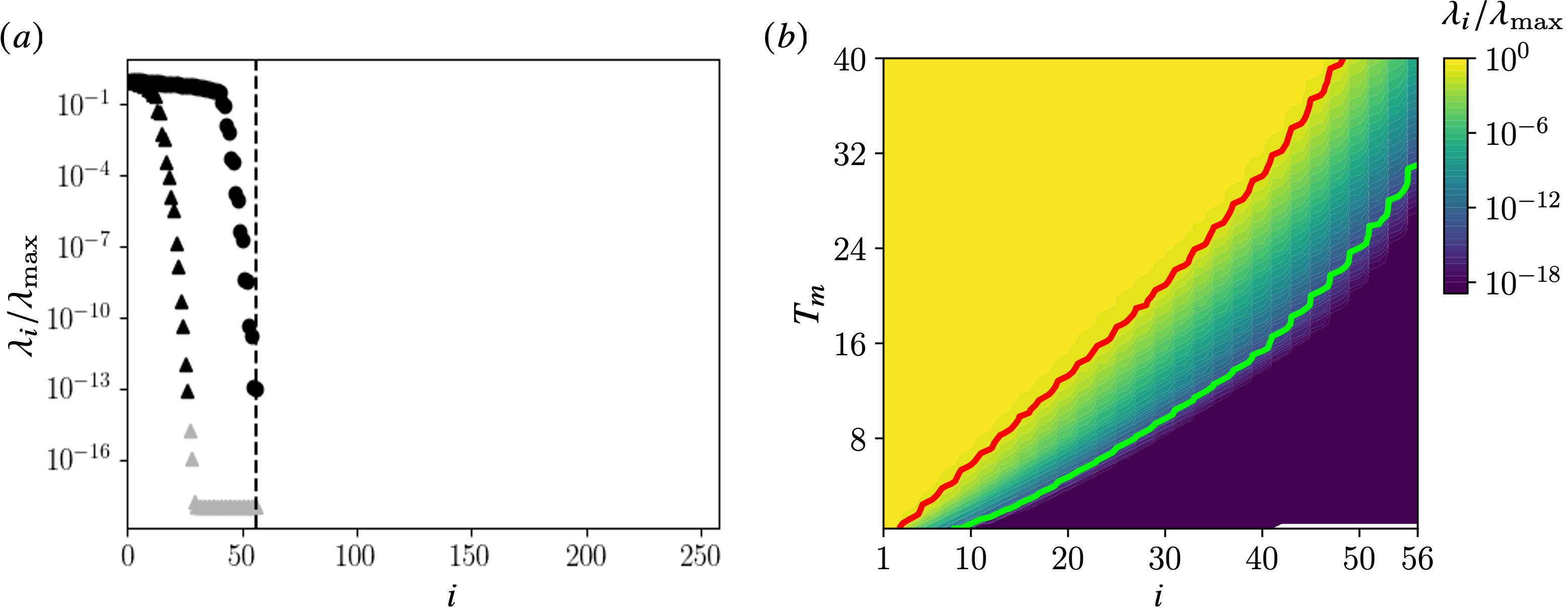}
    \caption{$(a)$ Eigenvalues $\lambda_i/\lambda_{\text{max}}$ for the wave spectrum $S_c$ with short (\mytriangle{black}) and long (\mycircle{black}) measurement time windows. Black and grey color represent the data informed and un-informed eigenvalues divided by the cutoff $\lambda_i/\lambda_{\text{max}}>10^{-14}$, respectively. Vertical dashed lines mark the dimension $r_0$ of the initial condition. $(b)$ The eigenvalues $\lambda_i/\lambda_{\text{max}}$ in a range of measurement window length $T_m$, with contour lines $\lambda_i/\lambda_{\text{max}}=10^{-1}$ (\fullred{}) and $\lambda_i/\lambda_{\text{max}}=10^{-14}$ (\fullgreenlight{}) }
    \label{fig: eigenvalues S_c}
\end{figure}

In this section, we numerically compute the eigendecomposition (\ref{eq: prior preconditioned Gramian eigendecomposition}), the solution split $\mathbf{u}^{(n)}=\mathbf{u}_{I}^{(n)}+\mathbf{u}_{u}^{(u)}$ and the posterior uncertainty of the respective parts, as described in \S\ref{subsec: uncertainty decomposition}. Since the normalized eigenvalues $\lambda_i/\lambda_{\mathrm{max}}$ are computed numerically, the zero ones usually have machine zero values (on the order of $10^{-15}\sim10^{-18}$) instead of being exactly zero. We therefore use a threshold $\lambda_i/\lambda_{\mathrm{max}}>10^{-14}$ to identify the eigenvalues and eigenvectors corresponding to the data-informed subspace, and define the space spanned by the rest of the eigenvectors as the uninformed subspace. This cutoff criteria reflects the most accurate identification of the informed and uninformed subspaces under the limitation of floating point arithmetic. 

With this criteria, we first examine the effect of measurement configuration on the decomposition. Two measurement time windows $4<t<36$ (long) and $16<t<24$ (short) are considered and visualized in figure \ref{fig: short long configuration}$(a)$. The relative eigenvalues $\lambda_i/\lambda_{\text{max}}$ of $\mathbf{G}$ for these two time windows with the same energy spectrum $S_c$ are shown in figure \ref{fig: eigenvalues S_c}. 
The total number of eigenvalues $\lambda_i$ is the number of dimensions where the waves are active (parameterized by $\boldsymbol{\xi}^{(0)}$). For the spectrum $S_c$ this dimension is $r_0=2(k_{\text{max}}-k_{\text{min}}+1)=56$, which is marked in figure \ref{fig: eigenvalues S_c}. With the longer time window, all of the eigenvalues $\lambda_i/\lambda_{\text{max}}$ are nonzero and correspond to data-informed space. In other words, the un-informed space is empty, $\mathbf{u}_{u}^{(n)}=0$ and $\mathbf{u}^{(n)}=\mathbf{u}_{I}^{(n)}$. In comparison, about half of the eigenvalues in the short time window case are zero numerically, which indicates that $\mathrm{span}(\mathbf{V}_u)$ is not empty and $\mathbf{u}_{u}^{(n)}$ is in general nonzero. 
In summary, more measurement data lead to more nonzero eigenvalues and a larger data-informed space. 

The influence of measurement window on the observability of the wave field is more clearly demonstrated in figure \ref{fig: eigenvalues S_c}$(b)$, where the eigenvalues $\lambda_i/\lambda_{\text{max}}$ is computed for a range of measurement window length $T_m$ (with the midpoint of the window fixed at $t=20$). 
The number of nonzero $\lambda_i/\lambda_{\text{max}}$  monotonically increase with $T_m$, which correspond to $\mathbf{u}_{I}^{(n)}$ progressively occupying more dimensions in the solution space. 
In addition, the isocontours of $\lambda_i/\lambda_{\text{max}}$ (see the two examples marked in \ref{fig: eigenvalues S_c}$(b)$) bend upwards and have positive curvature. The increase of informed dimensions introduced by increasing $T_m$ diminishes for large $T_m$. In other words, when $T_m$ is already large, introducing more measurement data only lead to marginal improvement of the observability of the wave field. 
The study in \ref{fig: eigenvalues S_c}$(b)$ only demonstrates one factor (which is the window length $T_m$) that can influence the observability of the wave field. Many other aspects of the data set, including sampling frequency, sample number and number of sensors, can have important effects on the informed/uninformed split of the solution space. These other factors could be important direction to investigate for future works.  

\begin{figure}
    \centering
    \includegraphics[width=1.0\textwidth]{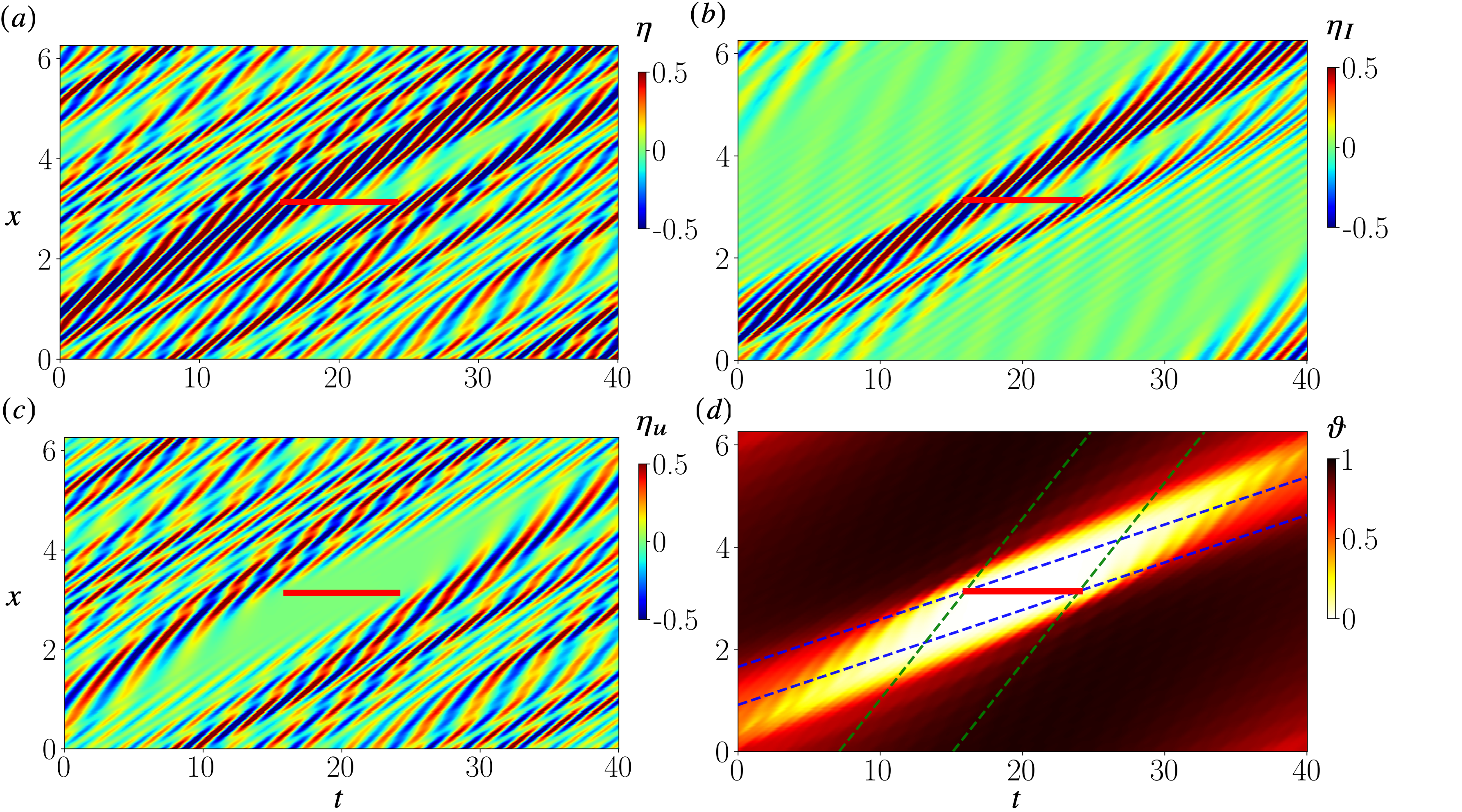}
    \caption{The $x$-$t$ iso-contour of sample surface elevations $\eta$ $(a)$, $\eta_{I}$ $(b)$, $\eta_{u}$ $(c)$ normalized by $H_s$, and $\vartheta$ $(d)$.}
    \label{fig: Sample informed uninformed sol}
\end{figure}

To intuitively understand the decomposition of $\mathbf{u}^{(n)}=\mathbf{u}_{I}^{(n)}+\mathbf{u}_{u}^{(u)}$, we compute a sample linear wave solution with $S_c$ spectrum and visualize the surface elevation parts $\eta_{I}$ and $\eta_{u}$ of the state vectors $\mathbf{u}_{I}^{(n)}$ and $\mathbf{u}_{u}^{(u)}$, respectively, for the short measurement window case. These solutions are visualized in figure \ref{fig: Sample informed uninformed sol}$(a$-$c)$, along with the posterior uncertainty $\vartheta$ (figure \ref{fig: Sample informed uninformed sol}$(d)$). The informed surface elevation $\eta_{I}$ is large near a band-shaped region that resembles the low-$\vartheta$ region, and becomes small outside that region. On contrary, $\eta_{u}$, the un-informed part, is almost zero near the measurement set, and becomes large and close to the total solution $\eta$ far away from the measurement set. From the definition of the split 
(equations (\ref{eq: prior preconditioned Gramian eigendecomposition})-(\ref{eq: u decomp})) as well as these visualizations, we can understand the $\eta_{I}$ as the surface elevation field that is visible to the measurement data, and $\eta_{u}$ as the field that cannot be `seen' from the measurements. 

\begin{figure}
    \centering
    \includegraphics[width=1.0\textwidth]{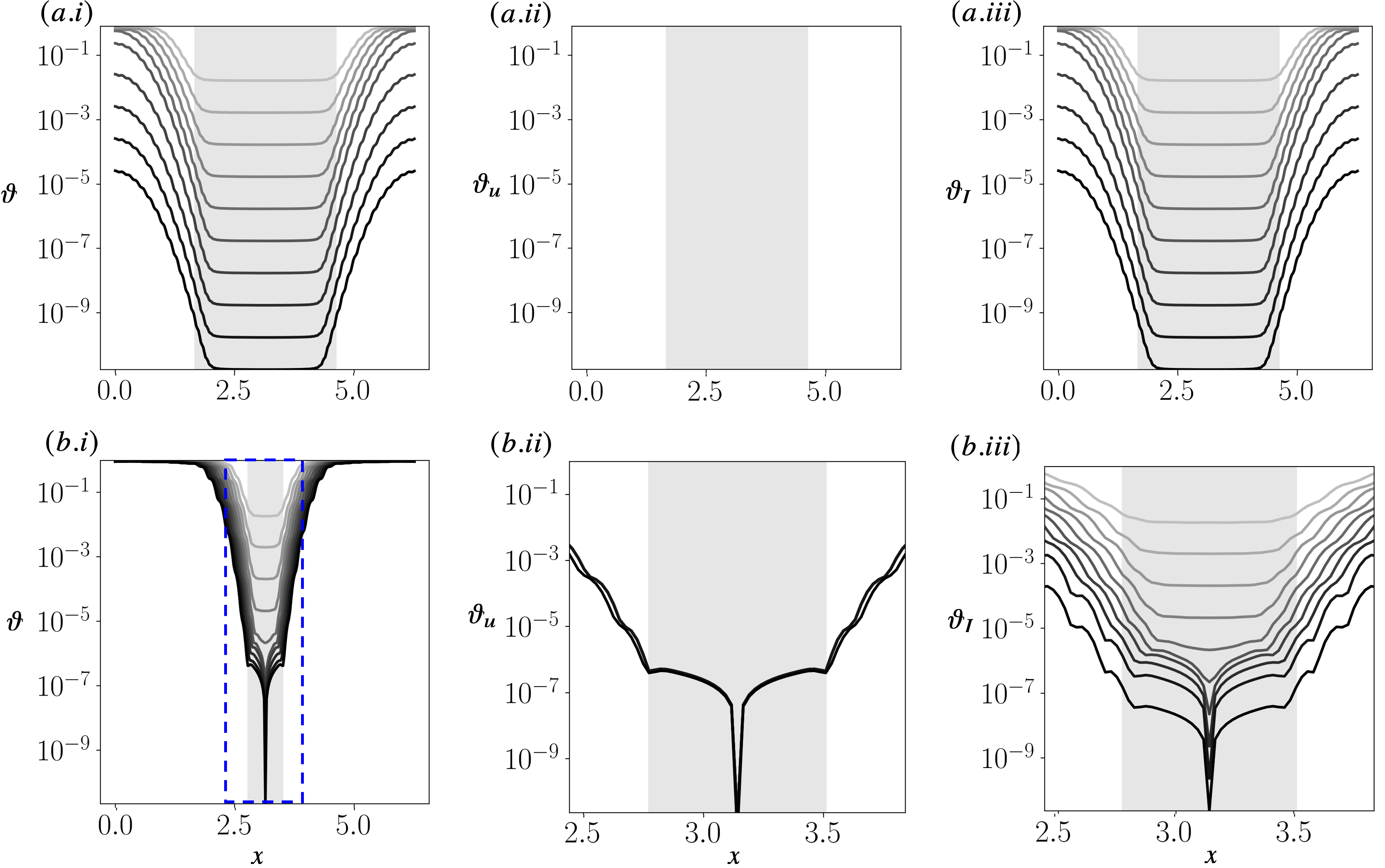}
    \caption{Decomposition of the total uncertainty $\vartheta$ (i) into $\vartheta_u$ (ii) and $\vartheta_{I}$ (iii) for $S_c$, using the long (a) and short (b) measurement time windows. Curves from light to dark show $\gamma=\{10^{-1},10^{-2},\ldots,10^{-10}\}$. The theoretical predictable zone is denoted by the gray area. Panels $(b.ii)$ and $(b.iii)$ are shown in the zoomed-in region (boxed by \dashedblue{}) of $(b.i)$}
    \label{fig: decomposition S_c}
\end{figure}

We return back to the computation of the posterior uncertainties $\tilde{\mathbf{\Sigma}}_{I}^{(n)}$ and $\tilde{\mathbf{\Sigma}}_{u}^{(n)}$, and their corresponding normalized posterior standard deviations, 
\begin{equation}\label{eq: pointwise uncertainty decomposition}
    \vartheta_{I}(x_j,t^{(n)})=\frac{\sqrt{\tilde{\Sigma}_{I,jj}^{(n)}}}{\varrho},\qquad
    \vartheta_u(x_j,t^{(n)})=\frac{\sqrt{\tilde{\Sigma}_{u,jj}^{(n)}}}{\varrho}.
\end{equation}
% YD: difficult to read here. Need to think about some guidence/simplifications. 
For both the long and short window cases, the decomposed relative uncertainties $\vartheta_{I}$ and $\vartheta_{u}$ are visualized at $t=20$ for prior spectrum $S_c$ in a wide range of noise level $\gamma$. 
For the long-window case (figure \ref{fig: decomposition S_c}$(a)$), $\vartheta_u=0$, and $\vartheta=\vartheta_{I}$ drops with $\gamma$ as expected. Therefore, as $\gamma\to 0$, the uncertainty $\vartheta$ of the full wave state $\mathbf{u}^{(n)}$ approaches zero, and $\mathbf{u}^{(n)}$ (including both surface elevation $\eta$ and surface potential $\psi$) can be exactly determined. In comparison, for the cases of short time window (figure \ref{fig: decomposition S_c}$(b)$), $\vartheta_u$ is nonzero and remains nearly unchanged in the considered range of $\gamma$, meanwhile $\vartheta_I$ approaches zero consistently. 
%This is because here the un-informed space is non-empty and $\mathbf{u}_{u}^{(n)}$ is in general nonzero. 
%The measurement does not help to reduce the uncertainty $\vartheta_{u}$ of $\mathbf{u}_{u}^{(n)}$. Consequently, no matter how accurate the measurements are (controlled by noise level $\gamma$), the uncertainty $\vartheta_u$ does not change.
%The data-informed part $\vartheta_{I}$ dominates the total uncertainty $\vartheta$ in the predictable zone for large noise level ($\gamma>10^{-5}$), and then become negligible with the reduction of $\gamma$. 
%Even though $\vartheta_{I}$ keep decreasing with $\gamma$, the un-informed part $\vartheta_I$ remain unchanged and acts as bottom floor of $\vartheta$ in the whole domain. 
At the limit of zero noise ($\gamma\to 0$),  $\mathbf{u}_{I}^{(n)}$ can be exactly determined ($\vartheta_{I}\to 0$), while $\mathbf{u}_{u}^{(n)}$ remains uncertain ($\vartheta_{u}\ne 0$). Therefore, the total wave state $\mathbf{u}^{(n)}$ has finite uncertainty at this limit. 
In summary, the measurement configuration (here the length of measurement time window) can affect the dimensions of the data-informed and un-informed subspaces, and consequently influence the limiting behavior of $\vartheta$ when $\gamma\to 0$. 
The case of short window is more relevant to realistic measurement of ocean surface waves, since measurement window can never be long enough to guarantee the exact recovery of the full wave state $\mathbf{u}^{(n)}$ on a infinitely large sea. In other words, $\mathbf{u}_{u}^{(n)}$ and $\vartheta_u$ are never zero on a large sea. 

Note that for the case of short measurement window, $\vartheta$ has low but nonzero value inside the theoretical predictable zone even when $\gamma \to 0$ (see figure \ref{fig: decomposition S_c}$(b.i)$). This observation seem to be in contradiction to the arguments held in \citet{qi2018predictable} that the wave inside this zone should be exactly predictable. An implicit assumption made in \citet{qi2018predictable} is that the wave amplitude $A(\omega)$ in the expression $\eta(x_m,t)=\text{Re}\left\{\int_{\omega_{\text{min}}}^{\omega_{\text{max}}}A(\omega)\mathrm{e}^{ik(\omega)x-i\omega t}\mathrm{d}\omega\right\}$ should be exactly solvable from the measurement data $\eta(x_m,t)$ taken in some finite time window. 
This, however, cannot be true because the precise computation of $A(\omega)$ using  temporal Fourier transform would require a infinitely long time sequence of $\eta(x_m,t)$, whereas the measurement time window is only finite. 
As a result, a small amount of uncertainty of $A(\omega)$ always exists when we compute it use the measurement data $\eta(x_m,t)$ in a finite time window. 
Consequently, the surface elevation field $\eta(x,t)$ computed from these estimated $A(\omega)$ would contains zero uncertainty even inside the predictable zone.  
%Such uncertainty of $A(\omega)$ produces uncertainty for the surface elevation $\eta(x,t)$ inside the predictable zone. 

\begin{figure}
    \centering
    \includegraphics[width=0.6\textwidth]{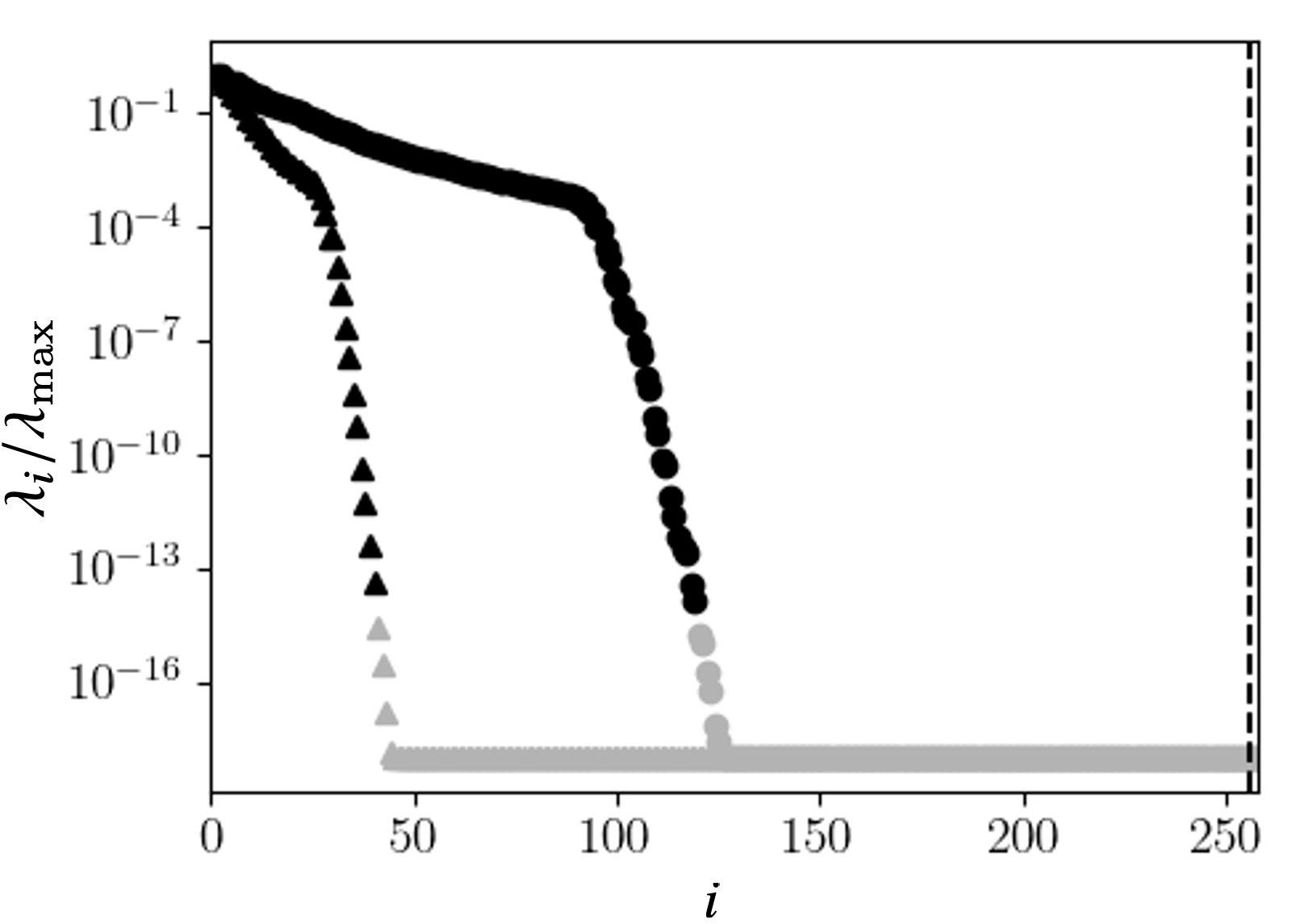}
    \caption{Eigenvalues $\lambda_i/\lambda_{\text{max}}$ for the wave spectrum $S_j$  with short (\mytriangle{black}) and long (\mycircle{black}) measurement time windows. Black and grey color represent the observable and un-observable modes divided by the cutoff $\lambda_i/\lambda_{\text{max}}>10^{-14}$, respectively. Vertical dashed lines mark the rank $r_0$ of the prior covariance.}
    \label{fig: eigenvalues S_j}
\end{figure}

\begin{figure}
    \centering
    \includegraphics[width=1.0\textwidth]{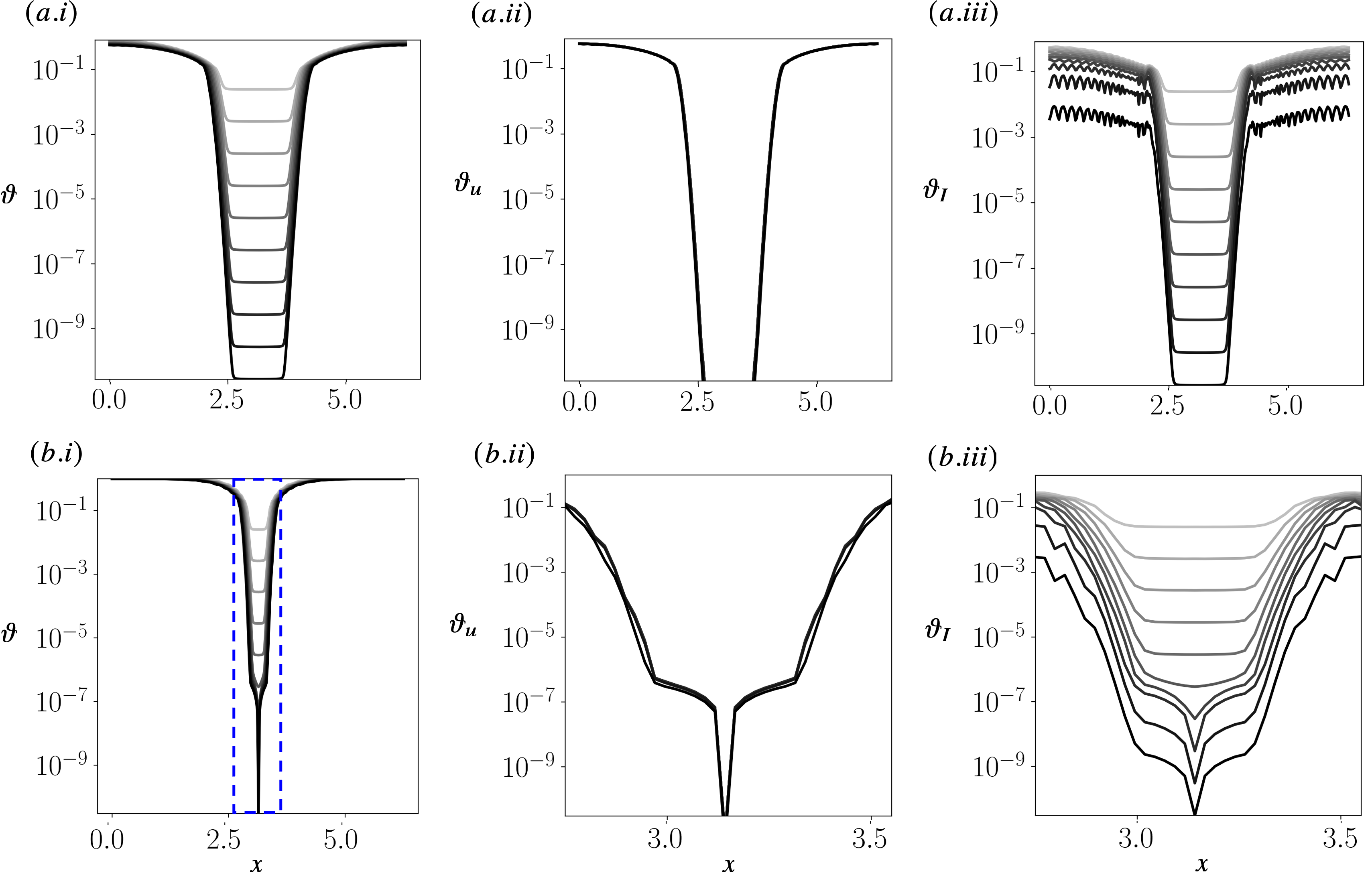}
    \caption{Decomposition of the total uncertainty $\vartheta$ (i) into $\vartheta_u$ (ii) and $\vartheta_{I}$ (iii) for $S_j$, using the long (a) and short (b) measurement time windows. Curves show $\gamma=\{10^{-1},10^{-2},\ldots,10^{-10}\}$ from light to dark. Panels $(b.ii)$ and $(b.iii)$ are shown in the zoomed-in region (boxed by \dashedblue{}) of $(b.i)$}
    \label{fig: decomposition S_j}
\end{figure}

We discuss another factor that influences the decomposition, which is the number of active wave modes in the prior distribution (or in the energy spectrum $S(k)$). We first recall that $r_0$, the length of $\boldsymbol{\xi}^{(0)}$,  represents the number of dimensions in which the wave is active at initial time.  Even for the same measurement configuration (which supplies the same amount of measurement data), a spectrum with larger $r_0$ or wider frequency support lead to more unknown waves to be inferred, and might result in a larger un-informed space. We now use the spectrum $S_j$, which has a larger dimension $r_0=256$ compared to that of $S_c$, to demonstrate this point. The relative eigenvalues $\lambda_i/\lambda_{\text{max}}$ of the observability Gramian $\mathbf{G}$ are shown in figure \ref{fig: eigenvalues S_j} for both time windows with $S_j$ spectrum. 
The long-measurement-window case for $S_j$ has more than half of its eigenvalues being zero (figure \ref{fig: eigenvalues S_j}), compared to the all-nonzero eigenvalues of $S_c$ case with the same measurement window (figure \ref{fig: eigenvalues S_c}). 
This is because the dimension $r_0$ for the $S_j$ spectrum is substantially larger than that of $S_c$, and there are more active wave modes to be inferred in the wave evolution. The same amount of data provided by the long-measurement-window configuration is enough to determine the wave evolution for the $S_c$ spectrum, but not enough for $S_j$.
%A non-empty un-informed space therefore appears by simply changing the wave spectrum from $S_c$ to $S_j$.  

For this prior spectrum $S_j$, we further examine some spatial profiles of $\vartheta$, $\vartheta_{I}$ and $\vartheta_{u}$  in a wide range of noise level $\gamma$ in figure \ref{fig: decomposition S_j}. The figures \ref{fig: decomposition S_j}$(a.i)$ and $(a.ii)$ confirms that a nonzero $\vartheta_{u}$ appears even in the long-measurement-window case, which corresponds to a part $\mathbf{u}_{u}^{(n)}$ in the wave state that is fundamentally unpredictable. Comparing figure \ref{fig: decomposition S_j} $(a.i)$ and $(b.i)$, a shorter measurement time window typically lead to a smaller spatial size of the low-uncertainty region and higher uncertainty near the measurement location. Note that some small scale oscillation appear in \ref{fig: decomposition S_j} $(a.iii)$, specifically at $x<2$ or $x >4.5$. This is usually because some very weakly informed eigenvectors (corresponding to small but nonzero $\lambda_i$) in $\mathbf{V}_I$ are oscillatory. These oscillatory components only contribute to a very small fraction of total uncertainty.

\section{Nonlinear waves}
\subsection{Nonlinear wave dynamics }\label{subsec: nonlinear dynamics}
To extend our framework to incorporate nonlinearity, we first construct a nonlinear operator $\mathcal{N}$ to be the realization of the evolution operator $\Phi$ in equation (\ref{eq:state}), i.e., $\mathbf{u}^{(n+1)}=\mathcal{N}(\mathbf{u}^{(n)})$. 
We introduce the governing equations of the nonlinear surface waves, which is the Zakharov form of the incompressible Euler's equation,
\begin{subequations}\label{eq:HOS}
    \begin{align}
 \frac{\partial \eta}{\partial t}+\frac{\partial \psi}{\partial x} \frac{\partial \eta}{\partial x}-\left[1+\frac{\partial \eta}{\partial x} \cdot \frac{\partial \eta}{\partial x}\right] \phi_z&=0, \\
\frac{\partial \psi}{\partial t}+\frac{1}{2} \frac{\partial \psi}{\partial x} \cdot \frac{\partial \psi}{\partial x}+\eta-\frac{1}{2}\left[1+\frac{\partial \eta}{\partial x}\frac{\partial \eta}{\partial x}\right] \phi_z^2&=0,
\end{align}
\end{subequations}
where $\psi(x,t)=\phi(x,\eta(x,t),t)$ and $\phi_z = \left.\frac{\partial \phi}{\partial z}\right\vert_{z=\eta}$ are the velocity potential and the vertical velocity evaluated at the free surface, respectively. 
The Dirichlet-to-Neumann problem from $\psi$ to $\phi_z$ is solved using an efficient pseudo-spectral method \citep{dommermuth1987high}. 
The operator $\mathcal{N}$ is therefore the one-step forward mapping of the $(\eta,\psi)$ field during a time step $\Delta t$, defined by the system (\ref{eq:HOS}) with a $4^{\text{th}}$-order Runga-Kutta temporal discretization.

\subsection{Approximate Bayesian inference for nonlinear wave dynamics}\label{subsec: method nonlinear}
We construct the Bayesian inversion framework for nonlinear wave fields in parallel to the linear version presented in \S (\ref{subsec: Bayesian linear}) by introducing the prior distribution $p(\boldsymbol{\xi}^{(0)})$ and the likelihood functions $p(\{\mathbf{m}^{(l)}\}_{l=N_{1},...,N_{2}}|\boldsymbol{\xi}^{(0)})$. For the prior distribution, we still adopt the standard Gaussian model (\ref{eq: prior}) for the initial Fourier coefficients $\boldsymbol{\xi}^{(0)}$. 

The previously discussed likelihood function (\ref{eq: linear likelihood}) and the subsequent posterior distribution (\ref{eq:linear posterior}) rely on the assumption of linear dynamics. For nonlinear dynamics such as (\ref{eq:HOS}), we first derive the exact expressions of the likelihood function and the posterior distribution, and then approximate the posterior using Laplace's method. With the nonlinear wave dynamics encoded in $\mathcal{N}$ (see \S \ref{subsec: nonlinear dynamics}), the likelihood function of the measurements $\{\mathbf{m}^{(l)}\}_{l=N_{1},...,N_{2}}$ conditioned on an initial Fourier coefficient vector  $\boldsymbol{\xi}^{(0)}$ is formed,
\begin{equation}\label{eq: nonlinear likelihood}
    p(\{\mathbf{m}^{(l)}\}_{l=N_{1},...,N_{2}}|\boldsymbol{\xi}^{(0)})\propto\mathrm{e}^{-\mathcal{J}_{\text{m}}(\boldsymbol{\xi}^{(0)})}, \qquad\mathcal{J}_\text{m}(\boldsymbol{\xi}^{(0)}) = \sum_{n=N_{1}}^{N_{2}}\frac{1}{2}\left\Vert\mathbf{M}\mathcal{N}^n(\mathbf{L}\boldsymbol{\xi}^{(0)})-\mathbf{m}^{(n)}\right\Vert^2_{\mathbf{R}^{-1}},
\end{equation}
where $\mathcal{N}^n(\cdot)$ represents the $n$-fold composition of $\mathcal{N}$.
Note that here we use $\mathbf{u}^{(0)} = \mathbf{L}\boldsymbol{\xi}^{(0)}$ to express the initial wave state $\mathbf{u}^{(0)}$ in terms of $\boldsymbol{\xi}^{(0)}$, which uses the analytical solution of linear waves (see (\ref{eq: L definition})). However, this is only a minor approximation at the initial time. The major effect of nonlinearity in the wave evolution is still included in the nonlinear operator $\mathcal{N}$. 
\del{We can compute the Hessian matrix of $\mathcal{J}_\text{m}(\boldsymbol{\xi}^{(0)})$,
\begin{equation}\label{eq: full Hessian nonlinear}
    \frac{\partial^2\mathcal{J}_\text{m}}{\partial (\boldsymbol{\xi}^{(0)})^2}=\sum_{n=N_{1}}^{N_{2}}\left(\frac{\partial\mathcal{N}^n}{\partial \mathbf{u}_{0}}\right)^{\dagger} \mathbf{M}^{\dagger}\mathbf{R}^{-1} \mathbf{M} \frac{\partial\mathcal{N}^n}{\partial \mathbf{u}_{0}}-\frac{\partial^2\mathcal{N}^n}{\partial (\mathbf{u}_{0})^2}\mathbf{R}^{-1}\left(\mathbf{M}\mathcal{N}^n(\boldsymbol{\xi}^{(0)})-\mathbf{m}^{(n)}\right)
\end{equation}
At the location of the true solution $\boldsymbol{\xi}^{(0)}_{*}$, $\mathbf{M}\mathcal{N}^n(\boldsymbol{\xi}^{(0)}_{*})-\mathbf{m}^{(n)}=0$, therefore we can define the `local observability Gramian' at the true solution to be,
\begin{equation}
    \mathbf{G}=\sum_{n=N_{1}}^{N_{2}}\left(\frac{\partial\mathcal{N}^n}{\partial \mathbf{u}_{0}}\right)^{\dagger} \mathbf{M}^{\dagger}\mathbf{R}^{-1} \mathbf{M} \frac{\partial\mathcal{N}^n}{\partial \mathbf{u}_{0}}
\end{equation}
As before, this matrix $\mathbf{G}$ quantifies how well the data constrains the solution at the vicinity of the truth. }
Using the Bayes formula, we multiply the prior $p(\boldsymbol{\xi}^{(0)})$ from (\ref{eq: prior}) with the likelihood $p(\{\mathbf{m}^{(l)}\}_{l=N_{1},...,N_{2}}|\boldsymbol{\xi}^{(0)})$ in (\ref{eq: nonlinear likelihood}) and arrive at a posterior distribution of $\boldsymbol{\xi}^{(0)}$, 
\begin{equation}\label{eq:nonlinear posterior}
    p(\boldsymbol{\xi}^{(0)}|\{\mathbf{m}^{(l)}\}_{l=N_1,...,N_2}) = \frac{1}{C}\mathrm{e}^{-\mathcal{J}(\boldsymbol{\xi}^{(0)})},\quad\mathcal{J}=\mathcal{J}_{\mathrm{m}}(\boldsymbol{\xi}^{(0)})+\mathcal{J}_0(\boldsymbol{\xi}^{(0)}), 
\end{equation}
where $C$ is the normalization constant that guarantees unit mass of the posterior distribution.
Since $\mathcal{N}$ is a nonlinear operator, the exponential part $-\mathcal{J}(\boldsymbol{\xi}^{(0)})$ is not quadratic with respect to $\boldsymbol{\xi}^{(0)}$. Therefore the posterior distribution (\ref{eq:nonlinear posterior}) is non-Gaussian and no closed-form posterior covariance can be derived in general. 

To estimate the posterior uncertainty, we use a Laplace approximation to represent the full posterior distribution $p(\boldsymbol{\xi}^{(0)}|\{\mathbf{m}^{(l)}\}_{l=N_1,...,N_2})$ locally by a Gaussian distribution centered at the MAP solution.
The MAP solution, denoted by $\boldsymbol{\xi}^{(0)}_{\star}$, contains the initial Fourier coefficients that maximize the posterior distribution, or equivalently minimize $\mathcal{J}$,
\begin{equation}\label{eq:MAP nonlinear}
\boldsymbol{\xi}^{(0)}_{\star}=\arg \min \mathcal{J}(\boldsymbol{\xi}^{(0)})
\end{equation}
Intuitively, the MAP solution $\boldsymbol{\xi}^{(0)}_{\star}$ contains the normalized initial Fourier coefficients that can produce a wave field that not only reproduces the measurement data but also remains consistent with the prescribed energy spectrum $S(k)$. Therefore, it produces the `best' wave field we can possibly reconstruct. We then expand $\mathcal{J}$ around $\boldsymbol{\xi}^{(0)}_{\star}$ up to second order. Since
$\left.\nabla_{\boldsymbol{\xi}}\mathcal{J}\right|_{\boldsymbol{\xi}^{(0)}=\boldsymbol{\xi}^{(0)}_{\star}}=0$ by the optimality of $\boldsymbol{\xi}^{(0)}_{\star}$,
the first-order term vanishes and we obtain
\begin{equation}\label{eq:laplace_taylor}
    \mathcal{J}(\boldsymbol{\xi}^{(0)})
    =
    \mathcal{J}(\boldsymbol{\xi}^{(0)}_{\star})
    +
    \frac{1}{2}
    (\boldsymbol{\xi}^{(0)}-\boldsymbol{\xi}^{(0)}_{\star})^{\dagger}
    \mathbf{H}_{\star}
    (\boldsymbol{\xi}^{(0)}-\boldsymbol{\xi}^{(0)}_{\star})+O(\left\Vert\boldsymbol{\xi}^{(0)}-\boldsymbol{\xi}^{(0)}_{\star}\right\Vert^3),
\end{equation}
where
\begin{equation}\label{eq:laplace_precision}
    \mathbf{H}_{\star}:=
    \left.
    \frac{\partial}{\partial\boldsymbol{\xi}^{(0)}}
    \left(\frac{\partial\mathcal{J}}
    {\partial(\boldsymbol{\xi}^{(0)})^{*}}\right)
    \right|_{\boldsymbol{\xi}^{(0)}=\boldsymbol{\xi}^{(0)}_{\star}}
    =
    \mathbf{I}_{r_0\times r_0}
    +
    \mathbf{H}_{\mathrm{m},\star}
\end{equation}
is the Hessian matrix of $\mathcal{J}$ evaluated at $\boldsymbol{\xi}^{(0)}_{\star}$. The detailed expression of the measurement Hessian $\mathbf{H}_{\mathrm{m},\star}$ is derived in appendix \S\ref{app: nonlinear posterior derivation}. By plugging (\ref{eq:laplace_taylor}) into (\ref{eq:nonlinear posterior}), omitting the third and higher order terms and enforcing the normalization (total probability being equal to $1$), an approximate Gaussian distribution $\tilde{p}$ is constructed as,
\begin{eqnarray}\label{eq:laplace_posterior}
    \tilde{p}(\boldsymbol{\xi}^{(0)}|\{\mathbf{m}^{(l)}\}_{l=N_1,\ldots,N_2})
    =
    \frac{\exp\left[
    -\frac{1}{2}
    (\boldsymbol{\xi}^{(0)}-\boldsymbol{\xi}^{(0)}_{\star})^{\dagger}
    \mathbf{H}_{\star}
    (\boldsymbol{\xi}^{(0)}-\boldsymbol{\xi}^{(0)}_{\star})
    \right]}{(2\pi)^{r_0/2}\det(\mathbf{H}_{\star})^{-1/2}}
\end{eqnarray}
From (\ref{eq:laplace_posterior}) we identify $\tilde{\mathbf{\Sigma}}^{(0)}_{\boldsymbol{\xi}}:=\mathbf{H}_{\star}^{-1}$ as the covariance matrix of the approximate posterior distribution $\tilde{p}$.  

Once the posterior covariance $\tilde{\mathbf{\Sigma}}^{(0)}_{\boldsymbol{\xi}}$ of the initial Fourier coefficient $\boldsymbol{\xi}^{(0)}$ is obtained, we propagate it through the solution trajectory. Let
$\mathbf{u}_{\star}^{(n)} := \mathcal{N}^n(\mathbf{L}\boldsymbol{\xi}_{\star}^{(0)})=\mathcal{N}^n(\mathbf{u}_{\star}^{(0)})$
be the MAP trajectory. 
A first-order expansion of a generic solution trajectory $\mathbf{u}^{(n)} = \mathcal{N}^n(\mathbf{L}\boldsymbol{\xi}^{(0)})$ around $\boldsymbol{\xi}^{(0)}_{\star}$ yields
\begin{equation}\label{eq:nonlinear_pushforward_linearization}
    \mathbf{u}^{(n)}
    -
    \mathbf{u}^{(n)}_{\star}=
    \mathbf{T}^{(n)}_{\star}\mathbf{L}
    \left(\boldsymbol{\xi}^{(0)}-\boldsymbol{\xi}^{(0)}_{\star}\right)
    +
    O(\|\boldsymbol{\xi}^{(0)}-\boldsymbol{\xi}^{(0)}_{\star}\|^2), 
\end{equation}
where
\begin{equation}
    \mathbf{T}^{(n)}_{\star}
    :=
    \left.
    \frac{\partial\mathcal{N}^n(\mathbf{u}^{(0)})}
    {\partial\mathbf{u}^{(0)}}
    \right|_{\mathbf{u}^{(0)}=\mathbf{u}^{(0)}_{\star}}
\end{equation}
is the tangent linear operator of the nonlinear evolution along the MAP trajectory from time step $0$ to time step $n$. Neglecting the higher-order terms,
(\ref{eq:nonlinear_pushforward_linearization}) maps the deviation
$\boldsymbol{\xi}^{(0)}-\boldsymbol{\xi}^{(0)}_{\star}$
linearly to the deviation
$\mathbf{u}^{(n)}-\mathbf{u}^{(n)}_{\star}$.
Therefore, the Gaussian posterior distribution
$\boldsymbol{\xi}^{(0)} \sim N(\boldsymbol{\xi}^{(0)}_{\star},
\tilde{\mathbf{\Sigma}}^{(0)}_{\boldsymbol{\xi}})$ centered around $\boldsymbol{\xi}^{(0)}_{\star}$
(see (\ref{eq:laplace_posterior})) is mapped to a Gaussian distribution of
$\mathbf{u}^{(n)}$ centered around $\mathbf{u}_{\star}^{(n)}$,
\begin{equation}\label{eq:posterior_time_n_gaussian}
    \mathbf{u}^{(n)} \sim
    N(\mathbf{u}_{\star}^{(n)},\tilde{\mathbf{\Sigma}}^{(n)}), 
\end{equation}
with covariance
\begin{equation}\label{eq:posterior_covariance_propagation}
    \tilde{\mathbf{\Sigma}}^{(n)}
    =
    \mathbf{T}^{(n)}_{\star}\mathbf{L}
    \tilde{\mathbf{\Sigma}}^{(0)}_{\boldsymbol{\xi}}\mathbf{L}^{\dagger}
    \left(
    \mathbf{T}^{(n)}_{\star}
    \right)^{\dagger}.
\end{equation}
%\codex{[Check: Standardize the covariance subscript as $\boldsymbol{\xi}$, consistent with the preceding definition; this equation uses $\xi$.]}
Similar to the linear case, here we also define the posterior standard deviation $\rho_j^{(n)}=\sqrt{(\tilde{\mathbf{\Sigma}}^{(n)})_{jj}}$ (for $1<j<J$) to describe the uncertainty level of the surface elevation $\eta$ at $x_j$, and define the normalized uncertainty $\vartheta^{(n)}_j = \rho_j^{(n)}/\varrho$ to quantify the reduction of posterior standard deviation $\rho_j^{(n)}$ compared with the prior standard deviation $\varrho$. Here we note that although $\varrho$ could be time-dependent along the nonlinear wave evolution, the time window we consider in the predictable zone problem is much smaller compared to the kinetic time scale, such that the energy spectrum and $\varrho$ are nearly constant in time. 
%\codex{[Check: In the nonlinear case, $\varrho$ is the stationary linear-prior standard deviation from (\ref{eq: normalized uncertainty}); nonlinear propagation need not preserve it. Define $\varrho$ as a fixed reference scale, or compute the nonlinear prior variance before interpreting $\vartheta$ as a posterior-to-prior reduction.]}

\subsection{Numerical methods}\label{subsec: numerical methods}
%For the numerical experiments, the measurement data are first generated from a reference solution sampled from the prior distribution. The posterior inference algorithm then takes these measurements as input and computes the MAP trajectory and posterior covariance.

\begin{algorithm}[h]
	\SetAlgoLined
	\textbf{Step 1}: Reference solution and measurements\;
	\Indp
	\textbf{Inputs}: $S(k)$, $\mathbf{L}$, $\mathbf{M}$, $\mathbf{R}$, $N_1$, $N_2$, $N_T$,  $\mathcal{N}$\;
	\textbf{Begin}\;
	\Indp
	Sample $\boldsymbol{\xi}_{\mathrm{ref}}^{(0)}$ from (\ref{eq: prior})\;
    Compute $\mathbf{u}_{\mathrm{ref}}^{(0)}=\mathbf{L}\boldsymbol{\xi}_{\mathrm{ref}}^{(0)}$ using (\ref{eq: L definition})\;
	\textbf{For} $n\in\{1,\ldots,N_T\}$:\\
	\Indp
	Compute $\mathbf{u}_{\mathrm{ref}}^{(n)}$ using (\ref{eq:HOS})\;
	\Indm
	\textbf{End}\\
	\textbf{For} $n\in\{N_1,\ldots,N_2\}$:\\
	\Indp
	Compute $\mathbf{m}^{(n)}$ from $\mathbf{u}_{\mathrm{ref}}^{(n)}$ using (\ref{eq:obs})\;
	\Indm
	\textbf{End}\\
	\Indm
	\textbf{End}\\
	\Indm

	\textbf{Step 2}: MAP estimation\;
	\Indp
	\textbf{Inputs}: $\{\mathbf{m}^{(n)}\}_{n=N_1}^{N_2}$, $\mathbf{L}$, $\mathbf{M}$, $\mathbf{R}$, $N_1$, $N_2$, $N_T$,  $\mathcal{N}$\;
	\textbf{Begin}\;
	\Indp
	Assemble $\mathcal{J}$ using (\ref{eq: nonlinear likelihood}) and (\ref{eq:nonlinear posterior})\;
	Compute $\boldsymbol{\xi}_{\star}^{(0)}$ using (\ref{eq:MAP nonlinear}) with L-BFGS\;
	\textbf{For} $n\in\{1,\ldots,N_T\}$:\\
	\Indp
	Compute $\mathbf{u}_{\star}^{(n)}$ from $\boldsymbol{\xi}_{\star}^{(0)}$ using (\ref{eq: L definition}) and (\ref{eq:HOS})\;
	\Indm
	\textbf{End}\\
	\Indm
	\textbf{End}\\
	\Indm

	\textbf{Step 3}: Laplace posterior covariance\;
	\Indp
	\textbf{Inputs}: $\boldsymbol{\xi}_{\star}^{(0)}$, $\{\mathbf{u}_{\star}^{(n)}\}_{n=0}^{N_T}$, $\mathbf{L}$, $\mathbf{M}$, $\mathbf{R}$, $N_1$, $N_2$, $N_T$,  $\mathcal{N}$\;
	\textbf{Begin}\;
	\Indp
	\textbf{For} $n\in\{1,\ldots,N_T\}$:\\
	\Indp
	Compute $\mathbf{T}_{\star}^{(n)}$, $(\mathbf{T}_{\star}^{(n)})^{\dagger}$, and the second-order derivative of $\mathcal{N}^n$ using automatic differentiation\;
	\Indm
	\textbf{End}\\
	Assemble $\mathbf{H}_{\star}$ using (\ref{eq:laplace_precision}) and (\ref{eq:hessian_nonlinear_exact}).

	Compute $\tilde{\mathbf{\Sigma}}^{(0)}_{\xi}=\mathbf{H}_{\star}^{-1}$\;
	\textbf{For} $n\in\{1,\ldots,N_T\}$:\\
	\Indp
	Compute $\tilde{\mathbf{\Sigma}}^{(n)}$ using (\ref{eq:posterior_covariance_propagation})\;
	\Indm
	\textbf{End}\\
	\Indm
	\textbf{End}\\
	\Indm

	\caption{Numerical computation of the posterior covariance}
	\label{alg:posterior_computation}
\end{algorithm}
%\codex{[Check: Steps 2--3 use $\mathbf{L}$, $N_T$, the measurements and the assimilation indices explicitly, but omit some of them from their input lists; Step 3 instead lists $\mathbf{\Sigma}^{(0)}$ without using it. Make the inputs consistent with the stated computation.]}
%\codex{[Check: Before forming $\mathbf{H}_{\star}^{-1}$, report the positive-definiteness check for $\mathbf{H}_{\star}$ (or the treatment of non-positive eigenvalues), since the residual-dependent Hessian term has no fixed sign.]}

We first generate the measurement data using a synthetic approach. 
A reference solution is computed on a spatial domain $x\in[0,2\pi]$ by sampling a initial condition from the prior distribution and marching it forward in time using the wave dynamics (\ref{eq:HOS}). The synthetic measurements $\{\mathbf{m}^{(n)}\}_{n=N_1,...,N_2}$ is then formed by taking data from the reference solution and contaminating it with noise. The reference solution is thereafter considered hidden and not accessible in the Bayesian inference computations below. 
%The wave equations (\ref{eq:HOS}) are discretized using a higher order spectral method \citep{dommermuth1987high}.

After the synthetic measurements are obtained, we compute the posterior covariance $\tilde{\mathbf{\Sigma}}^{(n)}$ numerically using the summarized Algorithm \ref{alg:posterior_computation}, which is also briefly explained as follows. 
%Since the Laplace posterior distribution (\ref{eq:laplace_posterior}) exactly reduces to the posterior distribution (\ref{eq:linear posterior}) for linear models when the nonlinear terms are removed, only the posterior inference formulations for the nonlinear system is numerically implemented. The results for the linear system is obtained by turning off the nonlinear terms in equation (\ref{eq:HOS}).
%using the methods discussed in \S \ref{subsec: Gramian and Bayesian posterior} and \ref{subsec: Extension to nonlinear dynamics} respectively.
%For the linear wave models, the state transition operator $\mathbf{A}$ is constructed numerically in wavenumber space for all Fourier coefficients by a phase shift, and transformed back to physical space using discrete Fourier transform. The posterior covariance $\tilde{\mathbf{\Sigma}}^{(n)}$ is then computed exactly using the matrix expressions shown in (\ref{eq: posterior mean cov}) and (\ref{eq: mean cov propagate linear}).
We first search for an MAP initial condition $\boldsymbol{\xi}^{(0)}_{\star}$ by minimizing $\mathcal{J}$ in (\ref{eq:nonlinear posterior}) using L-BFGS gradient optimization \citep{liu1989limited}. We then compute the posterior covariance $\tilde{\mathbf{\Sigma}}^{(0)}_{\boldsymbol{\xi}}$ by evaluating (\ref{eq:laplace_precision}) and (\ref{eq:hessian_nonlinear_exact}) and taking the inverse of $\mathbf{H}_{\star}$. The posterior covariance for the wave state $\tilde{\mathbf{\Sigma}}^{(n)}$ is obtained from $\tilde{\mathbf{\Sigma}}^{(0)}_{\boldsymbol{\xi}}$ by (\ref{eq:posterior_covariance_propagation}), and the normalized posterior uncertainty $\vartheta$ is subsequently extracted. 

We furthermore discuss a few technical details in the numerical implementation. The evaluation of (\ref{eq:hessian_nonlinear_exact}) requires the tangent operator $\mathbf{T}^{(n)}_{\star}$, its adjoint $\left(\mathbf{T}^{(n)}_{\star}\right)^{\dagger}$, and the second-order derivative of the nonlinear evolution along the MAP solution trajectory. The optimization of $\mathcal{J}$ requires the gradient $\nabla_{\boldsymbol{\xi}}\mathcal{J}$. These objects are all computed using automatic differentiation enabled by the JAX library \citep{jax2018github}.

\subsection{Numerical results}\label{subsec: results nonlinear}

\begin{figure}
    \centering
    \includegraphics[width=1.0\textwidth]{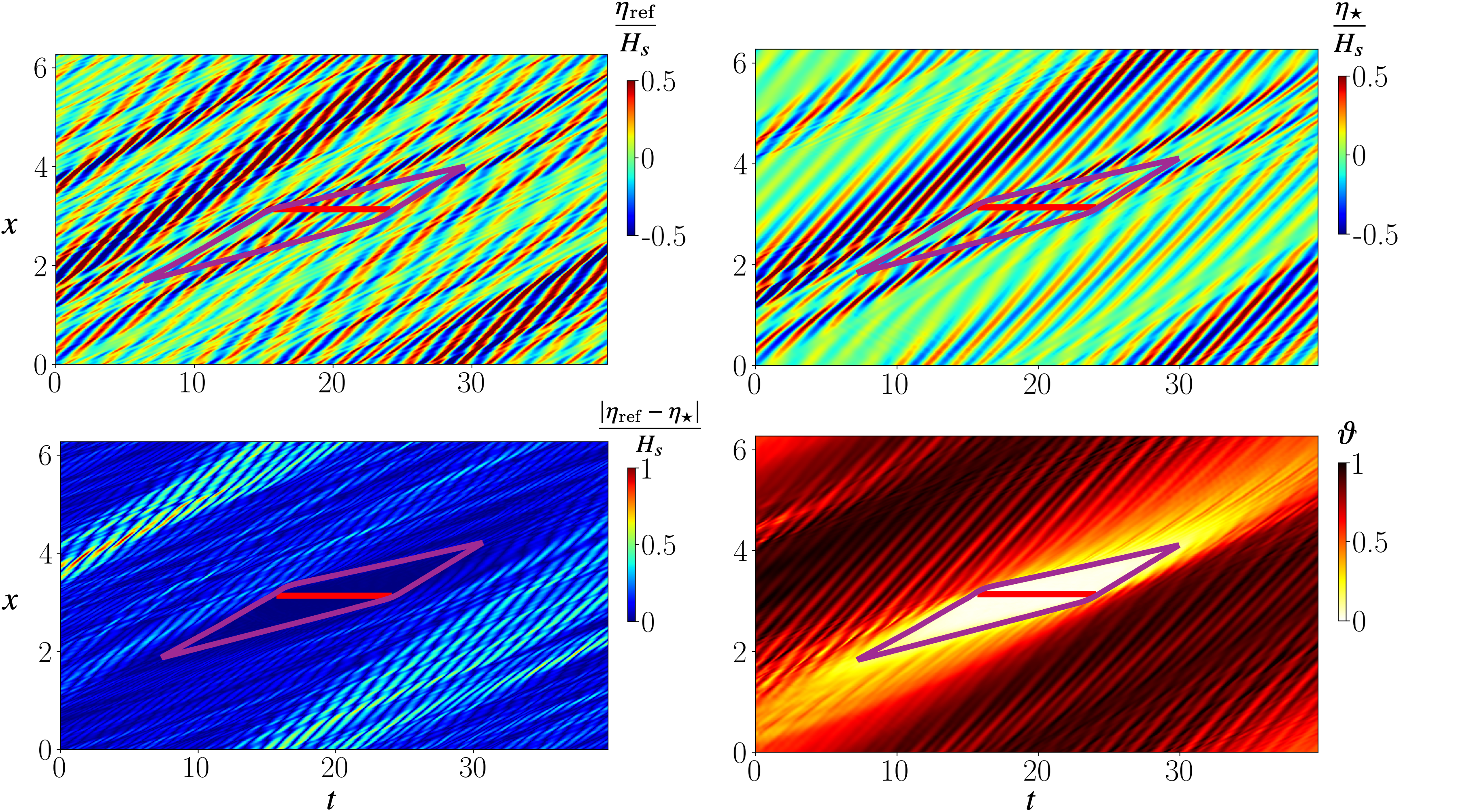}
    \caption{The $x$-$t$ iso-contour of $\eta_{\text{ref}}$ $(a)$, $\eta_{\star}$ $(b)$, $|\eta_{\text{ref}}-\eta_{\star}|$ $(c)$ normalized by $H_s$, and $\vartheta$ $(d)$. Each panel is marked with the  measurement set (\fullthickred{}) and a region where the $|\eta_{\text{ref}}-\eta_{\star}|/H_s\lessapprox 0.05$ (\fullpurple{}). }
    \label{fig: MAP solution}
\end{figure}

\begin{figure}
    \centering
    \includegraphics[width=1.0\textwidth]{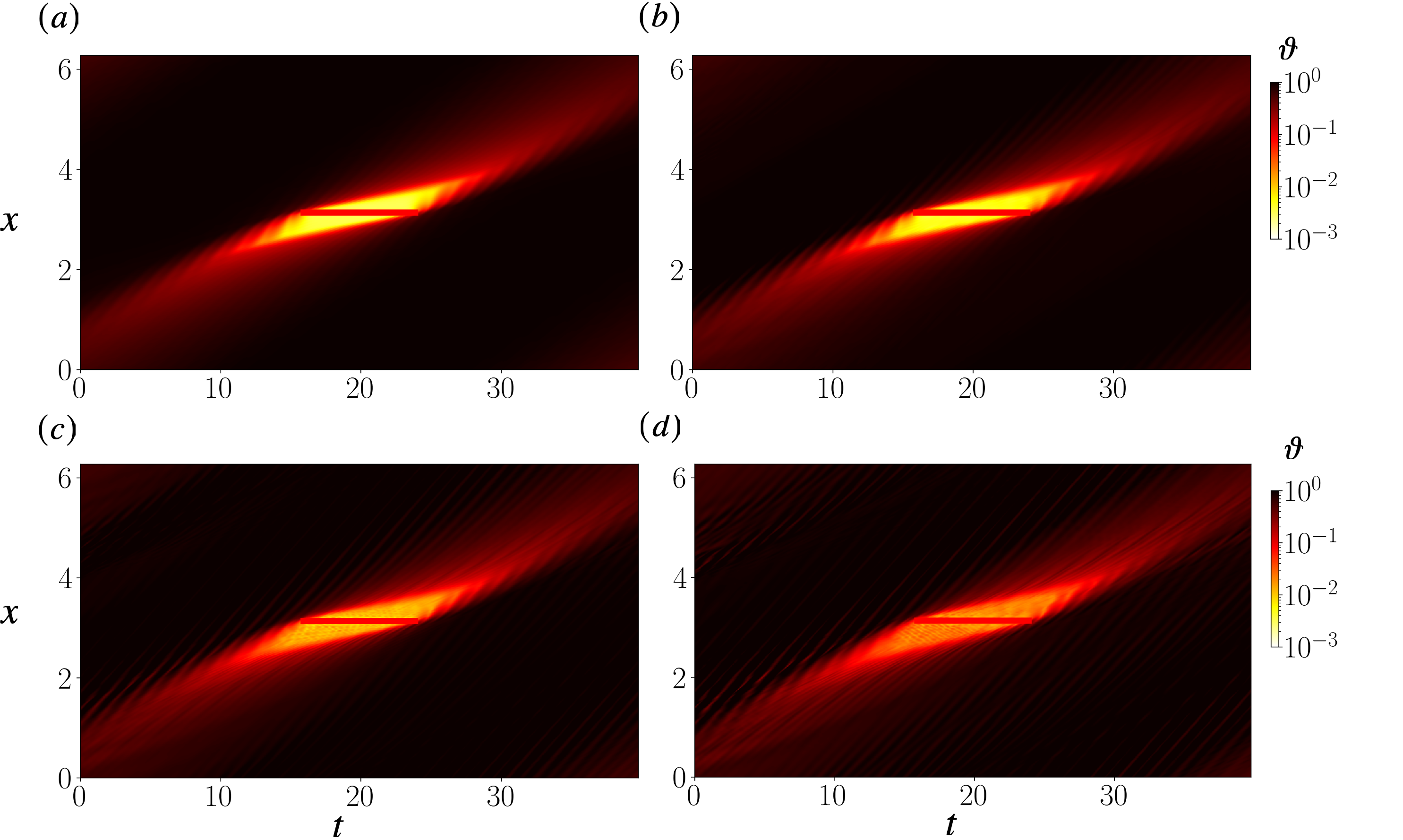}
    \caption{The $x$-$t$ map of $\vartheta$ with energy spectrum $S_j$ and $k_p=10$ for linear wave $(a)$ and $\epsilon=\{0.03,0.06,0.09\}$(panels $(b$-$d)$).}
    \label{fig: nonlinear rho jonswap x t}
\end{figure}

\begin{figure}
    \centering
    \includegraphics[width=1.0\textwidth]{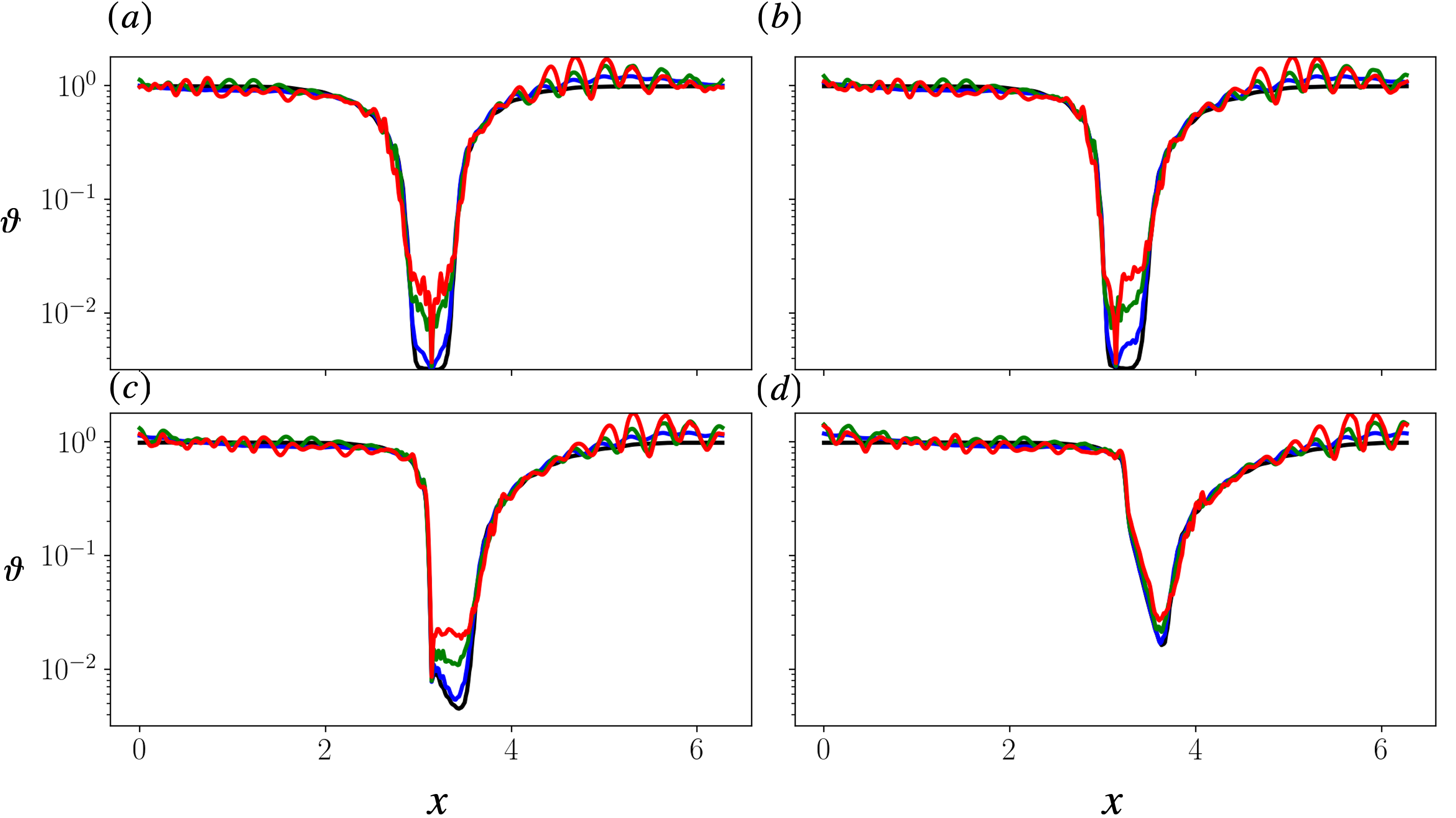}
    \caption{The spatial profile of $\vartheta$ with energy spectrum $S_j$ at $t = \{20, 22,24,26\}$ ($(a$-$d)$) with linear wave and three nonlinearity levels $\epsilon=\{0.03,0.06,0.09\}$ }
    \label{fig: nonlinear rho jonswap x 1d}
\end{figure}

We present the numerical results of the posterior uncertainty for nonlinear wave fields. The spatiotemporal domain setup and measurement configurations are replicated from \S \ref{subsec: results linear}, and the JONSWAP spectrum $S_j$ with $k_p=10$ is adopted in this section. The reference wave field is generated by marching forward an initial condition with $\epsilon:=H_sk_p/2=0.03$, and the surface elevation $\eta_{\text{ref}}(x,t)$ is visualized in figure \ref{fig: MAP solution}$(a)$. Using only the measurement data, we compute the MAP solution $\eta_{\star}(x,t)$ (figure \ref{fig: MAP solution}$(b)$), which is the optimal estimation of the surface elevation produced by the Bayesian framework. The pointwise deviation $|\eta_{\text{ref}}-\eta_{\star}|$ in figure \ref{fig: MAP solution}$(c)$ shows that the MAP reconstruction agrees closely with the reference solution only within a limited region surrounding the measurements. While $|\eta_{\text{ref}}-\eta_{\star}|$ can be evaluated here because $\eta_{\text{ref}}$ is known in the model problem, this deviation is unavailable in practical applications where $\eta_{\text{ref}}$ is unknown. Moreover, the theoretical predictable-zone \citep{morris1998short,qi2018nonlinear} is not directly applicable because of the unbounded frequency support of the $S_j$ spectrum and nonlinear wave dynamics. 
In contrast, the posterior standard deviation $\vartheta$, shown in figure \ref{fig: MAP solution}$(d)$, provides an uncertainty-based indication of where the MAP field $\eta_{\star}$ can be trusted without requiring knowledge of the true solution. In particular, the low-$\vartheta$ region closely coincides with the region of small $|\eta_{\text{ref}}-\eta_{\star}|$. We emphasize that the two quantities $\vartheta$ and $|\eta_{\text{ref}}-\eta_{\star}|$, although visually exhibit similar pattern, need not agree pointwise because they fundamentally have different meanings. In particular,  $\vartheta$ characterizes the local spread of the whole posterior distribution, whereas $|\eta_{\text{ref}}-\eta_{\star}|$ is the error between the reference field and the MAP estimation (which is one particular realization of the posterior distribution). Nevertheless, their agreement in the extent of the low-error region demonstrates that the Bayesian posterior uncertainty $\vartheta$ not only quantifies predictability, but also provides useful indication about the accuracy of the reconstructed MAP solution $\eta_{\star}$ relative to the true wave field $\eta_{\text{ref}}$.

Using the Bayesian framework, we investigate the effect of nonlinearity on the predictability of the surface wave field. For this purpose, we generate initial conditions with different levels of wave steepness $\epsilon:=H_sk_p/2=\{0.03,0.06,0.09\}$ by taking different values of $H_s$, march them forward to obtain the reference solution, and extract the measurements. The measurement noise level $\gamma=0.0025H_s$ is taken as a fixed percentage of $H_s$ such that the signal-to-noise ratio remain the same across different cases. The normalized posterior standard deviation $\vartheta$ is computed and shown on the spatiotemporal plane in figure \ref{fig: nonlinear rho jonswap x t}. For increasingly higher nonlinearity, the value of $\vartheta$ becomes higher in the low-uncertainty region (or the predictable zone)
%and becomes slightly lower outside of the predictable zone. 
This observation is made more clearly on the spatial profiles of $\vartheta$ at $t=\{20,22,24,26\}$ shown in figure \ref{fig: nonlinear rho jonswap x 1d}. These results indicate that with the increase of $\epsilon$, the wave field inside the predictable zone becomes less predictable (i.e., contains higher uncertainty), and the uncertainty difference between the inside and outside of the predictable zone becomes smaller. The concept of `predictable zone' itself becomes less well-defined, especially for panels \ref{fig: nonlinear rho jonswap x t}$(c,d)$ where there does not exist a sharp boundary that separates the high-uncertainty and low-uncertainty regions. 

We explain this phenomenon using a simple argument based on nonlinear modal interactions. In the linear case, the wave elevation $\eta(x_0,t_0)$ at any point $(x_0,t_0)$ in the predictable zone can be written in terms of the measured $\eta$ \citep{qi2018predictable}, with a slight uncertainty due to the reason of finite measurement time window we discussed in \S \ref{subsec: results linear}. We therefore can write a functional dependence $\eta(x_0,t_0)\approx F(\left\{\mathbf{m}^{(l)}\right\}_{l=N_1,...,N_2})$, where $F$ is a mapping from the measurement data to $\eta(x_0,t_0)$. In the nonlinear case, such simple functional dependence no longer holds. Due to the nonlinear modal interaction, the surface elevation $\eta(x_0,t_0)$ now have stronger dependence on unmeasured wave field outside of the predictable zone. Such dependence on unmeasured, outside wave field brings higher uncertainty into the predictable zone. Of course by a similar argument, the surface elevation $\eta(x_1,t_1)$ at a point $(x_1,t_1)$ outside of the predictable zone can have dependence on measured data, therefore the uncertainty of $\eta(x_1,t_1)$ could be slightly reduced. This latter effect is apparently very weak compared to the nearly $O(1)$ value of $\vartheta(x_1,t_1)$, therefore cannot be clearly observed in figures \ref{fig: nonlinear rho jonswap x t} and \ref{fig: nonlinear rho jonswap x 1d}. We recognize this effect as a redistribution of uncertainty due to nonlinear effect. 

\begin{figure}
    \centering
    \includegraphics[width=0.6\textwidth]{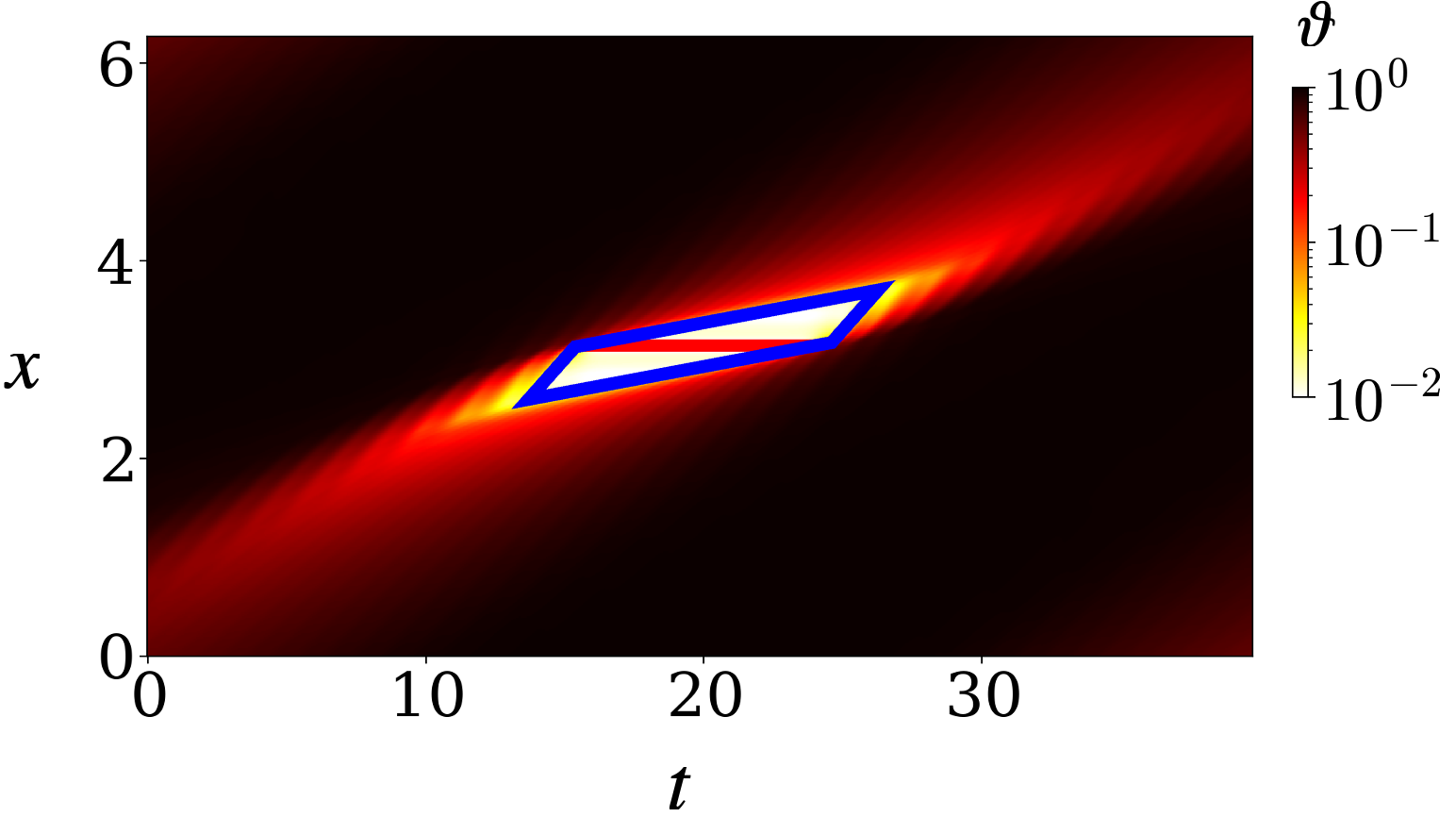}
    \caption{The numerical predictable zone, marked by the blue line contour.  }
    \label{fig: pred_zone_Z}
\end{figure}

\begin{figure}
    \centering
    \includegraphics[width=0.8\textwidth]{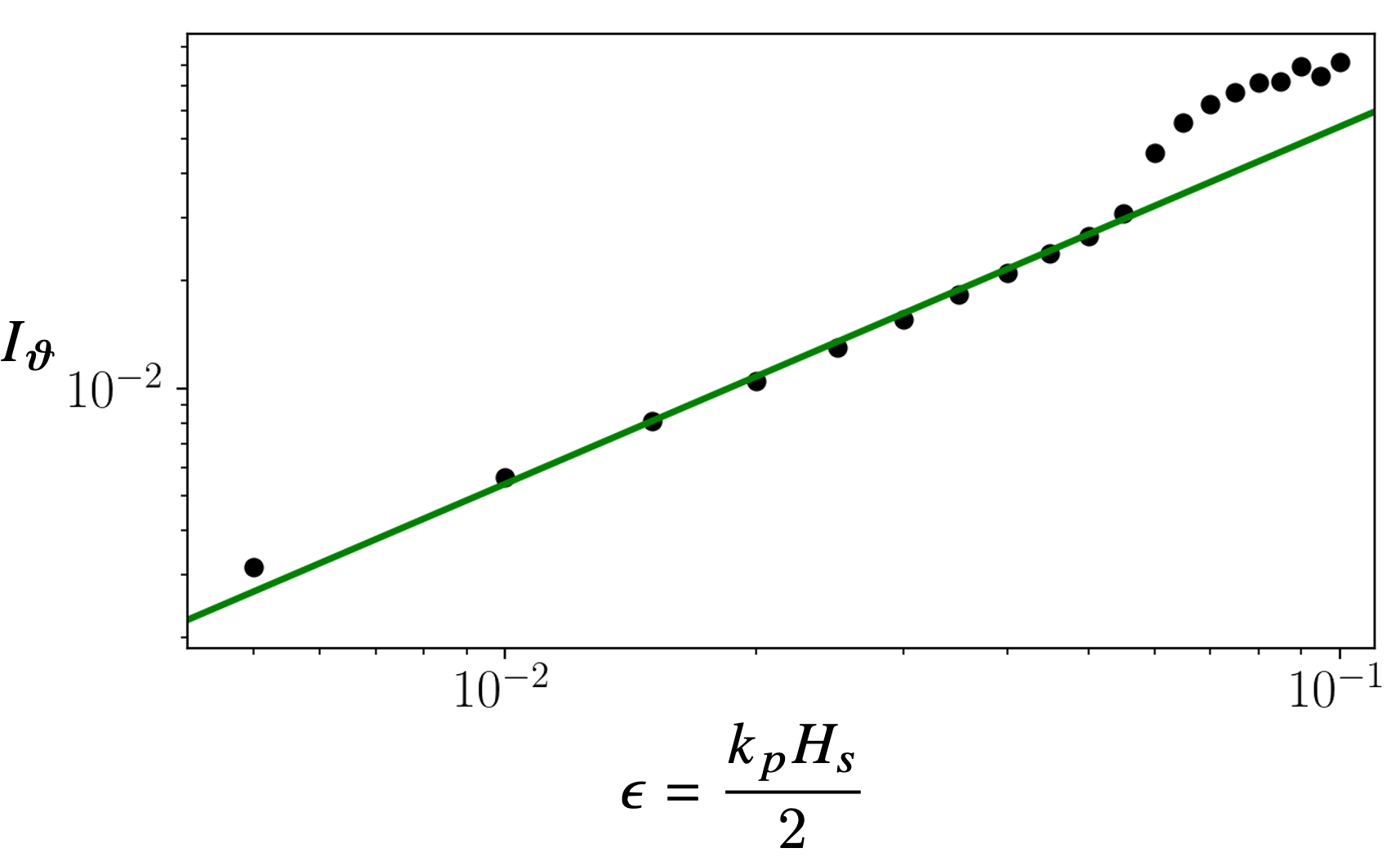}
    \caption{The uncertainty increment $I_{\vartheta}(\epsilon)-I_{\vartheta}(0)$ (\mycircle{black}) depending on steepness $\epsilon$, compared with linear scaling (\fullgreen{}). }
    \label{fig: integrated rho vs epsilon}
\end{figure}

To investigate the leading order effect that contributes to redistribution of uncertainty, We further quantify the scaling of this redistribution with respect to steepness $\epsilon$. We first define a `numerical predictable zone' $\mathcal{Z}$ for the linear case, which is the region on the $x$-$t$ plane that has $\vartheta<\vartheta_{c}$ . The threshold $\vartheta_{c}$ is chosen to be $0.01$, and the resulting $\mathcal{Z}$ is marked in figure \ref{fig: pred_zone_Z}. We can then use this $\mathcal{Z}$ to define the increase of uncertainty due to nonlinearity as, 
\begin{equation}\label{eq: integrated uncertainty}
    I_{\vartheta}=\sqrt{\int_{\mathcal{Z}}\vartheta^2(x,t;\epsilon)-\vartheta^2(x,t;0)\mathrm{d}x\mathrm{d}t}. 
\end{equation}
The quantity $I_{\vartheta}$ (figure \ref{fig: integrated rho vs epsilon}) characterizes the total change of uncertainty in this area $\mathcal{Z}$ due to the change of wave steepness $\epsilon$  (with $\mathcal{Z}$ obtained from the linear case and fixed for all $\epsilon$). The zero-steepness field $\vartheta(x,t;0)$ is obtained using linear wave dynamics. A clear linear growth of $I_{\vartheta}(\epsilon)$ with respect to $\epsilon$ is observed in the range $\epsilon\leq0.05$, indicating that the amount of uncertainty increase in $\mathcal{Z}$ is proportional to $\epsilon$. 
%\codex{[Check: Because $\epsilon=H_sk_p/2$ and $\vartheta=\rho/\varrho$, the literal zero-amplitude case has $\varrho=0$. Define $I_{\vartheta}(0)$ as a fixed-SNR $\epsilon\to0$ limit or as the linear-model baseline.]}
%\codex{[Check: State whether the normalized initial realization and noise realization are held fixed as $H_s$ (hence $\epsilon$) varies. Otherwise this sweep can mix steepness dependence with realization and noise dependence; alternatively report an ensemble statistic.]}
\del{Such linear dependence can be understood by considering the posterior covariance $\tilde{\mathbf{\Sigma}}^{(n)}=\mathbf{T}^{(n)}_{\star}(\epsilon)\mathbf{L}(\mathbf{I}+\mathbf{G}_{\star}(\epsilon))^{-1}\mathbf{L}^\dagger(\mathbf{T}^{(n)}_{\star}(\epsilon))^{\dagger}$ as a function of $\epsilon$.}
This linear growth can be understood by consider an asymptotic expansion of wave state vector $\mathbf{u}^{(n)}=\epsilon\mathbf{u}^{(n)}_0+\epsilon^2\mathbf{u}^{(n)}_1+\epsilon^3\mathbf{u}^{(n)}_2+O(\epsilon^4)$ with smallness parameter $\epsilon$, where $\mathbf{u}^{(n)}_0$ is the linear wave solution, and $\mathbf{u}^{(n)}_1$ and $\mathbf{u}^{(n)}_2$ are state vector corrections that correspond to quadratic and cubic nonlinearity. 
\del{In this expression, the leading-order term $\mathbf{T}^{(n)}_{\star,0}=\mathbf{A}^n$ is the tangent mapping of linear wave dynamics, and $\mathbf{T}^{(n)}_{\star,1}$ is the contribution from quadratic modal interactions. Expanding the full Hessian and the propagated covariance gives
$\vartheta_j^{(n)}(\epsilon)=\vartheta_j^{(n)}(0)+\epsilon Q_j^{(n)}+O(\epsilon^2)$, provided $\vartheta_j^{(n)}(0)>0$. The coefficient $Q_j^{(n)}$ includes both the first tangent correction and a residual-dependent contribution involving the second control derivative of the quadratic wave field. These contributions depend on the linear MAP solution and its measurement residuals, together with the prescribed spectrum and measurement configuration. Neither requires the second-order tangent coefficient $\mathbf{T}^{(n)}_{\star,2}$ (see appendix \ref{app: asymptotic expansion}).
}
Since the linear part $\epsilon \mathbf{u}^{(n)}_0$ does not cause $\vartheta$ to change, the approximately linear dependence in figure \ref{fig: integrated rho vs epsilon} should be primarily caused by the next term in the expansion, i.e. the quadratic nonlinearity corresponding to the non-resonant triad interactions. Beyond $\epsilon=0.05$ the data points in figure \ref{fig: integrated rho vs epsilon} deviates from the linear scaling. 
%There are two possible reasons for this deviation. First, 
The higher-order terms in the nonlinear wave evolution becomes non-negligible for larger $\epsilon$, therefore the dependence of the uncertainty increase $I_{\vartheta}(\epsilon)$ on $\epsilon$ is no longer linear.
%Second, the second term in the Hessian matrix (\ref{eq:hessian_nonlinear_exact}) might not be negligible because the second-order sensitivity along the dynamical system $\frac{\partial^2 \mathcal{N}^n}{\partial (\mathbf{u}^{(0)})^2}$ grows when the nonlinearity level increases.
%Second, the posterior distribution $\tilde{p}$ in equation (\ref{eq:laplace_posterior}) from Laplace approximation becomes less accurate when the nonlinearity level increases. 
Nevertheless, we have for the first time identified the physical effect that dominates the redistribution of uncertainty due to weak nonlinearity in the wave field.

%A closer examine of figures \ref{fig: nonlinear rho jonswap x t}$(b,c,d)$ shows that the low uncertainty streaks leaked from the predictable zone have slopes the same as the peak phase velocity $c_{p,\text{peak}}$. One possible explanation of these streaky structures can come from the second order bound waves generated from the $k_p$ mode. The wave field inside the predictable zone, including the peak mode, contains low uncertainty. This peak mode generates second order bound harmonics that also contain low uncertainty (since these bound harmonics fully depend on the peak mode). These bound modes are phase-locked with the peak mode, therefore propagate in the same phase speed as the peak mode and result in the low-uncertainty streaks in figures \ref{fig: nonlinear rho jonswap x t}$(b,c,d)$. 

%\codex{[Check: Add axis labels to figure \ref{fig: integrated rho vs epsilon}: horizontal $\epsilon$ and vertical $I_{\vartheta}(\epsilon)-I_{\vartheta}(0)$, and state that both axes are logarithmic if intended.]}

\section{Conclusions} \label{sec:conclusions}

In this work we consider the fundamental problem of predictable zone for ocean surface wave fields. The early theoretical construction of the predictable zone \citep{morris1998short,blondel2010deterministic,qi2018predictable} based on a group velocity argument has various limitations, including but not limited to (1) degeneracy for energy spectrum with unbounded frequency support (2) unable to characterize partially-predictable regions and (3) nonlinear effect not included. We establish a Bayesian inference framework to address all three limitations of existing works. We use a prior distribution to encode the existing knowledge about the wave field such as energy spectrum. By conditioning on the measurement data (possibly contaminated with noise), we infer the posterior distribution of the wave field, from which the posterior uncertainty of the surface elevation field is extracted and used as an indicator of local predictability. The numerical experiments not only verifies the consistency of our method with earlier theoretical construction of predictable zone, but also characterizes the predictability of partially-predictable regions and wave fields with unbounded frequency spectrum. 
Furthermore, we for the first time rigorously characterize the predictability of weakly nonlinear ocean surface wave fields.

With the newly developed Bayesian framework, we reach more refined understanding of predictable zones for one-dimensional ocean surface waves. For linear surface wave fields, there are two major effects that influence the predictability of the wave field (1) the amount of measurement data (2) the number of active modes in the frequency spectrum. We analyze these effects in detail in terms of the data-informed and un-informed solutions, with the former exactly determined and the latter being fundamentally uncertain at the limit of zero measurement noise. In particular, we find that the uncertainty of the wave field inside the classical predictable zone is in general nonzero. This nonzero uncertainty is due to the finite-time effect in the estimation of wave Fourier coefficients from measurement data. For nonlinear surface wave fields, we find that nonlinear effect redistributes uncertainty between the predictable zone and outside regions. The amount of uncertainty change scales linearly with wave steepness, which suggests that triad modal interaction is the dominant effect in this case. 

Our work opens multiple possibilities for future research. First, for real-life ocean wave measurements, it is necessary to extend the current framework to surface waves with two horizontal directions. Fast algorithms to compute the posterior covariance for such a problem with much larger state space could be meaningful directions to explore. Second, this framework could serve as a foundation for designing problems such as optimal sensor placements, where the goal could be defined as minimizing the posterior uncertainty of some `target quantity'. Third, the framework can be extended to analyze the wave predictability conditioned on more complex measurement data, such as ship body motions \citep{wang2025data} or LiDAR measurements \citep{wang2021phase}. 

\appendix

\section{Derivation of the posterior Gaussian form}
\label{app:posterior_gaussian}

We derive the Gaussian posterior stated in (\ref{eq:linear posterior}) by rewriting its exponent in the standard quadratic form of a Gaussian distribution. At time step $n$, the initial control vector $\boldsymbol{\xi}^{(0)}$ is first mapped to the wave state by $\mathbf{A}^{n}\mathbf{L}$ and then to the observed variables by $\mathbf{M}$. We denote this parameter-to-observation map by
\begin{equation}
    \mathbf{F}^{(n)} = \mathbf{M}\mathbf{A}^{n}\mathbf{L}.
\end{equation}
Thus, the observation predicted from $\boldsymbol{\xi}^{(0)}$ is $\mathbf{F}^{(n)}\boldsymbol{\xi}^{(0)}$, and its difference from the measurement $\mathbf{m}^{(n)}$ is the measurement residual $\mathbf{F}^{(n)}\boldsymbol{\xi}^{(0)}-\mathbf{m}^{(n)}$.

The prior for $\boldsymbol{\xi}^{(0)}$ is a zero-mean Gaussian with identity covariance $\mathbf{I}_{r_0\times r_0}$, while the measurement residual at each time step is Gaussian with covariance $\mathbf{R}$. Multiplying the prior density by the likelihoods according to Bayes' formula gives the posterior density, up to the normalization constant $C$:
\begin{eqnarray}
p(\boldsymbol{\xi}^{(0)}|\{\mathbf{m}^{(l)}\}_{l=N_{1},...,N_{2}})
&=&
\frac{1}{C}
\exp
\left[
-\frac{1}{2}
\left(\boldsymbol{\xi}^{(0)}\right)^{\dagger}
\boldsymbol{\xi}^{(0)}
\right.
\nonumber\\
&&
\left.
-\frac{1}{2}
\sum_{n=N_{1}}^{N_{2}}
\left(
\mathbf{F}^{(n)}\boldsymbol{\xi}^{(0)}-\mathbf{m}^{(n)}
\right)^{\dagger}
\mathbf{R}^{-1}
\left(
\mathbf{F}^{(n)}\boldsymbol{\xi}^{(0)}-\mathbf{m}^{(n)}
\right)
\right].
\label{eq:posterior_appendix_start}
\end{eqnarray}
To identify the posterior mean $\tilde{\boldsymbol{\xi}}^{(0)}$ and covariance $\tilde{\mathbf{\Sigma}}^{(0)}_{\boldsymbol{\xi}}$, we must express the exponent in (\ref{eq:posterior_appendix_start}) as one quadratic function of $\boldsymbol{\xi}^{(0)}$. For a single measurement time $n$, expansion of the negative measurement-misfit term gives
\begin{equation}
\begin{aligned}
&-\frac{1}{2}
\left(\mathbf{F}^{(n)}\boldsymbol{\xi}^{(0)}-\mathbf{m}^{(n)}\right)^{\dagger}
\mathbf{R}^{-1}
\left(\mathbf{F}^{(n)}\boldsymbol{\xi}^{(0)}-\mathbf{m}^{(n)}\right)
\\
&\qquad =
-\frac{1}{2}
\left(\boldsymbol{\xi}^{(0)}\right)^{\dagger}
\left(\mathbf{F}^{(n)}\right)^{\dagger}
\mathbf{R}^{-1}\mathbf{F}^{(n)}
\boldsymbol{\xi}^{(0)}
\\
&\qquad\quad +\Re\left[
\left(\boldsymbol{\xi}^{(0)}\right)^{\dagger}
\left(\mathbf{F}^{(n)}\right)^{\dagger}
\mathbf{R}^{-1}\mathbf{m}^{(n)}
\right]
\\
&\qquad\quad -\frac{1}{2}
\left(\mathbf{m}^{(n)}\right)^{\dagger}
\mathbf{R}^{-1}\mathbf{m}^{(n)}.
\label{eq:single_misfit_expansion}
\end{aligned}
\end{equation}
The first term on the right-hand side of (\ref{eq:single_misfit_expansion}) is quadratic in $\boldsymbol{\xi}^{(0)}$, the second is linear in $\boldsymbol{\xi}^{(0)}$, and the third is independent of $\boldsymbol{\xi}^{(0)}$. %The real-part operator $\Re[\cdot]$ appears because the two cross terms in the complex Fourier representation are conjugates of each other.

After summing (\ref{eq:single_misfit_expansion}) over all measurement times, the linear contribution contains the repeated sum
$\sum_{n=N_1}^{N_2}(\mathbf{F}^{(n)})^{\dagger}\mathbf{R}^{-1}\mathbf{m}^{(n)}$. We denote this specific sum by
\begin{equation}
    \mathbf{b} =
    \sum_{n=N_{1}}^{N_{2}}
    \left(\mathbf{F}^{(n)}\right)^{\dagger}
    \mathbf{R}^{-1}\mathbf{m}^{(n)},
\end{equation}
so that the linear contribution in (\ref{eq:single_misfit_expansion}) can be written compactly as
$\Re[(\boldsymbol{\xi}^{(0)})^{\dagger}\mathbf{b}]$. Adding the quadratic contribution from the prior, $-(\boldsymbol{\xi}^{(0)})^{\dagger}\boldsymbol{\xi}^{(0)}/2$, the terms in the exponent of the posterior (\ref{eq:posterior_appendix_start}) that depend on $\boldsymbol{\xi}^{(0)}$ are
\begin{eqnarray}
&&
-\frac{1}{2}
\left(\boldsymbol{\xi}^{(0)}\right)^{\dagger}
\left[
\mathbf{I}_{r_0\times r_0}
+
\sum_{n=N_{1}}^{N_{2}}
\left(\mathbf{F}^{(n)}\right)^{\dagger}
\mathbf{R}^{-1}
\mathbf{F}^{(n)}
\right]
\boldsymbol{\xi}^{(0)}
+
\Re
\left[
\left(\boldsymbol{\xi}^{(0)}\right)^{\dagger}
\mathbf{b}
\right]. 
\label{eq:posterior_appendix_expanded}
\end{eqnarray}
The remaining term
$-\sum_{n=N_{1}}^{N_{2}}(\mathbf{m}^{(n)})^{\dagger}\mathbf{R}^{-1}\mathbf{m}^{(n)}/2$
does not depend on $\boldsymbol{\xi}^{(0)}$. It changes the normalization of the posterior density but not its shape, and can therefore be absorbed into the normalization constant. Substituting $\mathbf{F}^{(n)}=\mathbf{M}\mathbf{A}^{n}\mathbf{L}$ into the quadratic coefficient in (\ref{eq:posterior_appendix_expanded}) gives
\begin{equation}
\mathbf{I}_{r_0\times r_0}+\mathbf{G},
\qquad
\mathbf{G}
=
\sum_{n=N_{1}}^{N_{2}}
\mathbf{L}^{\dagger}
(\mathbf{A}^n)^{\dagger}
\mathbf{M}^{\dagger}\mathbf{R}^{-1}\mathbf{M}
\mathbf{A}^n\mathbf{L},
\label{eq:posterior_precision_appendix}
\end{equation}
The matrix $\mathbf{G}$ is precisely the observability Gramian introduced in the main text. The identity matrix $\mathbf{I}_{r_0\times r_0}$ in (\ref{eq:posterior_precision_appendix}) is the prior precision, whereas $\mathbf{G}$ is the additional precision supplied by the measurements. Therefore, the total posterior precision is $\mathbf{I}_{r_0\times r_0}+\mathbf{G}$. Since the covariance of a Gaussian distribution is the inverse of its precision matrix, the posterior covariance is
\begin{equation}
\tilde{\mathbf{\Sigma}}^{(0)}_{\boldsymbol{\xi}}
=
\left(\mathbf{I}_{r_0\times r_0}+\mathbf{G}\right)^{-1}.
\label{eq:posterior_cov_appendix}
\end{equation}
It remains to determine the posterior mean $\tilde{\boldsymbol{\xi}}^{(0)}$. We do this by completing the square in the two terms of (\ref{eq:posterior_appendix_expanded}) that depend on $\boldsymbol{\xi}^{(0)}$. First, because
\begin{equation}
    \Re\left[\left(\boldsymbol{\xi}^{(0)}\right)^{\dagger}\mathbf{b}\right]
    =
    \frac{1}{2}\left(\boldsymbol{\xi}^{(0)}\right)^{\dagger}\mathbf{b}
    +
    \frac{1}{2}\mathbf{b}^{\dagger}\boldsymbol{\xi}^{(0)},
\end{equation}
the linear term can be represented by two conjugate cross terms. We then add and subtract the quantity
$\mathbf{b}^{\dagger}\tilde{\mathbf{\Sigma}}^{(0)}_{\boldsymbol{\xi}}\mathbf{b}/2$, which is independent of $\boldsymbol{\xi}^{(0)}$. This gives
\begin{equation}
\begin{aligned}
&-\frac{1}{2}
\left(\boldsymbol{\xi}^{(0)}\right)^{\dagger}
\left(\tilde{\mathbf{\Sigma}}^{(0)}_{\boldsymbol{\xi}}\right)^{-1}
\boldsymbol{\xi}^{(0)}
+\Re\left[\left(\boldsymbol{\xi}^{(0)}\right)^{\dagger}\mathbf{b}\right]
\\
&= -\frac{1}{2}\bigg[
\left(\boldsymbol{\xi}^{(0)}\right)^{\dagger}
\left(\tilde{\mathbf{\Sigma}}^{(0)}_{\boldsymbol{\xi}}\right)^{-1}
\boldsymbol{\xi}^{(0)}
-\left(\boldsymbol{\xi}^{(0)}\right)^{\dagger}\mathbf{b}
-\mathbf{b}^{\dagger}\boldsymbol{\xi}^{(0)}
+\mathbf{b}^{\dagger}\tilde{\mathbf{\Sigma}}^{(0)}_{\boldsymbol{\xi}}\mathbf{b}
\bigg]
\\
&\qquad
+\frac{1}{2}\mathbf{b}^{\dagger}
\tilde{\mathbf{\Sigma}}^{(0)}_{\boldsymbol{\xi}}\mathbf{b}
\\
&= -\frac{1}{2}
\left(
\boldsymbol{\xi}^{(0)}
-\tilde{\mathbf{\Sigma}}^{(0)}_{\boldsymbol{\xi}}\mathbf{b}
\right)^{\dagger}
\left(\tilde{\mathbf{\Sigma}}^{(0)}_{\boldsymbol{\xi}}\right)^{-1}
\left(
\boldsymbol{\xi}^{(0)}
-\tilde{\mathbf{\Sigma}}^{(0)}_{\boldsymbol{\xi}}\mathbf{b}
\right)
+\frac{1}{2}\mathbf{b}^{\dagger}
\tilde{\mathbf{\Sigma}}^{(0)}_{\boldsymbol{\xi}}\mathbf{b}.
\label{eq:posterior_appendix_compact}
\end{aligned}
\end{equation}
In the last equality, the four terms inside the square brackets have been factored into one quadratic expression, using the fact that the covariance $\tilde{\mathbf{\Sigma}}^{(0)}_{\boldsymbol{\xi}}$ is Hermitian and invertible. The final term
$\mathbf{b}^{\dagger}\tilde{\mathbf{\Sigma}}^{(0)}_{\boldsymbol{\xi}}\mathbf{b}/2$
does not depend on $\boldsymbol{\xi}^{(0)}$ and is absorbed into the normalization constant. The remaining quadratic expression depends on $\boldsymbol{\xi}^{(0)}$ only through its difference from
$\tilde{\mathbf{\Sigma}}^{(0)}_{\boldsymbol{\xi}}\mathbf{b}$. Therefore, the posterior mean is
\begin{equation}
\tilde{\boldsymbol{\xi}}^{(0)}
=
\tilde{\mathbf{\Sigma}}^{(0)}_{\boldsymbol{\xi}}\mathbf{b}
=
\tilde{\mathbf{\Sigma}}^{(0)}_{\boldsymbol{\xi}}
\sum_{n=N_{1}}^{N_{2}}
\mathbf{L}^{\dagger}
(\mathbf{A}^n)^{\dagger}
\mathbf{M}^{\dagger}\mathbf{R}^{-1}\mathbf{m}^{(n)}.
\label{eq:posterior_mean_appendix}
\end{equation}
Substituting the completed-square expression (\ref{eq:posterior_appendix_compact}) into (\ref{eq:posterior_appendix_start}) yields
\begin{eqnarray}
p(\boldsymbol{\xi}^{(0)}|\{\mathbf{m}^{(l)}\}_{l=N_{1},...,N_{2}})
&=&
\frac{1}{\tilde{C}}
\exp\left[
-\frac{1}{2}
\left(
\boldsymbol{\xi}^{(0)}-\tilde{\boldsymbol{\xi}}^{(0)}
\right)^{\dagger}
\left(\tilde{\mathbf{\Sigma}}^{(0)}_{\boldsymbol{\xi}}\right)^{-1}
\left(
\boldsymbol{\xi}^{(0)}-\tilde{\boldsymbol{\xi}}^{(0)}
\right)
\right],
\label{eq:posterior_appendix_standard}
\end{eqnarray}
The terms independent of $\boldsymbol{\xi}^{(0)}$ are included in the new normalization constant $\tilde{C}$. Equation (\ref{eq:posterior_appendix_standard}) is therefore the Gaussian form stated in (\ref{eq: linear posterior centered})-(\ref{eq: posterior mean cov}), with mean (\ref{eq:posterior_mean_appendix}) and covariance (\ref{eq:posterior_cov_appendix}).

\section{Derivation of the uncertainty decomposition }\label{app: uncertainty decomposition}

Here we derive the prior and posterior distribution of the wave states $\mathbf{u}^{(n)}_{I}$ and $\mathbf{u}^{(n)}_{u}$, then quantitatively demonstrate the properties of the covariances of these distributions mentioned in \S\ref{subsec: uncertainty decomposition}.

The complete eigenvector matrix $\mathbf{V}=[\mathbf{V}_{I}\ \mathbf{V}_{u}]$ is unitary. Its two blocks therefore satisfy
\begin{equation}
    \mathbf{V}_{I}^{\dagger}\mathbf{V}_{I}=\mathbf{I}_{r},
    \qquad
    \mathbf{V}_{u}^{\dagger}\mathbf{V}_{u}=\mathbf{I}_{r_0-r},
    \qquad
    \mathbf{V}_{I}^{\dagger}\mathbf{V}_{u}=\mathbf{0},
\end{equation}
and the associated orthogonal projection matrices are
\begin{equation}
    \mathbf{P}_{I}=\mathbf{V}_{I}\mathbf{V}_{I}^{\dagger},
    \qquad
    \mathbf{P}_{u}=\mathbf{V}_{u}\mathbf{V}_{u}^{\dagger},
    \qquad
    \mathbf{P}_{I}+\mathbf{P}_{u}=\mathbf{I}_{r_0\times r_0}.
\end{equation}
Consequently, the initial control vector $\boldsymbol{\xi}^{(0)}$ has the orthogonal decomposition
\begin{equation}
    \boldsymbol{\xi}^{(0)}=\boldsymbol{\xi}_{I}+\boldsymbol{\xi}_{u},
    \qquad
    \boldsymbol{\xi}_{I}=\mathbf{P}_{I}\boldsymbol{\xi}^{(0)},
    \qquad
    \boldsymbol{\xi}_{u}=\mathbf{P}_{u}\boldsymbol{\xi}^{(0)},
\end{equation}
which is the decomposition stated in (\ref{eq: xi decomp}).

We will repeatedly use the covariance rule for a linear transformation stated as follow. If a random vector $\mathbf{x}$ has mean $\boldsymbol{\mu}_{\mathbf{x}}$ and covariance $\mathbf{\Sigma}_{\mathbf{x}}$, and $\mathbf{B}$ is a deterministic matrix, then $\mathbf{y}=\mathbf{B}\mathbf{x}$ has covariance
\begin{equation}\label{eq: covariance linear transformation}
\begin{aligned}
    \operatorname{Cov}(\mathbf{y})=\operatorname{Cov}(\mathbf{B}\mathbf{x})
    &=
    \mathbb{E}\left[
    \mathbf{B}(\mathbf{x}-\boldsymbol{\mu}_{\mathbf{x}})
    (\mathbf{x}-\boldsymbol{\mu}_{\mathbf{x}})^{\dagger}
    \mathbf{B}^{\dagger}
    \right]\\
    &=
    \mathbf{B}\mathbf{\Sigma}_{\mathbf{x}}\mathbf{B}^{\dagger}.
\end{aligned}
\end{equation}
More generally, two transformations $\mathbf{B}\mathbf{x}$ and $\mathbf{C}\mathbf{x}$ have cross-covariance
$\operatorname{Cov}(\mathbf{B}\mathbf{x},\mathbf{C}\mathbf{x})=\mathbf{B}\mathbf{\Sigma}_{\mathbf{x}}\mathbf{C}^{\dagger}$.

Equation (\ref{eq: u decomp}) gives the direct linear mapping from $\boldsymbol{\xi}^{(0)}$ to $\mathbf{u}^{(n)}_{I}$ and $\mathbf{u}^{(n)}_{u}$. Under the rule (\ref{eq: covariance linear transformation}), the the prior covariance $\mathbf{I}_{r_0\times r_0}$ of $\boldsymbol{\xi}^{(0)}$ is transformed to the prior covariance of $\mathbf{u}^{(n)}_{I}$ and $\mathbf{u}^{(n)}_{u}$ as, 
\begin{subequations}\label{eq: prior cov split state appendix}
\begin{align}
    \mathbf{\Sigma}_{I}^{(n)}
    &=
    \mathbf{A}^{n}\mathbf{L}\mathbf{V}_{I}
    \mathbf{I}_{r}
    \mathbf{V}_{I}^{\dagger}\mathbf{L}^{\dagger}
    \left(\mathbf{A}^{\dagger}\right)^{n},\\
    \mathbf{\Sigma}_{u}^{(n)}
    &=
    \mathbf{A}^{n}\mathbf{L}\mathbf{V}_{u}
    \mathbf{I}_{r_0-r}
    \mathbf{V}_{u}^{\dagger}\mathbf{L}^{\dagger}
    \left(\mathbf{A}^{\dagger}\right)^{n},\\
    \mathbf{\Sigma}^{(n)}
    &=
    \mathbf{\Sigma}_{I}^{(n)}+\mathbf{\Sigma}_{u}^{(n)}.
\end{align}
\end{subequations}
The last equality contains no cross-covariance terms because
\begin{equation}
\begin{aligned}
    \operatorname{Cov}(\mathbf{u}_{I}^{(n)},\mathbf{u}_{u}^{(n)})
    &=
    \mathbf{A}^{n}\mathbf{L}
    \mathbf{P}_{I}\mathbf{P}_{u}^{\dagger}
    \mathbf{L}^{\dagger}(\mathbf{A}^{\dagger})^{n}
    =\mathbf{0}.
\end{aligned}
\end{equation}

We now similarly derive the posterior covariance of $\mathbf{u}^{(n)}_{I}$ and $\mathbf{u}^{(n)}_{u}$. Equation (\ref{eq: posterior mean cov}) gives the posterior covariance of $\boldsymbol{\xi}^{(0)}$ as
\begin{equation}
    \tilde{\mathbf{\Sigma}}^{(0)}_{\boldsymbol{\xi}}
    =
    \left(\mathbf{I}_{r_0\times r_0}+\mathbf{G}\right)^{-1}.
\end{equation}
Because $\mathbf{V}$ diagonalizes $\mathbf{G}$ and is unitary, the $\mathbf{I}_{r_0\times r_0}+\mathbf{G}$ before taking inversion can be written as
\begin{equation}
    \mathbf{I}_{r_0\times r_0}+\mathbf{G}
    =
    \mathbf{V}
    \begin{bmatrix}
        \mathbf{I}_{r}+\boldsymbol{\Lambda}_{I} & \mathbf{0}\\
        \mathbf{0} & \mathbf{I}_{r_0-r}+\boldsymbol{\Lambda}_{u}
    \end{bmatrix}
    \mathbf{V}^{\dagger}.
\end{equation}
Inverting this matrix amounts to inverting the two diagonal blocks, and therefore
\begin{equation}\label{eq: whitened posterio cov 2}
    \tilde{\mathbf{\Sigma}}^{(0)}_{\boldsymbol{\xi}}
    =
    \mathbf{V}_{I}
    \left(\mathbf{I}_{r}+\boldsymbol{\Lambda}_{I}\right)^{-1}
    \mathbf{V}_{I}^{\dagger}
    +
    \mathbf{V}_{u}
    \left(\mathbf{I}_{r_0-r}+\boldsymbol{\Lambda}_{u}\right)^{-1}
    \mathbf{V}_{u}^{\dagger}.
\end{equation}
Under the rule (\ref{eq: covariance linear transformation}) and the linear mappings (\ref{eq: u decomp}), the posterior covariance $\tilde{\mathbf{\Sigma}}^{(0)}_{\boldsymbol{\xi}}$ of $\boldsymbol{\xi}^{(0)}$ is transformed to the posterior covariance of $\mathbf{u}^{(n)}_{I}$ and $\mathbf{u}^{(n)}_{u}$ as
\begin{subequations}\label{eq: posterior cov split 1}
\begin{align}
    \tilde{\mathbf{\Sigma}}_{I}^{(n)}
    &=
    \mathbf{A}^{n}\mathbf{L}\mathbf{V}_{I}
    \left(\mathbf{I}_{r}+\boldsymbol{\Lambda}_{I}\right)^{-1}
    \mathbf{V}_{I}^{\dagger}\mathbf{L}^{\dagger}
    \left(\mathbf{A}^{\dagger}\right)^{n},\\
    \tilde{\mathbf{\Sigma}}_{u}^{(n)}
    &=
    \mathbf{A}^{n}\mathbf{L}\mathbf{V}_{u}
    \left(\mathbf{I}_{r_0-r}+\boldsymbol{\Lambda}_{u}\right)^{-1}
    \mathbf{V}_{u}^{\dagger}\mathbf{L}^{\dagger}
    \left(\mathbf{A}^{\dagger}\right)^{n},\\
    \tilde{\mathbf{\Sigma}}^{(n)}
    &=
    \tilde{\mathbf{\Sigma}}_{I}^{(n)}
    +
    \tilde{\mathbf{\Sigma}}_{u}^{(n)}.
\end{align}
\end{subequations}
%\codex{The posterior cross-covariance of the wave-state components is $\mathbf{A}^{n}\mathbf{L}\mathbf{P}_{I}\tilde{\mathbf{\Sigma}}^{(0)}_{\boldsymbol{\xi}}\mathbf{P}_{u}^{\dagger}\mathbf{L}^{\dagger}(\mathbf{A}^{\dagger})^{n}=\mathbf{0}$ because (\ref{eq: whitened posterio cov 2}) is block diagonal in the two eigenspaces. Thus, their covariances add without cross terms.}

The formulae (\ref{eq: posterior cov split 1}) and (\ref{eq: prior cov split state appendix}) now show explicitly how the measurements affect the uncertainty of $\mathbf{u}_{I}^{(n)}$ and $\mathbf{u}_{u}^{(n)}$. Since $\boldsymbol{\Lambda}_{u}=\mathbf{0}$,
$(\mathbf{I}_{r_0-r}+\boldsymbol{\Lambda}_{u})^{-1}=\mathbf{I}_{r_0-r}$, and therefore
$\tilde{\mathbf{\Sigma}}_{u}^{(n)}=\mathbf{\Sigma}_{u}^{(n)}$: the data-uninformed covariance is unchanged by conditioning on the measurements. In contrast, every eigenvalue in $\boldsymbol{\Lambda}_{I}$ is positive, so
\begin{equation}
    \left(\mathbf{I}_{r}+\boldsymbol{\Lambda}_{I}\right)^{-1}
    \preceq \mathbf{I}_{r}.
\end{equation}
Consequently, $\tilde{\mathbf{\Sigma}}_{I}^{(n)}\preceq\mathbf{\Sigma}_{I}^{(n)}$ \footnote{Semidefinite order is preserved by congruence transformation, i.e. if $\mathbf{A}\preceq \mathbf{B}$, then $\mathbf{C}^{\dagger}\mathbf{A}\mathbf{C}\preceq \mathbf{C}^{\dagger}\mathbf{B}\mathbf{C}$.}. Thus, the measurements reduce uncertainty in the data-informed component $\mathbf{u}_{I}^{(n)}$ and leave the data-uninformed component $\mathbf{u}_{u}^{(n)}$ unchanged.

The same eigenvalue representation explains the zero-noise limit. If $\mathbf{R}=\gamma^2\mathbf{I}$, then $\mathbf{G}(\gamma)=\gamma^{-2}\mathbf{G}_{1}$ is a function of $\gamma$, and $\mathbf{G}_{1}$ is the observability Gramian computed using $\gamma=1$. Each positive eigenvalue $\lambda_i(\gamma)$ of $\mathbf{G}(\gamma)$ can therefore be written as $\lambda_i(\gamma)=\gamma^{-2}\lambda_i(1)$, where $\lambda_i(1)>0$ are the eigenvalues of $\mathbf{G}_{1}$. The diagonal elements of $\left(\mathbf{I}_{r}+\boldsymbol{\Lambda}_{I}\right)^{-1}$ in (\ref{eq: posterior cov split 1}$(a)$) are, 
\begin{equation}
    \left(1+\lambda_i(\gamma)\right)^{-1}
    =
    \frac{\gamma^2}{\gamma^2+\lambda_i(1)}
    \rightarrow 0
    \qquad
    \text{as }\gamma\rightarrow 0.
\end{equation}
Consequently, the informed covariance $\tilde{\mathbf{\Sigma}}_{I}^{(n)}$ tends to zero as the measurement noise vanishes. The uninformed covariance $\tilde{\mathbf{\Sigma}}_{u}^{(n)}=\mathbf{\Sigma}_{u}^{(n)}$ remains unchanged. Noiseless measurements therefore remove uncertainty only from the informed wave-state component $\mathbf{u}_{I}^{(n)}$; they provide no information about $\mathbf{u}_{u}^{(n)}$.

\section{Derivation of the matrix exponential}\label{app: matrix exponential}

We finally derive the single-wavenumber propagator $\exp(\mathbf{Q}_{k}\Delta t)$ used to construct the discrete-time linear evolution operator $\mathbf{A}$. Because the Fourier modes $\hat{\eta}_{k}$ and $\hat{\psi}_{k}$ evolve independently of other wavenumbers under the linearized wave equations, it is sufficient to consider one wavenumber $k$. Equation (\ref{eq:linear wave dynamics}) for that mode can be written as the constant-coefficient system
\begin{equation}
    \frac{\mathrm{d}}{\mathrm{d}t}
    \begin{pmatrix}
        \hat{\eta}_{k}\\
        \hat{\psi}_{k}
    \end{pmatrix}
    =
    \mathbf{Q}_{k}
    \begin{pmatrix}
        \hat{\eta}_{k}\\
        \hat{\psi}_{k}
    \end{pmatrix},
    \qquad
    \mathbf{Q}_{k}
    =
    \begin{pmatrix}
        0 & |k|\\
        -g & 0
    \end{pmatrix}.
\end{equation}
The exact update over one time interval $\Delta t$ is generated by the matrix exponential $\exp(\mathbf{Q}_{k}\Delta t)$. Rather than diagonalizing $\mathbf{Q}_{k}$, we use the simple identity
\begin{equation}
    \mathbf{Q}_{k}^{2}
    =
    -g|k|\mathbf{I}
    =
    -\omega_k^2\mathbf{I},
    \qquad
    \omega_k=\sqrt{g|k|}.
\end{equation}
For the active positive wavenumbers $k$ in the independent control vector $\boldsymbol{\xi}^{(0)}$, $|k|=k$, so $\omega_k=\sqrt{gk}$ is the deep-water dispersion relation used in the main text. The identity $\mathbf{Q}_{k}^{2}=-\omega_k^2\mathbf{I}$ implies that every even power of $\mathbf{Q}_{k}$ is proportional to $\mathbf{I}$ and every odd power is proportional to $\mathbf{Q}_{k}$. Separating these powers in the exponential series and recognizing the cosine and sine series gives
\begin{eqnarray}
\exp(\mathbf{Q}_{k}\Delta t)
&=&
\sum_{j=0}^{\infty}
\frac{(\mathbf{Q}_{k}\Delta t)^{2j}}{(2j)!}
+
\sum_{j=0}^{\infty}
\frac{(\mathbf{Q}_{k}\Delta t)^{2j+1}}{(2j+1)!}
\nonumber\\
&=&
\mathbf{I}\cos(\omega_k\Delta t)
+
\frac{\mathbf{Q}_{k}}{\omega_k}\sin(\omega_k\Delta t),
\qquad k\ne0.
\end{eqnarray}
Substituting the entries of the single-mode evolution matrix $\mathbf{Q}_{k}$ into this expression yields the explicit one-step propagator $\exp(\mathbf{Q}_{k}\Delta t)$:
\begin{equation}
\exp(\mathbf{Q}_{k}\Delta t)
=
\begin{pmatrix}
\cos(\omega_k\Delta t)
&
\dfrac{|k|}{\omega_k}\sin(\omega_k\Delta t)
\\
-\dfrac{g}{\omega_k}\sin(\omega_k\Delta t)
&
\cos(\omega_k\Delta t)
\end{pmatrix}.
\end{equation}
Comparing the four entries of $\exp(\mathbf{Q}_{k}\Delta t)$ with the parameterization in (\ref{eq: propagator single k}) identifies the propagation coefficients $c_k$, $b_k$, and $d_k$:
\begin{equation}
c_k=\cos(\omega_k\Delta t),
\qquad
b_k=\frac{|k|}{\omega_k}\sin(\omega_k\Delta t),
\qquad
d_k=-\frac{g}{\omega_k}\sin(\omega_k\Delta t),
\end{equation}
which recovers (\ref{eq: linear evolution coeffs}). The derivation above assumes $k\ne0$ because it divides by $\omega_k$. If the zero wavenumber is retained, $\mathbf{Q}_{0}^{2}=\mathbf{0}$, so the exponential series terminates after its linear term and
$\exp(\mathbf{Q}_{0}\Delta t)=\mathbf{I}+\mathbf{Q}_{0}\Delta t$.

\section{Derivation of the measurement Hessian for nonlinear wave dynamics}
\label{app: nonlinear posterior derivation}

This appendix derives the measurement Hessian $\mathbf{H}_{\mathrm{m},\star}$ in (\ref{eq:laplace_precision}). Write $\mathbf{a}=[a_{k_{\mathrm{min}}},\ldots,a_{k_{\mathrm{max}}}]^{\mathsf T}$, so that $\boldsymbol{\xi}^{(0)}=[\mathbf{a};\mathbf{a}^{*}]$ and $(\boldsymbol{\xi}^{(0)})^{*}=[\mathbf{a}^{*};\mathbf{a}]$. We treat $\mathbf{a}$ and $\mathbf{a}^{*}$ as independent when differentiating, and restore their conjugate pairing after evaluation. For a real-valued scalar $f$, define
\begin{equation}
    \nabla_{\boldsymbol{\xi}} f
    :=
    \frac{\partial f}{\partial(\boldsymbol{\xi}^{(0)})^{*}}
    =
    \begin{bmatrix}
        \partial f/\partial\mathbf{a}^{*}\\
        \partial f/\partial\mathbf{a}
    \end{bmatrix},
    \qquad
    \nabla^2_{\boldsymbol{\xi}} f
    :=
    \frac{\partial}{\partial\boldsymbol{\xi}^{(0)}}
    \left(\frac{\partial f}{\partial(\boldsymbol{\xi}^{(0)})^{*}}\right).
\end{equation}
With this augmented convention, the gradient, chain rule, and product rule have the same matrix form as in real-valued calculus. In particular, the prior in (\ref{eq: prior}) satisfies $\nabla_{\boldsymbol{\xi}}\mathcal{J}_0=\boldsymbol{\xi}^{(0)}$ and $\nabla^2_{\boldsymbol{\xi}}\mathcal{J}_0=\mathbf{I}_{r_0\times r_0}$.

For convenience, we reproduce the measurement objective here:
\begin{equation}
    \mathcal{J}_{\mathrm{m}}(\boldsymbol{\xi}^{(0)})
    =\frac{1}{2}\sum_{n=N_1}^{N_2}
    \left[
    \mathbf{M}\mathcal{N}^{n}(\mathbf{L}\boldsymbol{\xi}^{(0)})
    -\mathbf{m}^{(n)}
    \right]^{\dagger}
    \mathbf{R}^{-1}
    \left[
    \mathbf{M}\mathcal{N}^{n}(\mathbf{L}\boldsymbol{\xi}^{(0)})
    -\mathbf{m}^{(n)}
    \right].
\end{equation}
Applying the chain rule to this expression gives
\begin{equation}\label{eq:gradient_nonlinear_appendix_wirtinger}
\begin{aligned}
    \nabla_{\boldsymbol{\xi}}\mathcal{J}_{\mathrm{m}}
    ={}&\sum_{n=N_1}^{N_2}
    \mathbf{L}^{\dagger}
    (\mathbf{T}^{(n)}_{\star})^{\dagger}
    \mathbf{M}^{\dagger}\mathbf{R}^{-1}
    \left[
    \mathbf{M}\mathcal{N}^{n}(\mathbf{L}\boldsymbol{\xi}^{(0)})
    -\mathbf{m}^{(n)}
    \right].
\end{aligned}
\end{equation}
Differentiating (\ref{eq:gradient_nonlinear_appendix_wirtinger}) once more then gives
\begin{equation}\label{eq:hessian_nonlinear_exact}
    \mathbf{H}_{\mathrm{m},\star}
    =\sum_{n=N_1}^{N_2}\left[
    \mathbf{L}^{\dagger}(\mathbf{T}^{(n)}_{\star})^{\dagger}
    \mathbf{M}^{\dagger}\mathbf{R}^{-1}\mathbf{M}
    \mathbf{T}^{(n)}_{\star}\mathbf{L}
    +\mathbf{S}^{(n)}_{\star}
    \right].
\end{equation}
The first term is the positive-semidefinite Gauss--Newton contribution. The second is the residual-dependent curvature contribution, 
\begin{equation}\label{eq:S_nonlinear}
    [\mathbf{S}^{(n)}_{\star}]_{ij}
    =
    \left[
    \left.
    \frac{\partial^2\mathcal{N}^{n}(\mathbf{L}\boldsymbol{\xi}^{(0)})}
    {\partial(\xi_i^{(0)})^{*}\partial\xi_j^{(0)}}
    \right|_{\boldsymbol{\xi}^{(0)}=\boldsymbol{\xi}^{(0)}_{\star}}
    \right]^{\dagger}
    \mathbf{M}^{\dagger}\mathbf{R}^{-1}
    \left[
    \mathbf{M}\mathcal{N}^{n}(\mathbf{L}\boldsymbol{\xi}^{(0)}_{\star})
    -\mathbf{m}^{(n)}
    \right].
\end{equation}
The Laplace approximation requires $\mathbf{H}_{\star}=\mathbf{I}_{r_0\times r_0}+\mathbf{H}_{\mathrm{m},\star}$ to be positive definite, which is satisfied at the numerically obtained MAP solutions in this work.

\del{
\section{Asymptotic expansion of $\vartheta$ with respect to $\epsilon$}\label{app: asymptotic expansion}

\codex{We derive the small-steepness expansion of the normalized posterior standard deviation used in \S\ref{subsec: results nonlinear}. The initial Laplace covariance is obtained from the full Hessian in (\ref{eq:hessian_nonlinear_exact}), including the residual-dependent term $\mathbf{S}^{(n)}_{\star}$, and is propagated by the tangent operator. Both the inference of the initial covariance and its subsequent propagation depend on steepness and must be expanded.}

\subsection{Amplitude normalization and the tangent expansion}

Consider a family of spectra with fixed shape and fixed $k_p$, for which only $H_s$ varies. Since $\epsilon=H_sk_p/2$, the spectrum, initial-state map and prior standard deviation satisfy
\begin{equation}\label{eq: asym amplitude scaling}
    S(k;\epsilon)=\epsilon^2\overline{S}(k),\qquad
    \mathbf{L}(\epsilon)=\epsilon\overline{\mathbf{L}},\qquad
    \varrho(\epsilon)=\epsilon\overline{\varrho},
    \qquad \epsilon>0.
\end{equation}
Here the barred quantities are independent of $\epsilon$, and
$\overline{\varrho}^{\,2}=(\overline{\mathbf{L}}\,\overline{\mathbf{L}}^{\dagger})_{jj}$ for a surface-elevation index $1\leq j\leq J$. The prescribed noise level $\gamma=0.01\varrho$ likewise gives
\begin{equation}\label{eq: asym noise scaling}
    \mathbf{R}(\epsilon)=\epsilon^2\overline{\mathbf{R}},
    \qquad \overline{\mathbf{R}}=(0.01\overline{\varrho})^2\mathbf{I}.
\end{equation}
Thus the signal-to-noise ratio is fixed in this limit. 

We assume that the amplitude-normalized measurements and the corresponding MAP solution form a smooth family as $\epsilon\to0$. This can be achieved for synthetic data by fixing the normalized initial realization and the normalized noise realization while varying $\epsilon$. The sensor configuration, observation times, spectral support and discretization are held fixed. 

We start from the usual weakly nonlinear expansion of the surface elevation and surface velocity potential \citep{mei2005theory},
\begin{equation}\label{eq: asym wave field series}
    \begin{pmatrix}\eta\\\psi\end{pmatrix}
    =\epsilon\begin{pmatrix}\eta_0\\\psi_0\end{pmatrix}
    +\epsilon^2\begin{pmatrix}\eta_1\\\psi_1\end{pmatrix}
    +\epsilon^3\begin{pmatrix}\eta_2\\\psi_2\end{pmatrix}
    +O(\epsilon^4).
\end{equation}
\codex{The leading fields $(\eta_0,\psi_0)$ satisfy the linear wave equations, while $(\eta_1,\psi_1)$ are the second-order fields generated by quadratic modal interactions. To connect this field expansion with the evolution map, define $\mathbf{u}^{(n)}_a$ by collecting the values of $\eta_a$ and $\psi_a$ at the grid points in exactly the same order as in $\mathbf{u}^{(n)}$ (\S\ref{sec: problem setup}). Equation (\ref{eq: asym wave field series}) then gives
\begin{equation}\label{eq: asym stacked state series}
    \mathbf{u}^{(n)}=\epsilon\mathbf{u}^{(n)}_0
    +\epsilon^2\mathbf{u}^{(n)}_1
    +\epsilon^3\mathbf{u}^{(n)}_2+O(\epsilon^4).
\end{equation}
For the initial state $\mathbf{u}^{(0)}=\epsilon\overline{\mathbf{u}}^{(0)}$, the evolved state is $\mathcal{N}^n(\epsilon\overline{\mathbf{u}}^{(0)})$, and each coefficient $\mathbf{u}^{(n)}_a$ is a function of $\overline{\mathbf{u}}^{(0)}$. Making this dependence explicit and dividing (\ref{eq: asym stacked state series}) by $\epsilon$ yields}

\begin{equation}\label{eq: asym state field series}
    \epsilon^{-1}\mathcal{N}^n(\epsilon\overline{\mathbf{u}}^{(0)})
    =\mathbf{u}^{(n)}_0(\overline{\mathbf{u}}^{(0)})
    +\epsilon\mathbf{u}^{(n)}_1(\overline{\mathbf{u}}^{(0)})
    +\epsilon^2\mathbf{u}^{(n)}_2(\overline{\mathbf{u}}^{(0)})
    +O(\epsilon^3),
\end{equation}
where $\overline{\mathbf{u}}^{(0)}=\mathbf{u}^{(0)}/\epsilon$ is the amplitude-normalized initial state and
$\mathbf{u}^{(n)}_0(\overline{\mathbf{u}}^{(0)})=\mathbf{A}^n\overline{\mathbf{u}}^{(0)}$.
The subscript indicates the order in the normalized expansion, whereas the superscript $(n)$ retains its meaning as the time-step index.

Differentiating (\ref{eq: asym state field series}) with respect to $\overline{\mathbf{u}}^{(0)}$ gives the tangent operator: the amplitude factors on the left-hand side cancel by the chain rule. Evaluating this derivative at the normalized MAP initial condition
$\overline{\mathbf{u}}^{(0)}_{\star}(\epsilon)
=\overline{\mathbf{L}}\boldsymbol{\xi}^{(0)}_{\star}(\epsilon)$ yields
\begin{equation}\label{eq: asym tangent series}
    \mathbf{T}^{(n)}_{\star}(\epsilon)
    =\mathbf{T}^{(n)}_{\star,0}
    +\epsilon\mathbf{T}^{(n)}_{\star,1}
    +\epsilon^2\mathbf{T}^{(n)}_{\star,2}+O(\epsilon^3),
    \qquad \mathbf{T}^{(n)}_{\star,0}=\mathbf{A}^n,
\end{equation}
with
\begin{equation}\label{eq: asym first tangent coefficient}
    \mathbf{T}^{(n)}_{\star,1}
    =\left.
    \frac{\partial\mathbf{u}^{(n)}_1}
         {\partial\overline{\mathbf{u}}^{(0)}}
    \right|_{\overline{\mathbf{u}}^{(0)}=\overline{\mathbf{u}}^{(0)}_{\star}(0)}
\end{equation}
\codex{corresponding to the first tangent correction that accounts for the quadratic, non-resonant triad interactions. The expansion is assumed to hold together with its initial-state derivatives over the fixed time interval considered. Since the leading tangent $\mathbf{A}^n$ is independent of the base state, the first-order change in the normalized MAP initial condition enters the tangent expansion only at order $\epsilon^2$.}

\codex{\subsection{Expansion of the residual-dependent Hessian term}

To evaluate $\mathbf{S}^{(l)}_{\star}$, the second control derivatives and the measurement residual must both be expanded. All control derivatives below are taken in independent real coordinates for the Fourier coefficients, or equivalently in their conjugate-paired representation with the corresponding adjoints. Write
\begin{equation}\label{eq: asym data map expansion}
\begin{aligned}
    \epsilon^{-1}\mathbf{m}^{(l)}(\epsilon)
    &=\overline{\mathbf{m}}^{(l)}_0
    +\epsilon\overline{\mathbf{m}}^{(l)}_1+O(\epsilon^2),\\
    \boldsymbol{\xi}^{(0)}_{\star}(\epsilon)
    &=\boldsymbol{\xi}^{(0)}_{\star,0}
    +\epsilon\boldsymbol{\xi}^{(0)}_{\star,1}+O(\epsilon^2).
\end{aligned}
\end{equation}
The normalized MAP residual is
\begin{equation}\label{eq: asym residual series}
    \overline{\mathbf{r}}^{(l)}_{\star}(\epsilon)
    :=\epsilon^{-1}\left[
    \mathbf{M}\mathcal{N}^l(\epsilon\overline{\mathbf{L}}
    \boldsymbol{\xi}^{(0)}_{\star}(\epsilon))
    -\mathbf{m}^{(l)}(\epsilon)\right]
    =\overline{\mathbf{r}}^{(l)}_0
    +\epsilon\overline{\mathbf{r}}^{(l)}_1+O(\epsilon^2),
\end{equation}
where (\ref{eq: asym state field series}) gives
\begin{equation}\label{eq: asym residual coefficients}
\begin{aligned}
    \overline{\mathbf{r}}^{(l)}_0
    &=\mathbf{M}\mathbf{A}^l\overline{\mathbf{L}}
    \boldsymbol{\xi}^{(0)}_{\star,0}-\overline{\mathbf{m}}^{(l)}_0,\\
    \overline{\mathbf{r}}^{(l)}_1
    &=\mathbf{M}\mathbf{A}^l\overline{\mathbf{L}}
    \boldsymbol{\xi}^{(0)}_{\star,1}
    +\mathbf{M}\mathbf{u}^{(l)}_1
    (\overline{\mathbf{L}}\boldsymbol{\xi}^{(0)}_{\star,0})
    -\overline{\mathbf{m}}^{(l)}_1.
\end{aligned}
\end{equation}
The leading MAP residual need not vanish, since the MAP solution balances measurement misfit against the prior penalty.

For brevity in the second-derivative contraction, define the state vectors
\begin{equation}\label{eq: asym second control derivatives}
    \mathbf{U}^{(l)}_{a,ij}
    :=\left.
    \frac{\partial^2\mathbf{u}^{(l)}_a
          (\overline{\mathbf{L}}\boldsymbol{\xi}^{(0)})}
    {\partial\xi_i^{(0)}\partial\xi_j^{(0)}}
    \right|_{\boldsymbol{\xi}^{(0)}=\boldsymbol{\xi}^{(0)}_{\star,0}},
    \qquad a=1,2.
\end{equation}
For the amplitude-scaled initial condition used here, $\mathbf{u}^{(l)}_1$ is quadratic in the initial state, so its second control derivative is independent of that state. The linear coefficient $\mathbf{u}^{(l)}_0$ has zero second derivative. Differentiating the physical state expansion and evaluating at the moving MAP point therefore gives
\begin{equation}\label{eq: asym physical second derivative}
    \left.
    \frac{\partial^2\mathcal{N}^l
    (\epsilon\overline{\mathbf{L}}\boldsymbol{\xi}^{(0)})}
    {\partial\xi_i^{(0)}\partial\xi_j^{(0)}}
    \right|_{\boldsymbol{\xi}^{(0)}=\boldsymbol{\xi}^{(0)}_{\star}(\epsilon)}
    =\epsilon^2\mathbf{U}^{(l)}_{1,ij}
    +\epsilon^3\mathbf{U}^{(l)}_{2,ij}+O(\epsilon^4).
\end{equation}
There is no order-$\epsilon^3$ contribution from shifting the evaluation point of the first term, because the third control derivative of the quadratic field is zero. The MAP shift does contribute through $\overline{\mathbf{r}}^{(l)}_1$ in (\ref{eq: asym residual coefficients}).

Substituting (\ref{eq: asym physical second derivative}), (\ref{eq: asym residual series}) and $\mathbf{R}^{-1}=\epsilon^{-2}\overline{\mathbf{R}}^{-1}$ into (\ref{eq:S_nonlinear}) yields
\begin{equation}\label{eq: asym residual hessian series}
    \mathbf{S}^{(l)}_{\star}(\epsilon)
    =\epsilon\mathbf{S}^{(l)}_{\star,1}
    +\epsilon^2\mathbf{S}^{(l)}_{\star,2}+O(\epsilon^3),
    \qquad \mathbf{S}^{(l)}_{\star,0}=\mathbf{0},
\end{equation}
with
\begin{equation}\label{eq: asym residual hessian coefficients}
\begin{aligned}
    [\mathbf{S}^{(l)}_{\star,1}]_{ij}
    &=(\mathbf{U}^{(l)}_{1,ij})^{\dagger}
    \mathbf{M}^{\dagger}\overline{\mathbf{R}}^{-1}
    \overline{\mathbf{r}}^{(l)}_0,\\
    [\mathbf{S}^{(l)}_{\star,2}]_{ij}
    &=(\mathbf{U}^{(l)}_{2,ij})^{\dagger}
    \mathbf{M}^{\dagger}\overline{\mathbf{R}}^{-1}
    \overline{\mathbf{r}}^{(l)}_0
    +(\mathbf{U}^{(l)}_{1,ij})^{\dagger}
    \mathbf{M}^{\dagger}\overline{\mathbf{R}}^{-1}
    \overline{\mathbf{r}}^{(l)}_1.
\end{aligned}
\end{equation}
Thus the residual-dependent term is generally $O(\epsilon)$ at fixed relative noise: its factors scale as $\epsilon^2$, $\epsilon^{-2}$ and $\epsilon$, respectively. If the leading residual vanishes, its first-order coefficient vanishes, but this is not assumed in the derivation.

The coefficient $\boldsymbol{\xi}^{(0)}_{\star,1}$ needed at second order can be obtained by expanding the MAP stationarity condition. The zeroth-order equation is the linear MAP equation, while the first-order equation gives
\begin{equation}\label{eq: asym map shift}
\begin{aligned}
    (\mathbf{I}+\mathbf{G}_{\star,0})
    \boldsymbol{\xi}^{(0)}_{\star,1}
    =-\sum_{l=N_1}^{N_2}\big\{
    &\overline{\mathbf{L}}^{\dagger}(\mathbf{A}^l)^{\dagger}
    \mathbf{M}^{\dagger}\overline{\mathbf{R}}^{-1}
    [\mathbf{M}\mathbf{u}^{(l)}_1
    (\overline{\mathbf{L}}\boldsymbol{\xi}^{(0)}_{\star,0})
    -\overline{\mathbf{m}}^{(l)}_1]\\
    &+\overline{\mathbf{L}}^{\dagger}
    (\mathbf{T}^{(l)}_{\star,1})^{\dagger}
    \mathbf{M}^{\dagger}\overline{\mathbf{R}}^{-1}
    \overline{\mathbf{r}}^{(l)}_0\big\}.
\end{aligned}
\end{equation}
Here $\mathbf{G}_{\star,0}$ is the linear likelihood Hessian specified below. No unknown nonlinear MAP derivative is needed to determine the first-order uncertainty coefficient.}

\subsection{Expansion of the inferred and propagated covariances}

\codex{Combining the tangent-product and residual-dependent contributions in (\ref{eq:hessian_nonlinear_exact}) gives the full likelihood-Hessian expansion
\begin{equation}\label{eq: asym gramian series}
    \mathbf{G}_{\star}(\epsilon)
    =\mathbf{G}_{\star,0}+\epsilon\mathbf{G}_{\star,1}
    +\epsilon^2\mathbf{G}_{\star,2}+O(\epsilon^3).
\end{equation}
The factors $\epsilon$ from the two initial-state maps cancel the factor $\epsilon^{-2}$ from the noise precision in the tangent-product term. The complete coefficients are consequently
\begin{equation}\label{eq: asym gramian coefficients}
\begin{aligned}
    \mathbf{G}_{\star,a}
    =\sum_{l=N_1}^{N_2}\bigg[\sum_{b=0}^{a}
    &\overline{\mathbf{L}}^{\dagger}
    (\mathbf{T}^{(l)}_{\star,b})^{\dagger}
    \mathbf{M}^{\dagger}\overline{\mathbf{R}}^{-1}\mathbf{M}
    \mathbf{T}^{(l)}_{\star,a-b}\overline{\mathbf{L}}
    +\mathbf{S}^{(l)}_{\star,a}\bigg],
    \quad a=0,1,2.
\end{aligned}
\end{equation}
In particular,
\begin{equation}\label{eq: asym first gramian coefficient}
\begin{aligned}
    \mathbf{G}_{\star,1}
    =\sum_{l=N_1}^{N_2}\big\{\overline{\mathbf{L}}^{\dagger}\big[
    &(\mathbf{T}^{(l)}_{\star,1})^{\dagger}
    \mathbf{M}^{\dagger}\overline{\mathbf{R}}^{-1}\mathbf{M}\mathbf{A}^{l}\\
    &+(\mathbf{A}^{l})^{\dagger}
    \mathbf{M}^{\dagger}\overline{\mathbf{R}}^{-1}\mathbf{M}
    \mathbf{T}^{(l)}_{\star,1}\big]\overline{\mathbf{L}}
    +\mathbf{S}^{(l)}_{\star,1}\big\}.
\end{aligned}
\end{equation}
The first-order coefficient therefore includes both the change in tangent sensitivity and the contraction of the quadratic wave-field curvature with the leading measurement residual, summed over the observation window.}

The initial posterior covariance has the expansion
\begin{equation}\label{eq: asym control covariance series}
    \tilde{\mathbf{\Sigma}}^{(0)}_{\boldsymbol{\xi}}(\epsilon)
    =\tilde{\mathbf{\Sigma}}^{(0)}_{\boldsymbol{\xi},0}
    +\epsilon\tilde{\mathbf{\Sigma}}^{(0)}_{\boldsymbol{\xi},1}
    +\epsilon^2\tilde{\mathbf{\Sigma}}^{(0)}_{\boldsymbol{\xi},2}
    +O(\epsilon^3).
\end{equation}
Equating coefficients in
$(\mathbf{I}+\mathbf{G}_{\star})\tilde{\mathbf{\Sigma}}^{(0)}_{\boldsymbol{\xi}}=\mathbf{I}$ gives
\begin{equation}\label{eq: asym inverse matching}
\begin{aligned}
    (\mathbf{I}+\mathbf{G}_{\star,0})
    \tilde{\mathbf{\Sigma}}^{(0)}_{\boldsymbol{\xi},0}&=\mathbf{I},\\
    (\mathbf{I}+\mathbf{G}_{\star,0})
    \tilde{\mathbf{\Sigma}}^{(0)}_{\boldsymbol{\xi},a}
    +\sum_{b=1}^{a}\mathbf{G}_{\star,b}
    \tilde{\mathbf{\Sigma}}^{(0)}_{\boldsymbol{\xi},a-b}&=\mathbf{0},
    \qquad a=1,2.
\end{aligned}
\end{equation}
Consequently,
\begin{equation}\label{eq: asym inverse coefficients}
\begin{aligned}
    \tilde{\mathbf{\Sigma}}^{(0)}_{\boldsymbol{\xi},0}
    &=(\mathbf{I}+\mathbf{G}_{\star,0})^{-1},\\
    \tilde{\mathbf{\Sigma}}^{(0)}_{\boldsymbol{\xi},1}
    &=-\tilde{\mathbf{\Sigma}}^{(0)}_{\boldsymbol{\xi},0}
    \mathbf{G}_{\star,1}\tilde{\mathbf{\Sigma}}^{(0)}_{\boldsymbol{\xi},0},\\
    \tilde{\mathbf{\Sigma}}^{(0)}_{\boldsymbol{\xi},2}
    &=\tilde{\mathbf{\Sigma}}^{(0)}_{\boldsymbol{\xi},0}\mathbf{G}_{\star,1}
    \tilde{\mathbf{\Sigma}}^{(0)}_{\boldsymbol{\xi},0}\mathbf{G}_{\star,1}
    \tilde{\mathbf{\Sigma}}^{(0)}_{\boldsymbol{\xi},0}
    -\tilde{\mathbf{\Sigma}}^{(0)}_{\boldsymbol{\xi},0}\mathbf{G}_{\star,2}
    \tilde{\mathbf{\Sigma}}^{(0)}_{\boldsymbol{\xi},0}.
\end{aligned}
\end{equation}
The order of the matrix factors is retained throughout. The identity contribution from the prior makes $\mathbf{I}+\mathbf{G}_{\star,0}$ positive definite even when the linear Gramian is rank deficient.

For the state covariance, we use an overbar solely to denote amplitude normalization,
$\overline{\tilde{\mathbf{\Sigma}}}^{(n)}(\epsilon)
=\epsilon^{-2}\tilde{\mathbf{\Sigma}}^{(n)}(\epsilon)$.
The covariance propagation formula (\ref{eq:posterior_covariance_propagation}) then gives
\begin{equation}\label{eq: asym state series}
    \overline{\tilde{\mathbf{\Sigma}}}^{(n)}(\epsilon)
    =\overline{\tilde{\mathbf{\Sigma}}}^{(n)}_0
    +\epsilon\overline{\tilde{\mathbf{\Sigma}}}^{(n)}_1
    +\epsilon^2\overline{\tilde{\mathbf{\Sigma}}}^{(n)}_2+O(\epsilon^3),
\end{equation}
where the coefficients through second order can be written as
\begin{equation}\label{eq: asym state coefficients}
    \overline{\tilde{\mathbf{\Sigma}}}^{(n)}_a
    =\sum_{b+c+d=a}
    \mathbf{T}^{(n)}_{\star,b}\overline{\mathbf{L}}
    \tilde{\mathbf{\Sigma}}^{(0)}_{\boldsymbol{\xi},c}
    \overline{\mathbf{L}}^{\dagger}(\mathbf{T}^{(n)}_{\star,d})^{\dagger},
    \qquad a=0,1,2.
\end{equation}
Here the sum is over non-negative integers $b,c,d$. In particular, the leading covariance and its first correction are
\begin{subequations}\label{eq: asym first state coefficients}
\begin{align}
    \overline{\tilde{\mathbf{\Sigma}}}^{(n)}_0
    ={}&\mathbf{A}^{n}\overline{\mathbf{L}}
    \tilde{\mathbf{\Sigma}}^{(0)}_{\boldsymbol{\xi},0}
    \overline{\mathbf{L}}^{\dagger}(\mathbf{A}^{n})^{\dagger},\\
    \overline{\tilde{\mathbf{\Sigma}}}^{(n)}_1
    ={}&\mathbf{T}^{(n)}_{\star,1}\overline{\mathbf{L}}
    \tilde{\mathbf{\Sigma}}^{(0)}_{\boldsymbol{\xi},0}
    \overline{\mathbf{L}}^{\dagger}(\mathbf{A}^{n})^{\dagger}\nonumber\\
    &+\mathbf{A}^{n}\overline{\mathbf{L}}
    \tilde{\mathbf{\Sigma}}^{(0)}_{\boldsymbol{\xi},0}
    \overline{\mathbf{L}}^{\dagger}(\mathbf{T}^{(n)}_{\star,1})^{\dagger}\nonumber\\
    &-\mathbf{A}^{n}\overline{\mathbf{L}}
    \tilde{\mathbf{\Sigma}}^{(0)}_{\boldsymbol{\xi},0}\mathbf{G}_{\star,1}
    \tilde{\mathbf{\Sigma}}^{(0)}_{\boldsymbol{\xi},0}
    \overline{\mathbf{L}}^{\dagger}(\mathbf{A}^{n})^{\dagger}.
\end{align}
\end{subequations}
\codex{The first two terms in the correction describe the change in covariance propagation. The third describes the change in the initial posterior covariance, with $\mathbf{G}_{\star,1}$ now containing both contributions in (\ref{eq: asym first gramian coefficient}). In particular, retaining the residual-dependent Hessian adds
\begin{equation}\label{eq: asym residual covariance correction}
    -\mathbf{A}^{n}\overline{\mathbf{L}}
    \tilde{\mathbf{\Sigma}}^{(0)}_{\boldsymbol{\xi},0}
    \left(\sum_{l=N_1}^{N_2}\mathbf{S}^{(l)}_{\star,1}\right)
    \tilde{\mathbf{\Sigma}}^{(0)}_{\boldsymbol{\xi},0}
    \overline{\mathbf{L}}^{\dagger}(\mathbf{A}^{n})^{\dagger}
\end{equation}
to the first-order propagated covariance relative to the Gauss--Newton approximation at the same MAP solution. The inverse and propagation formulas retain their form because they apply to the complete Hessian coefficients.}

\subsection{Pointwise and integrated uncertainty}

Let $\overline{\tilde{\Sigma}}^{(n)}_{a,jj}$ denote the $j$th diagonal entry of $\overline{\tilde{\mathbf{\Sigma}}}^{(n)}_a$, with $1\leq j\leq J$ selecting surface elevation. The amplitude factors cancel in the normalized standard deviation defined in (\ref{eq: normalized uncertainty}), so that
\begin{equation}\label{eq: asym sqrt expansion}
\begin{aligned}
    \vartheta_j^{(n)}(\epsilon)
    &=\frac{\sqrt{\overline{\tilde{\Sigma}}^{(n)}_{0,jj}
    +\epsilon\overline{\tilde{\Sigma}}^{(n)}_{1,jj}
    +\epsilon^2\overline{\tilde{\Sigma}}^{(n)}_{2,jj}+O(\epsilon^3)}}
    {\overline{\varrho}}\\
    &=\vartheta_j^{(n)}(0)+\epsilon Q_j^{(n)}
    +\frac{\epsilon^2}{\overline{\varrho}}\left[
    \frac{\overline{\tilde{\Sigma}}^{(n)}_{2,jj}}
    {2\sqrt{\overline{\tilde{\Sigma}}^{(n)}_{0,jj}}}
    -\frac{(\overline{\tilde{\Sigma}}^{(n)}_{1,jj})^2}
    {8(\overline{\tilde{\Sigma}}^{(n)}_{0,jj})^{3/2}}\right]
    +O(\epsilon^3),
\end{aligned}
\end{equation}
where
\begin{equation}\label{eq: asym Q coefficient}
    \vartheta_j^{(n)}(0)
    =\frac{\sqrt{\overline{\tilde{\Sigma}}^{(n)}_{0,jj}}}{\overline{\varrho}},
    \qquad
    Q_j^{(n)}
    =\frac{\overline{\tilde{\Sigma}}^{(n)}_{1,jj}}
    {2\overline{\varrho}^{\,2}\vartheta_j^{(n)}(0)}.
\end{equation}
\codex{Equations (\ref{eq: asym residual hessian coefficients}), (\ref{eq: asym first gramian coefficient}), (\ref{eq: asym inverse coefficients}) and (\ref{eq: asym first state coefficients}) determine $Q_j^{(n)}$ from the linear propagator, the first tangent correction, the second control derivative of the quadratic wave field, and the leading MAP residual. The prescribed spectrum and measurement configuration enter through $\overline{\mathbf{L}}$, $\mathbf{M}$ and $\overline{\mathbf{R}}$. No second-order tangent coefficient $\mathbf{T}^{(l)}_{\star,2}$ or cubic wave-field coefficient $\mathbf{u}^{(l)}_2$ enters $Q_j^{(n)}$. A second derivative with respect to the control thus contributes at first order in steepness and must not be confused with a second-order steepness correction.

The square-root expansion assumes $\overline{\tilde{\Sigma}}^{(n)}_{0,jj}>0$. When it holds uniformly on the fixed region $\mathcal{Z}$, integration gives
\begin{equation}\label{eq: asym integrated uncertainty}
    I_{\vartheta}(\epsilon)-I_{\vartheta}(0)
    =\epsilon\int_{\mathcal{Z}}Q(x,t)\,\mathrm{d}x\,\mathrm{d}t
    +O(\epsilon^2).
\end{equation}
Retaining the full Hessian therefore changes the coefficient of the leading uncertainty correction without generally changing its order in steepness. Both first-order contributions arise from the quadratic wave field, consistent with the interpretation in terms of non-resonant triad interactions. Their integral need not be positive or nonzero; its sign and magnitude depend on the MAP solution, measurement residuals and observation configuration.}
}

\bibliographystyle{jfm}
\bibliography{reference}

\end{document}